\documentclass[%
 pof,
 sd,%
 amsmath,amssymb,
reprint,%
longbibliography,
]{revtex4-1}

\usepackage{graphicx}% Include figure files
\usepackage{dcolumn}% Align table columns on decimal point
\usepackage{bm}% bold math
\usepackage{subcaption}
\usepackage{stmaryrd}

\usepackage[linktocpage=true,colorlinks=true,linkcolor=blue,citecolor=blue]{hyperref}
\draft % marks overfull lines with a black rule on the right

\begin{document}

\title{Influence of thermal effects on stability of nanoscale films and filaments on thermally conductive
substrates} %Title of paper

\author{Ivana Seric}
\author{Shahriar Afkhami}
\author{Lou Kondic}

\affiliation{Department of Mathematical Sciences, New Jersey Institute of Technology, Newark, NJ, USA}

\date{\today}

\begin{abstract}
We consider films and filaments of nanoscale thickness on thermally conductive substrates 
 exposed to external heating.   Particular 
focus is on metal films exposed to laser irradiation.  Due to 
short length scales involved, the absorption of heat in the metal is directly coupled to the film 
evolution, since the absorption length and the film thickness are comparable.  
Such a setup requires self-consistent consideration of fluid mechanical and 
thermal effects.   We approach the problem via  Volume-of-Fluid
based simulations that include destabilizing liquid metal-solid substrate interaction 
potentials.   These simulations  couple fluid dynamics directly with the spatio-temporal 
evolution of the temperature field both in the fluid and in the substrate.    We focus  on the influence of 
the temperature variation of material parameters, in particular 
of surface tension and viscosity.    Regarding 
variation of surface tension with temperature, the main finding is that while Marangoni effect
may not play a significant role in the considered setting,  the temporal variation of surface tension 
(modifying normal stress balance) is significant and could lead to complex 
evolution including oscillatory evolution of the liquid metal-air interface.   Temperature
variation of film viscosity is also found to be relevant.  Therefore, the variations of surface tensions
and viscosity could both influence the emerging wavelengths in experiments.   
In contrast, the filament geometry is found to be much less sensitive to a variation of 
material parameters with temperature. 
\end{abstract}


\maketitle 
\section{Introduction}

Metal films of nanoscale thickness are of interest in numerous applications  including solar cells, plasmonics related applications, 
sensing and detection among others. These applications include various geometries: particles, films, filaments and more complicated 
shapes. For a recent application-centered review, see~\cite{hughes17,Makarov2016} for recent application-centered reviews.  These films are typically exposed to a heat source (pulsed nanosecond laser) and, while in the liquid phase, evolve on a time scale measured in nanoseconds.  This evolution is typically unstable 
due to the presence of destabilizing forces, in particular involving liquid metal-solid substrate
interactions~\cite{bischof_prl96}.  In addition to its scientific interest, 
understanding these instabilities and the subsequent dynamics is further motivated by their potential to drive various self- and directed-assembly 
mechanisms in a variety of contexts; only some examples are cited here~\cite{ross10,favazza_apl06,chou_09,krishna_nanotech10,Nanolett14,reinhardt13}. 

A significant progress has been reached in understanding the instability 
mechanisms by considering essentially isothermal models that assume films and
other geometries under isothermal conditions~\cite{kondic2009pre,lang13,gdk_jfm13,Kondic2015,POF2016}.
In particular for the film geometry, the film-substrate
interaction forces are
crucial: a destabilizing force is needed for the instability to develop.   In our earlier
works, such a force has been included in the Navier-Stokes solver~\cite{Mahady2015a,Mahady2015b}, 
and the implemented approach is used in the present work as well.   For filaments or other 
geometries that are characterized by limited spatial extent and the presence of contact lines, capillary effects are known 
to be dominant.   In particular, for the commonly considered filament geometry, 
it has been shown that simply considering Rayleigh-Plateau instability mechanism
with appropriately chosen material parameters (contact angle in particular) leads
to satisfactory results, see, e.g.~\cite{dk_pof09}.

Clearly, the evolution of metal films and other geometries exposed to laser radiation
is more complicated than that of an isothermal film.  A laser pulse leads to a 
significant variation of the temperature field both in the film and in the substrate, resulting 
in  phase change, and in variation of material parameters.   The influence of such variation of 
material parameters with temperature on the stability of films and other metal geometries has been 
considered only to the limited extend so far~\cite{trice_prb07,trice_prl08, atena09,krishna_nanotech10,DK2016}.  
Furthermore, the existing studies, some of which we discuss next, have 
focused mostly on Marangoni effects, related to 
spatial variation of the film surface tension with temperature.  We are not aware of any works in the present 
context that focus on the influence of temporal variation of surface 
tension (that is clearly relevant in the case of a time dependent laser pulse), or on the
influence of  variation of the film viscosity with temperature.   These effects are among those discussed
in the present work. 

The Marangoni effect results from the spatial variation of tangential stresses due to 
temperature dependence of surface tension. 
In \cite{trice_prl08},  Marangoni effect is claimed to be responsible for the change of
the average distances between the drops that form during pulse irradiation
of metal films.
For computing the temperature field, Ref.~\cite{trice_prl08} used the model, that we reference in the 
remaining part of this paper as `reduced' model, that includes two important assumptions (i) that 
the temperature field is
slaved to the film thickness, meaning essentially that the temperature is completely 
defined by the current value of the film thickness, and (ii) that the heat flux in the 
plane of the substrate can be ignored.    Next, in a recent modeling and computational study~\cite{DK2016}, the Marangoni effect was
considered (within  long wave limit)  in a setup that relaxed the assumption (i) above, but still used the
assumption (ii).   In that work, it was found that the results were dramatically
different compared to the ones obtained if the assumption (i) was used.   In particular, a regime
characterized by an oscillatory instability development has been found.   This is in 
contrast to usual monotonic nature of instability evolution.    

To summarize, the influence of thermal effects on the evolution of metal films and other 
geometries has not been studied extensively, and the results of the existing works 
are not always consistent.    This motivates the present paper that focuses on providing further 
insight by carrying out careful and self-consistent 
simulations of the evolution of metal films and filaments.   The considered approach 
is mainly computational and is based on a Volume-of-Fluid (VoF) method for solving the 
fluid mechanical problem, coupled with a thermal solver that computes the temperature
field in the metal and in the substrate.    Thin film (long wave) limit is used only for the
purpose of obtaining basic insight via linear stability analysis, and the thermal problem 
is solved fully and self-consistently with the Navier-Stokes equations governing thin film 
evolution.  In particular, the thermal problem considers both the in-plane and out-of-the-plane
heat transport: we will see that for the setup considered, considering the in-plane heat
diffusion is crucial.    The VoF solver includes  the 
interaction between metal and the substrate modeled via disjoining pressure approach~\cite{Mahady2015b}, as well
as efficient calculation of tangential stresses and the resulting Marangoni effect that has 
been developed recently~\cite{seric2017}.   We note that in the present work, we discuss the phase 
change effects only indirectly, by limiting the metal film and filament evolution
to the times for which metal temperature is above melting.    Furthermore, we do not 
consider possible phase change of the substrate itself~\cite{font2017}.     Inclusion of these 
effects is left for future work.   

The remainder of this paper is organized as follows.   
First, in Section~\ref{sec:equations} we formulate the governing equations coupling 
fluid mechanics with the thermal effects.  
We outline two temperature solutions used in the study of film breakup: first in 
Section~\ref{sec:Trice}, we present a reduced model that ignores the in-plane heat 
conduction, as well as temporal evolution of the film or filament (referred to 
as the `reduced'  model' from now on) \cite{trice_prb07}; and second in Section~\ref{sec:dns_model}, we 
present the numerical 2D temperature solution computed using {\sc Gerris} 
(referred to as the `complete' model'). 
In Section~\ref{sec:exact}, we outline an analytical solution for a flat film, 
which we use for validating and comparing the two models above.
For definiteness, throughout the paper we use the parameters
corresponding to nickel at the melting temperature, if not specified differently.  

Section~\ref{sec:results} presents the main findings.   First, in Section~\ref{sec:LSA}, 
using linear stability analysis (LSA) (in long wave limit), we find 
that the spatial temperature variations in the film can have a stabilizing or
destabilizing effect depending on the film thickness:
for films thinner than a critical value $h_c$, the 
temperature variations have a stabilizing effect, and for the films of thickness larger 
than $h_c$, the temperature variations have a destabilizing effect;  such a 
critical value appears due to the nature of absorption of the laser energy by the film.
Second, in Section~\ref{sec:film}, using the direct numerical simulations, 
we consider the influence of temperature variation of surface tension on stability  of 
a film, for the reduced and complete temperature models. We find that for the reduced model the 
relevance of Marangoni effect is exaggerated due to the lack of in-plane
heat conduction.  Furthermore, we find that, compared to the 
{\it spatial} variations leading to Marangoni effect, the {\it temporal} variation of surface tension has 
significantly stronger effect on the film stability;  in other words, the balance of normal stresses (and its dependence
on temperature) play much more important role than the variation of tangential stresses that lead to 
Marangoni effect. The temporal variation of the temperature  is found also to influence the film evolution 
via temperature dependence of viscosity; this effect is 
discussed in Section~\ref{sec:viscosity}.   Section~\ref{sec:discussion} is devoted to a brief
discussion of the expected influence of temperature variation of material properties in 
physical experiments.
Finally, in Section~\ref{sec:filaments}, we consider the influence of the 
thermal effects on the breakup of the liquid metal filaments. We find that the 
influence of thermal effects is weak compared to the capillary ones governing
 the Rayleigh-Plateau type of instability.  The paper is concluded by 
Section~\ref{sec:conclusions}, where we also discuss some directions for the future work.

\section{Governing Equations and Numerical Methods}
\label{sec:equations}

\begin{figure}
\centering
\includegraphics[width=0.45\textwidth]{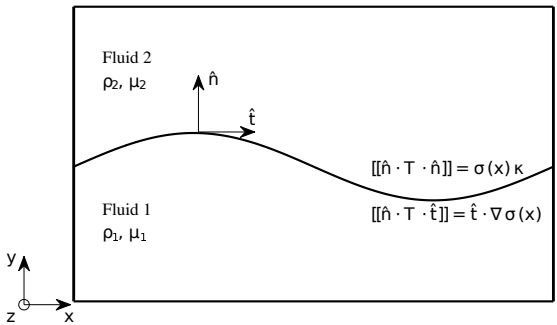}
\caption{Schematic of a system with two immiscible fluids and the corresponding boundary conditions.}
\label{fig:setup}
\end{figure}

We represent the two-phase flow by the Navier-Stokes equations, where the material 
properties are phase dependent; additional energy equation is discussed later in this
section.   The surface forces at the interface between two 
fluids are represented by a body force using the Continuum Surface Force (CSF) method \cite{Brackbill92,FCDKSW2006,Afkhami2008,Popinet2009a}. 
Hence, the equations governing the flow are 
\begin{gather}
\label{eq:Fma}
  \rho(\partial_t \textbf{u} + \textbf{u}\cdot \nabla \textbf{u}) = 
     - \nabla p + \nabla \cdot \left( 2 \mu D\right)
     + \textbf{F}
     , \\
  \nabla\cdot\textbf{u} = 0,
\end{gather}
and the advection of the phase-dependent density $\rho\left( \chi \right)$
\begin{equation}
  \label{eq:adv_rho}
  \partial_t \rho + (\textbf{u} \cdot \nabla)\rho = 0, \,
\end{equation}
where $\textbf{u} = (u, v, w)$ is the fluid velocity, $p$ is the pressure, $ \rho
(\chi) = \chi \rho_1 + (1-\chi)\rho_2$ and  $\mu (\chi) = \chi \mu_1 + 
(1-\chi)\mu_2$  are the phase dependent density and viscosity respectively.
$D$ is the rate of deformation tensor, $D = \left(\nabla \textbf{u} + 
\nabla\textbf{u}^T\right)/2$.
Subscripts $1$ and $2$ correspond to the fluids $1$ and $2$, respectively (see Fig.~\ref{fig:setup}).
Here, $\chi$ is the characteristic function, such that $\chi = 1$ in 
the fluid $1$, and $\chi = 0$ in the fluid $2$. Note that any body force can be 
included in $\textbf{F}$. The characteristic 
function is advected with the flow, thus
\begin{equation}
 \label{eq:chi}
 \partial_t \chi + (\textbf{u} \cdot \nabla)\chi = 0.
\end{equation}
Note that solving Eq.~\eqref{eq:chi} is equivalent to solving Eq.~\eqref{eq:adv_rho}.

The presence of an interface gives rise to the stress boundary conditions, see 
Fig.~\ref{fig:setup}. The normal stress boundary condition at the interface defines the stress jump \citep{landau1987, levich1969}
\begin{gather}
 \llbracket \hat{\mathbf{n}}\cdot \mathbf{T}\cdot \hat{\mathbf{n}}\rrbracket = \sigma \left( \mathbf{x} \right) \kappa,
\end{gather} 
where $ \mathbf{T} = -p \mathbf{I} + \mu \left(\nabla \textbf{u} + \nabla\textbf{u}^T\right)$
is the total stress tensor, $\sigma \left( \mathbf{x} \right)$ is the surface tension,
$\kappa$ is the curvature of the interface, and $\hat{\mathbf{n}}$ 
is the unit normal at the interface pointing out of the fluid $1$.
The variation of surface tension  results in the tangential 
stress jump at the interface
\begin{equation}
\llbracket \hat{\mathbf{n}}\cdot \mathbf{T}\cdot \hat{\mathbf{t}}\rrbracket = 
		 \hat{\mathbf{t}}\cdot\mathbf{\nabla} \sigma \left( \mathbf{x} \right),
\end{equation}
which drives the flow from the regions of low surface tension to the ones with high surface tension.
Here, $\hat{\mathbf{t}}$ is the unit tangent vector in two dimensions (2D); 
in three dimensions (3D) there are two linearly independent unit tangent vectors.
Using the CSF method \citep{Brackbill1992}, the forces
resulting from the normal and tangential stress jump at the interface can be 
included in the body force $\textbf{F} = \textbf{F}_{sn} + \textbf{F}_{st}$, 
defined as  
\begin{equation}
\label{eq:surf_fn}
 \textbf{F}_{sn} = \sigma \left( \mathbf{x} \right) \kappa \delta_s \hat{\mathbf{n}},
\end{equation}
and 
\begin{equation}
\label{eq:surf_ft_nabla}
 \textbf{F}_{st} =\mathbf{\nabla}_s \sigma \left( \mathbf{x} \right) \delta_s,
\end{equation}
where $\delta_s$ is the Dirac delta function centered at the interface, 
$ \delta_s \hat{\mathbf{n}} = \nabla \chi $, and $\mathbf{\nabla}_s$ is the surface gradient.
The details of the computations of the interfacial curvature and normals in the VoF method 
can be found in \citep{Popinet2009a}, and the implementation of the surface gradients
of the surface tension in \citep{seric2017}. 

The destabilizing mechanism leading to breakup of nanoscale films is 
modeled by the fluid-solid interaction in the form of a disjoining 
pressure \cite{POF2016}. The disjoining pressure can be included in the Navier-Stokes 
equations \eqref{eq:Fma} as a body force specified by 
\begin{gather}
\label{eq:Fwdv}
     \mathbf{F_{vdw}} \left( y \right) = K_{\pi} 
     \left[ \left( \frac{h^*}{y} \right)^m - \left( \frac{h^*}{y} \right)^n \right] 
\hat{\mathbf{n}}\delta_s, \\
\label{eq:K_pi}
    K_\pi = \frac{\sigma_0(1 - \cos{\theta_{eq})}}{M h_*},\\ 
\label{eq:M_vdw}
    M = \frac{n-m}{(m-1)(n-1)} 
\end{gather}
where $\sigma_0$ is the surface tension  of nickel at the melting temperature, and
$\theta_{eq} = 70^\circ$ is the prescribed equilibrium contact angle.  We 
use the exponents $m = 3$ and $n = 2$ as in~\cite{POF2016};  see also~\cite{lang13,ajaev_pof03} 
and the references therein for further discussion regarding disjoining pressure models for metal 
films.    The equilibrium film thickness used is $h_* = 1.5$ nm, 
comparable to the values discussed in~\cite{POF2016}.
Hence, the complete Navier-Stokes equations with the surface forces are as follows
\begin{multline}
\label{eq:Fma_vdw}
 \rho(\partial_t \textbf{u} + \textbf{u}\cdot \nabla \textbf{u}) = 
     - \nabla p + \nabla \cdot \left( 2 \mu D\right)
     + \sigma  \kappa \delta_s \hat{\mathbf{n}}  + \\
      \mathbf{\nabla}_s \sigma \delta_s + \mathbf{F_{vdw}} \left( y \right) 
\end{multline}
The details of the implementation of the disjoining pressure in the VoF method 
can be found in \citep{Mahady2015a}.

\subsection{Temperature Models}
\label{sec:temp_models}
The variations of the temperature of a liquid metal film exposed to a pulsed 
laser can be caused by spatial variability of the pulse itself.
However, since the spatial scale of the pulse is typically few orders of magnitude larger than 
any other relevant length scale~\cite{lang13}, we consider a spatially homogenous pulse 
where a temperature variation may result from variable film thickness.  
This is due to fact that the optical (and energy absorption) 
properties of the metal depend on the film thickness~\cite{Heavens}, as we will discuss 
in what follows.   

We start this section by considering a simplified (`reduced') model
that ignores various aspects of the heat flow, such as heat transport in the in-plane 
direction, as well as the convective effects.  In Section~\ref{sec:Trice} a numerical solution 
of such a model, that has been previously used in the context of metal films~\cite{trice_prb07,trice_prl08} is discussed, 
and in Section~\ref{sec:exact}, an analytical solution of the underlying one-dimensional model for heat transport is presented.  
Complete numerical solution of the energy equation is given in Section~\ref{sec:dns_model}. 
The comparison of the results obtained using
these approaches will allow to gain better insight into the most important factors determining  
temperature distribution.

\subsubsection{Reduced Model for the Temperature of a Thin Film }
\label{sec:Trice}

\begin{figure}[tbh]
\centering
  \includegraphics[width = 0.45\textwidth]{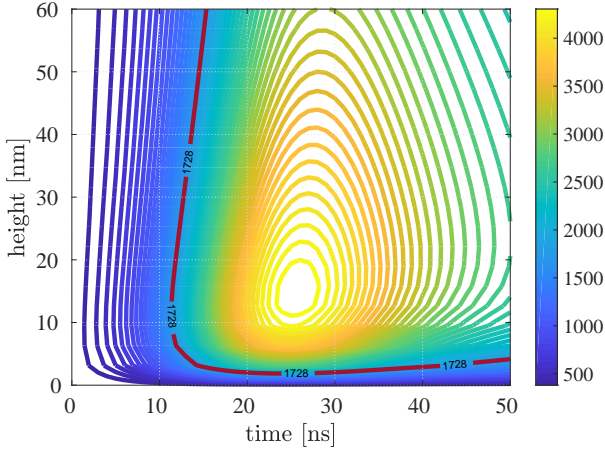}
  \caption{ The average temperature of a metal film, $T_m^*$, as a function of 
  film thickness and time. The red highlighted curve represents the melting temperature 
  of nickel, $T_M = 1728\, K$.  The parameters used are specified in Tab.~\ref{tab:thermal_params}.
  }
\label{fig:Trice_temp}
\end{figure}

In this section,  we outline the reduced model for the metal temperature; a version 
of such model is discussed in \citep{trice_prb07}.
The film--substrate bilayer is assumed to be infinitely wide in the in-plane
directions.  
The substrate layer is assumed to be thick compared to the film thickness and it is 
modeled as a semi-infinite medium $0 \le y < -\infty$. 
Assuming that any variation of film thickness occurs on the scale which is much larger
than the film thickness, one could argue that the heat conduction in the in--plane direction 
in the film is negligible compared to the conduction in the out-of-plane direction. Although the same argument
does not apply to a (thick) substrate, it is still assumed to hold.   
Hence, within this reduced model, the heat conduction in the bilayer is 
described by the one-dimensional heat equation in each layer,
\begin{align}
\label{eq:heat1}
 \left(\rho C_p\right)_m \frac{\partial T_m}{\partial t} &= k_m  \frac{\partial^2 T_m}{\partial y^2} + S(y, t)
\,\,\,\,\, \text{in the fluid,} \\
\label{eq:heat2}
 \left(\rho C_p\right)_s \frac{\partial T_s}{\partial t} &= k_s \frac{\partial^2 T_s}{\partial y^2} 
 \,\,\,\,\,\,\, \text{in the substrate,}
\end{align}
where $C_p$ is the effective heat capacity and $k$ is the thermal 
conductivity. The subscripts $s$ and $m$ correspond to substrate and metal, respectively. 
The source term can be written as 
\begin{equation}
\label{eq:source}
 S (y, t) =  \frac{E_0 f \left( t \right)}{\sqrt{2\pi} \sigma_{\rm tp} }
 \left[1 - r_0 \left(1 - e^{-a_r h} \right) 
\right]  e^{-\alpha_a \left( y + h \right) }, 
\end{equation} 
where the first factor represents the incident energy from the laser source,
second factor accounts for the reflectance of the metal film, and the last
factor  represents the energy absorbed by the film.  In
Eq.~\eqref{eq:source}, $E_0$ is the intensity of  the incident radiation,
$\sigma_{\rm tp} = t_p \left(2\sqrt{2\ln{2}}\right)^{-1}$ is the width of the Gaussian  laser pulse at half maximum, and $f(t)$ gives the
temporal profile of the laser fluence, 
\begin{equation}
 f(t) = \exp{\left[ {- (t-t_p)^2/(2
\sigma_{\rm tp}^2)} \right]}. 
\label{eq:source1}
\end{equation}
More details regarding the
derivation of the source term and the explanation of the parameters are given in Appendix \ref{app:laser}; the
parameters themselves are specified in Tab.~\ref{tab:thermal_params}.  
The boundary conditions are as  follows
\begin{align}
\label{eq:heatBC1}
 \frac{\partial T_m}{\partial y} = 0&
 \,\,\,\,\,\,\, \text{ at 
 } y = h(t,x),\\
 \label{eq:heatBC2}
 k_m \frac{\partial T_m}{\partial y} = k_s \frac{\partial T_s}{\partial y}&
 \,\,\,\,\,\,\, \text{ at 
 } y = 0,\\
 \label{eq:heatBC3}
 T_m = T_s &
 \,\,\,\,\,\,\, \text{ at  } y = 0,\\
 \label{eq:heatBC4}
 T_s \to T_0 & \,\,\,\,\,\,\, \text{ as } y \to - \infty ,
\end{align}
where $y = h(x, t)$ corresponds to the film-air  interface, and $y = 0$ is the 
film-substrate interface.
\begin{table*}
\caption{The values of the parameters used in simulations.  The material parameters come from 
 \cite{Smithells},  the source term properties are as in~\cite{hartnett2017}, and the parameters related to disjoining pressure are the ones used in~\cite{POF2016}. 
 }\label{tab:thermal_params}
\begin{tabular}{|c c c|} 
 \hline \hline
 Description & Notation & Value/Expression  \\ [0.5ex] 
 \hline
 Density of the metal & $\rho_m$ & $ 7900$ kg/m$^3$ \\  
 Density of the substrate & $\rho_s$ & $ 2200$ kg/m$^3$ \\  
 Room temperature  & $T_0$ & $ 300$ K  \\ 
 Melting temperature of the metal & $T_M$ & $ 1728$ K  \\ 
 Viscosity of the metal at $T_M$  & $ \mu_m $ & $ 4.61 \times 10^{-3}$ Pa \, s \\  
 Surface tension  & $ \sigma (T)$  &  $ \sigma_0 + \sigma_T(T - T_M) $  \\  
 Reference surface tension  & $ \sigma_0 $  &  $ 1.778$ N/m   \\
 Change of $\sigma$ with respect to temperature   & $\sigma_T $ & $ -3.3 \times 10^{-4}$ N/m K   \\
 Conductivity of the metal & $k_m $ & $90$ W/m K \\
 Conductivity of the substrate & $k_s $ & $ 1.4$ W/m K\\
  Heat capacity of the metal & $ (C_p)_m $ &  $0.44 \times 10^3$  J/kg K \\
  Heat capacity of the substrate & $ (C_p)_s $ &  $0.712 \times 10^3$ J/kg K  \\
  Laser fluence &  $E_0$ & $ 2500$ J/m$^2 $ \\
 Time of maximum fluence & $ t_p $ & $18 \times 10^{-9}$ s \\
  Absorption length & $\alpha_a$ & $0.11688 \times 10^{-9}$  m$^{-1}$ \\
  Fit parameter for reflectance & $r_0$ & $ 0.459363 $ \\
  Fit parameter for reflectance & $a_r$ & $ (8.0\times 10^{-9}$ m)$^{-1}$ \\
  Equilibrium contact angle of metal with substrate & $\theta_{eq}$ & $70^\circ$ \\
  Exponents in in the disjoining pressure model & $(n,m)$ & $(2,3)$ \\
  Precursor film thickness & $h_*$ & $ 1.5 \times 10^{-9}$ m\\
  \hline \hline
\end{tabular}
\end{table*}

The spatial variation of the temperature in the metal film
is expected to be small due to the small film 
thickness and high thermal conductivity of the metal. Hence, within this model, it can be assumed that
the temperature only weakly depends on the $y$-coordinate, and the 
temperature of the film-air interface can be approximated by the average film 
temperature, $T^*_m  = \frac{1}{h} \int^h_{0} T_m\, dy $. Integrating 
Eq.~\eqref{eq:heat1} with respect to $y$, from $y = 0$ to $y = h$, and using the
boundary conditions \eqref{eq:heatBC1} and \eqref{eq:heatBC2}, gives the equation 
for the average temperature of the film, $T^*_m$ 
\begin{align}
\label{eq:ode_film}
    \frac{\partial T^*_m}{\partial t} &=  S^* \left(h, t\right)
- \frac{1}{h} \frac{q_s(t)}{ (\rho C_p)_m},  \\
\label{eq:source_average}
 S^*(h, t) &=  {K_S f(t)\over h} \left[1 - r_0 \left(1 - e^{-a_r h} \right) 
\right] \left[1 - e^{-\alpha_a h  } \right],
\end{align}
where 
\[q_s(t) = {\partial T_s}/{\partial y} |_{y=0}\quad {\rm and} \quad K_S = {E_0}/\left({\sqrt{2\pi} \sigma_{\rm tp}  (\rho C_p)_m} \right). \] 
The heat equation for the 
substrate \eqref{eq:heat2} can be solved using Green's functions or Laplace 
transform. Using the boundary conditions \eqref{eq:heatBC2}, 
\eqref{eq:heatBC3} and \eqref{eq:heatBC4}, the average temperature of the film 
is found to be
\begin{multline}
\label{eq:triceIntegral}
    T^*_m \left( h, t \right) = T_0 +  S^*\, e^{-\frac{t_p^2 }{2 \sigma_{\rm tp}^2}} 
    \int_0^t\operatorname{exp}\left( - \frac{\left( t-u \right)^2}{2 \sigma_{\rm tp}^2} + \right. \\
    \left.
  \frac{t_p}{\sigma_{\rm tp}^2}\left( t- u \right)\right) e^{ K^2 u} 
  \operatorname{erfc}(K\sqrt{u}) \, du ,
\end{multline}
where $\operatorname{erfc}(u)$ is the complementary error function and 
\begin{equation*}
  K \left( h \right) = \frac{\sqrt{(\rho C_p k)_s}}{(\rho C_p)_m h} .
\end{equation*}

Figure \ref{fig:Trice_temp} shows the contour plot of the average temperature of
the metal film, $T^*_m(h,t)$, as given by Eq.~\eqref{eq:triceIntegral}. 
The red highlighted curve represents the melting temperature of nickel $(T_M = 1728\,
\text{K})$.   We note that the energy absorption depends on the  film thickness in a
non-monotonic manner. For a
small film thickness, $h < h_c \approx 14.3\, $nm, only a part of the laser energy is absorbed,
which leads to low film temperature. For film thicknesses $h > h_c$ the film
absorbs most of the laser pulse energy. Hence, the film of thickness 
$h \approx h_c$
reaches the highest temperature, and for the film of thicknesses $h > h_c$, the
temperature decreases as $h$ grows due to the larger amount of material that
needs to be heated.
Later in Sections~\ref{sec:LSA} and \ref{sec:film}, we 
study the dynamics of the films with the film thickness either smaller or larger than $h_c$.

A known temperature at the interface which is expressed as a function of the
film thickness, $h$, and time, $t$ only, is convenient for implementing the
Marangoni force in the VoF solver. The surface gradients of the surface tension,
$\nabla_s \sigma$, can be evaluated directly as
\begin{equation}
 \label{eq:temp_height_grad}
\nabla_s \sigma = \frac{\sigma_T \frac{\partial T}{\partial 
h} \frac{\partial h}{\partial x}}{\text{d}s}\ \mathbf{\hat{t}}, 
\end{equation}
where $\text{d}s = \sqrt{1 + \left({\partial \mathcal{H}}/{\partial x} 
\right)^2 }$ is the arc length, and $\mathcal{H}$ is the height function \cite{Francois2010,seric2017}. 
As discussed earlier, $T$ is approximated by the average temperature $T_m^*$, and 
therefore the gradient of the temperature with respect to
the film thickness is approximated by  $\partial T_m^*/\partial h$ that can be computed analytically
from Eq.~\eqref{eq:triceIntegral} as follows
\begin{widetext}
\begin{multline}
 \label{eq:dT_dh}
    \frac{\partial T^*_m}{\partial h} \left( h, t \right) = 
      \frac{\partial S^*}{\partial h} \, e^{-\frac{t_p^2 }{2 \sigma_{\rm tp}^2}} \int_0^t 
      \operatorname{exp}\left( - \frac{\left( t-u \right)^2}{2 \sigma_{\rm tp}^2} + 
      \frac{t_p}{\sigma_{\rm tp}^2}\left( t- u \right)\right) e^{ K^2 u} \operatorname{erfc}(K\sqrt{u}) \, du + \\ 
      2 S^*\, e^{-\frac{t_p^2 }{2 \sigma_{\rm tp}^2}} \int_0^t \frac{d K}{d h} 
      \operatorname{exp}\left( - \frac{\left( t-u \right)^2}{2 \sigma_{\rm tp}^2} + 
      \frac{t_p}{\sigma_{\rm tp}^2}\left( t- u \right)\right)
    \left[  K e^{ K^2 u} \operatorname{erfc}(K\sqrt{u}) - \sqrt{\frac{u}{\pi}}  \right] \, du 
\end{multline}
\begin{equation*}
  \frac{\partial S^*}{\partial h} = K_{S} \left\{ \left[- r_0 a_r  e^{-a_rh} \right)] \left[ 1 - e^{-\alpha_a h} \right]\frac{1}{h} + \right. \\ 
   \left.  \left[1 - r_0 \left( 1 - e^{-a_rh} \right) \right] \left[ \left[ \alpha_a e^{-\alpha_a h} \right]\frac{1}{h} - \left[ 1 - e^{-\alpha_a h} \right]\frac{1}{h^2} \right] \right\}.
\end{equation*}
\end{widetext}
Note that ${\partial h}/{\partial x}$ in Eq.~\eqref{eq:temp_height_grad}, 
is equivalent to ${\partial \mathcal{H}}/{\partial x}$, i.e., the derivative of the 
height function.
For small film thicknesses, $T^*_m$ and $\partial T^*_m/ \partial h$ need to be 
carefully computed to ensure that the integrals in Eqs.~\eqref{eq:triceIntegral} and 
\eqref{eq:dT_dh} converge, see Appendix \ref{app:trice_small_h}.
When used in our simulations, both $T^*_m$ and $\partial T^*_m/\partial h$ are evaluated 
for an array of $t$ and $h$ values before the start of the simulations. During the simulation, we use
bilinear interpolation to find the temperature at each interfacial cell and each
time step. This makes the computations significantly faster, since we do not
need to use the 
numerical integration to compute the integrals in Eqs.~\eqref{eq:triceIntegral}
and \eqref{eq:dT_dh} for each interfacial cell at each time step. 
\begin{figure*}[htb]
\centering
  \begin{subfigure}{0.40\textwidth}
    \includegraphics[width = \textwidth]{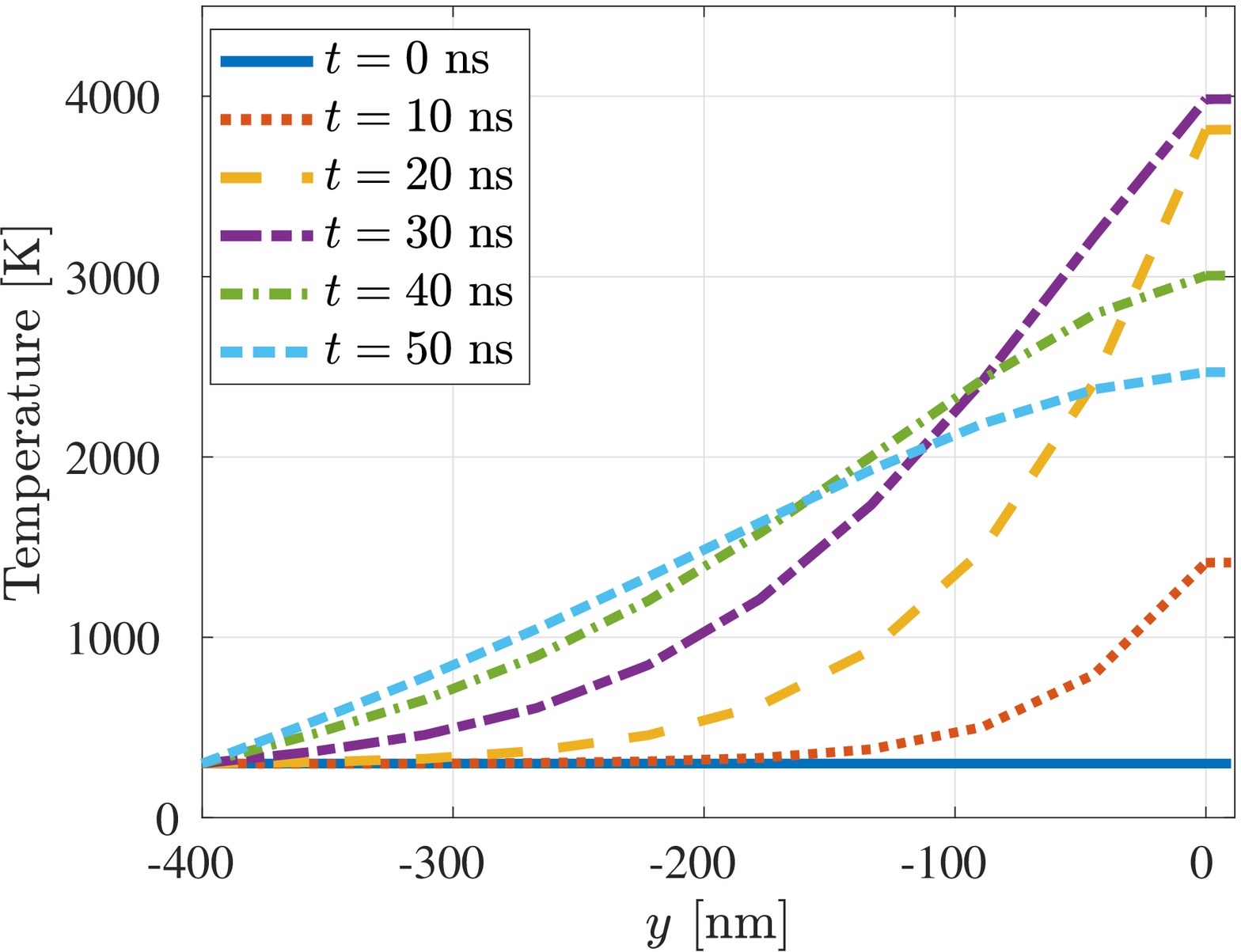}
    \caption{$h_0 = 10$ nm} 
  \end{subfigure}\hspace{0.1in}
  \begin{subfigure}{0.40\textwidth}
    \includegraphics[width = \textwidth]{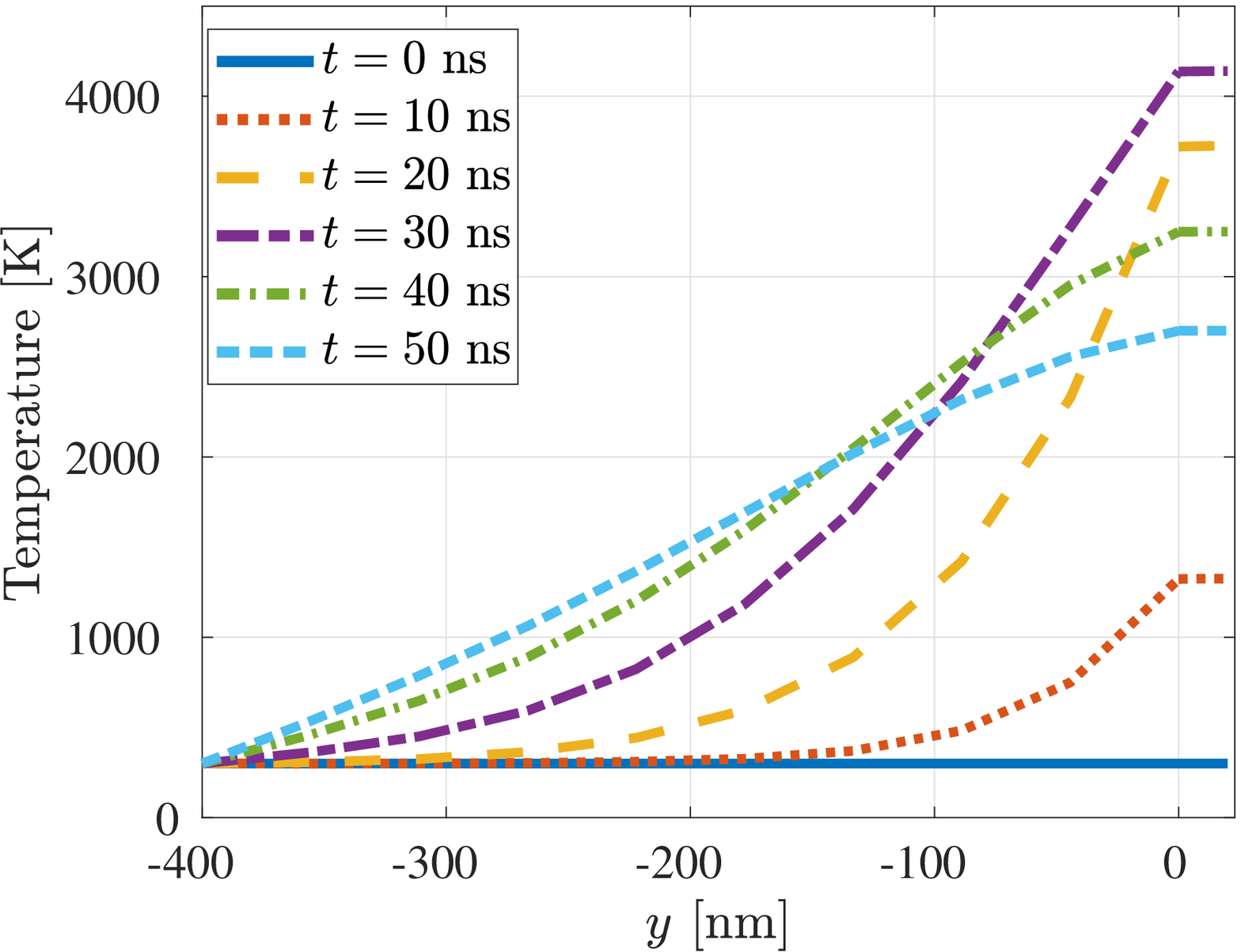}
  \caption{$h_0 = 20$ nm}
  \end{subfigure}
  \vspace{-0.1in}
  \caption{The analytical solution for the temperature of the film ($y>0$) and the 
  substrate ($y<0$).  
The lines show different times; the time specified in the legends is in ns.
Note very small variation of the temperature across the film.
}
  \label{fig:exact_eg}
\end{figure*}

\subsubsection{Analytical Solution of the Heat Equation in a Film--Substrate System}
\label{sec:exact}
The Eqs.~\eqref{eq:heat1} and 
\eqref{eq:heat2}, including the spatial variations in the $y$-direction in both 
the metal and the substrate, can be solved analytically using separation of 
variables, following the technique given in \citep{ozicsik1989}. 
The analytical solution presented here is used for the verification of the reduced 
model given in Eq.~\eqref{eq:triceIntegral} and the complete model 
is presented in Section~\ref{sec:dns_model}. 

In contrast to the reduced model presented in Section~\ref{sec:Trice}, for computing 
the analytical solution, we assume that the substrate is of finite depth.  
However, when comparing the solutions resulting from different models, we use substrate 
depth large enough so that the 
temperature solution is converged with increasing substrate thickness, and the solution is  
equivalent to that for the semi-infinite substrate (see Fig.~\ref{fig:analytical_subs_conv} in Appendix \ref{app:exact}).
Therefore, instead of Eq.~\eqref{eq:heatBC4}, here we use the following boundary condition
\begin{equation}
 \label{eq:heatBC4_b}
	T_s  = T_0 \,\,\,\,\,\,\,\, \text{at the bottom of the substrate } y = - b,
\end{equation}
where $b$ is the substrate thickness.

The solution can be found using separation of variables
\citep{ozicsik1989}, and it can be written compactly in terms of Green's
functions as
\begin{align}
\label{eq:exact_m}
    T_m \left(y, t\right) &= \int_0^t \int_a^b G_{12} \left( y, t; \xi, \tau \right) \frac{\mathcal{D}_m}{k_m} S \left( \xi, \tau \right) \text{d} \xi \text{d}\tau \\
\label{eq:exact_s}
    T_s \left(y, t\right) &= \int_0^t \int_a^b G_{22} \left( y, t; \xi, \tau \right) \frac{\mathcal{D}_s}{k_s} S \left( \xi, \tau \right) \text{d} \xi \text{d}\tau
\end{align}
where $G_{12}$ and $G_{22}$ are given in Appendix \ref{app:exact} along with the details of the solution.  

Figure \ref{fig:exact_eg} shows the analytical temperature solution in the metal 
film ($y>0$) and the substrate ($y<0$).  
The temperature variation across the film thickness is small compared to the variation
in the substrate. 
Therefore, ignoring temperature gradients across the film, as used in the reduced 
model, is justified. We note, however, that such a conclusion can be reached only for 
stationary flat films.  As we will see later, using the reduced model for nonuniform films, or for the time dependent films, in general cannot be justified.  On a different note, we point out that since there is 
no in-plane dependence in the source term, the 1D 
analytical solution presented here holds for 2D or 3D flat stationary films.

\subsubsection{Computational Model for the Temperature of the Film--Substrate System}
\label{sec:dns_model}

Next we consider the outlined problem via direct numerical 
simulations. We implement our numerical methods into the open source {\sc Gerris}
software~\cite{Gerris2}.    We denote the top subdomain 
containing metal and air by $\Omega_f$, and the bottom one containing the substrate by $\Omega_s$.
In general, the temperature in $\Omega_f$, denoted $T_f$, satisfies the advection-diffusion
equation, and the temperature in $\Omega_s$, $T_s$, satisfies the diffusion equation,
\begin{align}
\label{eq:heat1_full}
 \rho  C_p  \left[ \partial_t T_f + 
 \right(\mathbf{u}\cdot\nabla \left) T_f  \right] & = \nabla \cdot \left( k  \nabla T_f \right) + S_n(\mathbf{x}, t)
\,\,\,\,\, \text{in } \Omega_f \\
\label{eq:heat2_full}
 \left(\rho C_p\right)_s \partial_t T_s &= k_s \nabla^2 T_s 
 \,\,\,\,\,\,\, \text{in } \Omega_s 
\end{align}
where $\rho = \rho \left(\chi \right)$, $C_p = C_p\left(\chi \right) $ and 
$ k = k \left(\chi \right)$ are the phase dependent density, heat capacity and the 
conductivity of the metal and air, defined as the volume fraction weighted average 
of the metal and air properties. 
 \begin{figure}[tbh]
 \centering
 \includegraphics[width = 0.4\textwidth]{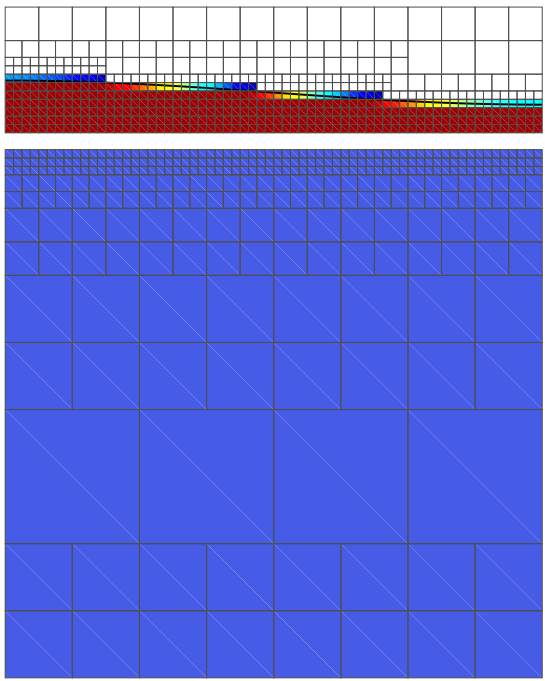}
 \caption{ The fluid--substrate setup used in the direct numerical simulations. }
 \label{fig:solid}
\end{figure}

The simulation setup has to address the following issue: using {\sc Gerris}
we cannot solve for the temperature inside of
the solid substrate directly, since except on the boundaries, the implementation 
of the solid entities does not contain the computational cells.
Hence, in order to solve for the temperature in the fluid and the substrate using
{\sc Gerris}, we treat the solid domain -- the substrate -- as an immobile fluid.  
Furthermore, in order to impose the no-slip boundary condition on the metal-substrate boundary,
we separate the two phases by a thin solid plate, see Fig.~\ref{fig:solid}. Later in this section we 
show by direct comparison with the analytical solution that this setup, referred to as the complete model, 
produces correct results.  

\begin{figure*}[thb]
\centering
  \begin{subfigure}{0.40\textwidth}
    \includegraphics[width = \textwidth]{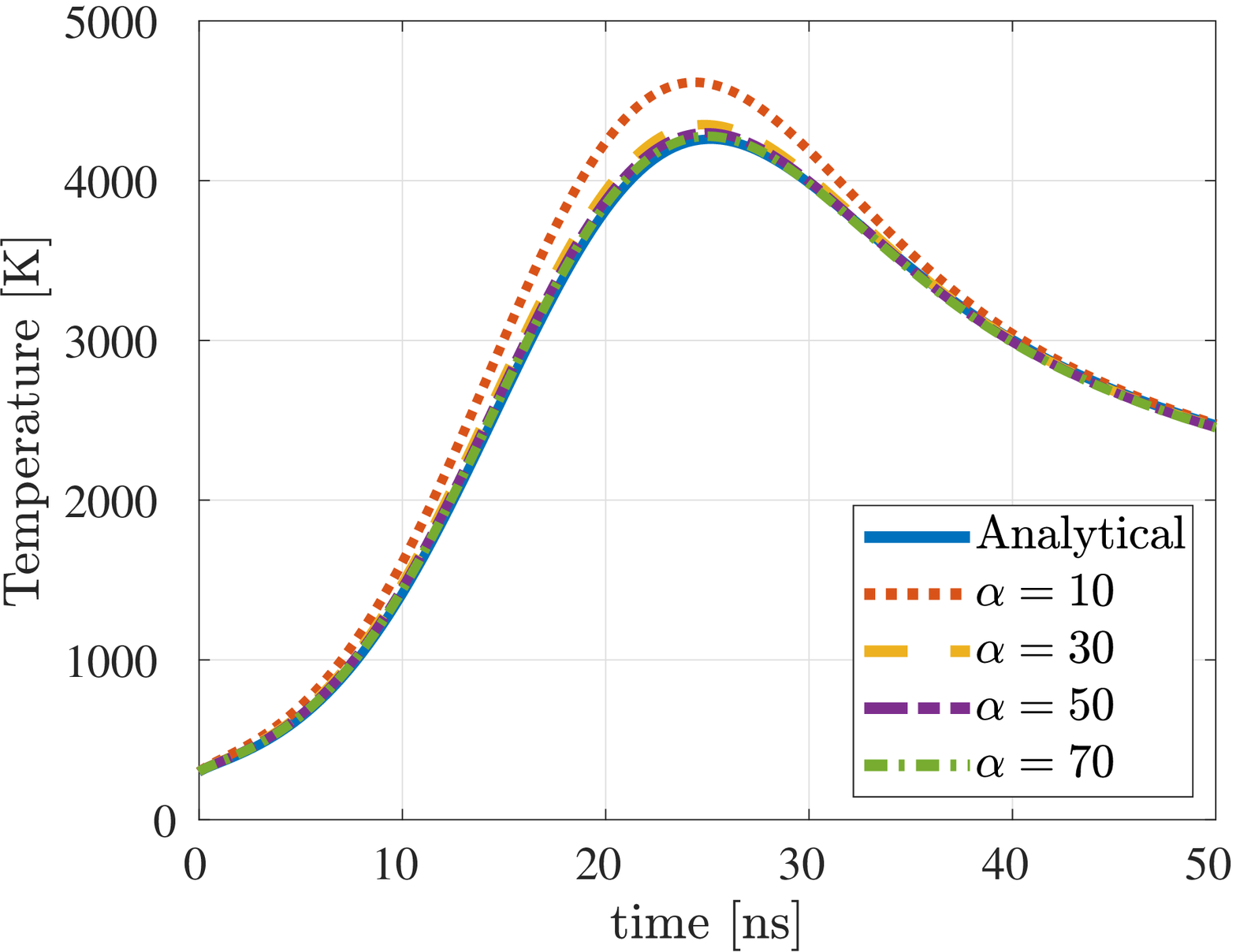}
    \caption{$h_0 = 10$ nm} 
  \end{subfigure}\hspace{0.1in}
  \begin{subfigure}{0.40\textwidth}
    \includegraphics[width = \textwidth]{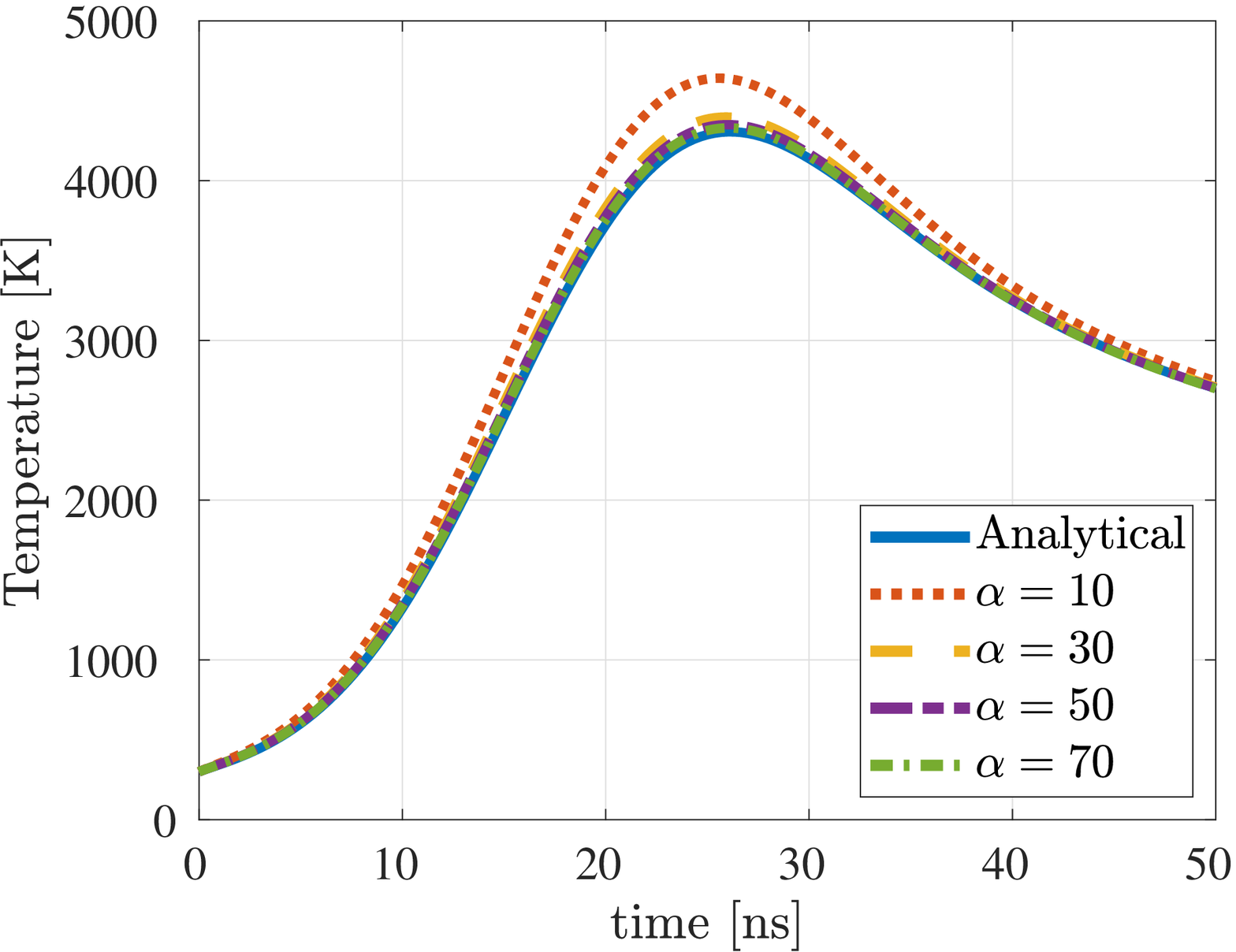}
  \caption{$h_0 = 20$ nm}
  \end{subfigure}
  \vspace{-0.1in}
  \caption{
  The convergence of the average temperature in the metal film using the 
  numerical solution of the complete model with increasing 
  $\alpha$. The units 
  of $\alpha$ in the legend are in W~m$^{-2} \text{K}^{-1}$.  Note that the source term is a Gaussian centered at $t = t_p = 18$ ns, 
  and of the width $\sigma_{tp}$, see Eq.~(\ref{eq:source1}).   
   }
  \label{fig:dns_alpha}
\end{figure*}
\begin{figure*}[thb]
\centering
  \begin{subfigure}{0.40\textwidth}
    \includegraphics[width = \textwidth]{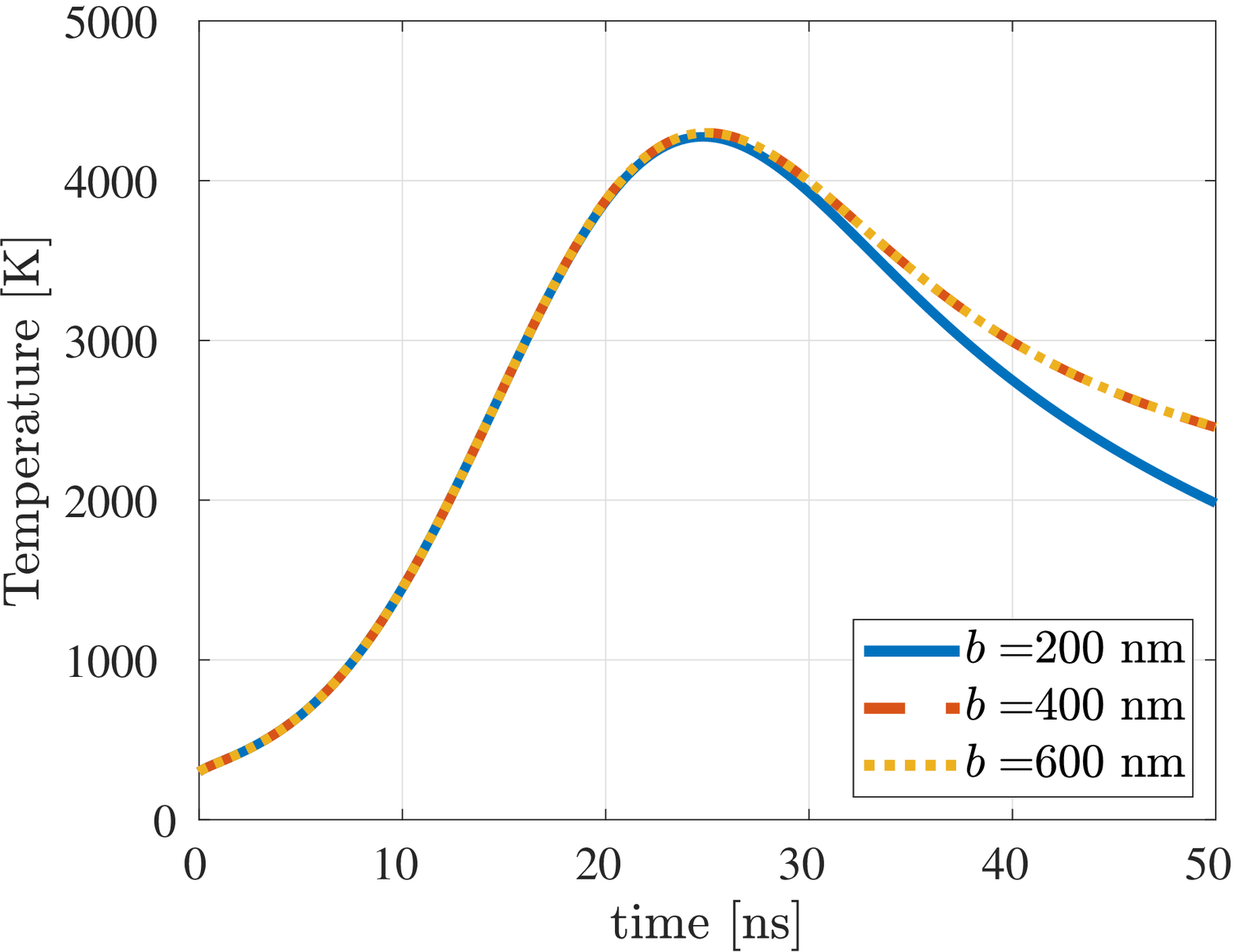}
    \caption{$h_0 =10$ nm} 
  \end{subfigure}\hspace{0.1in}
  \begin{subfigure}{0.40\textwidth}
    \includegraphics[width = \textwidth]{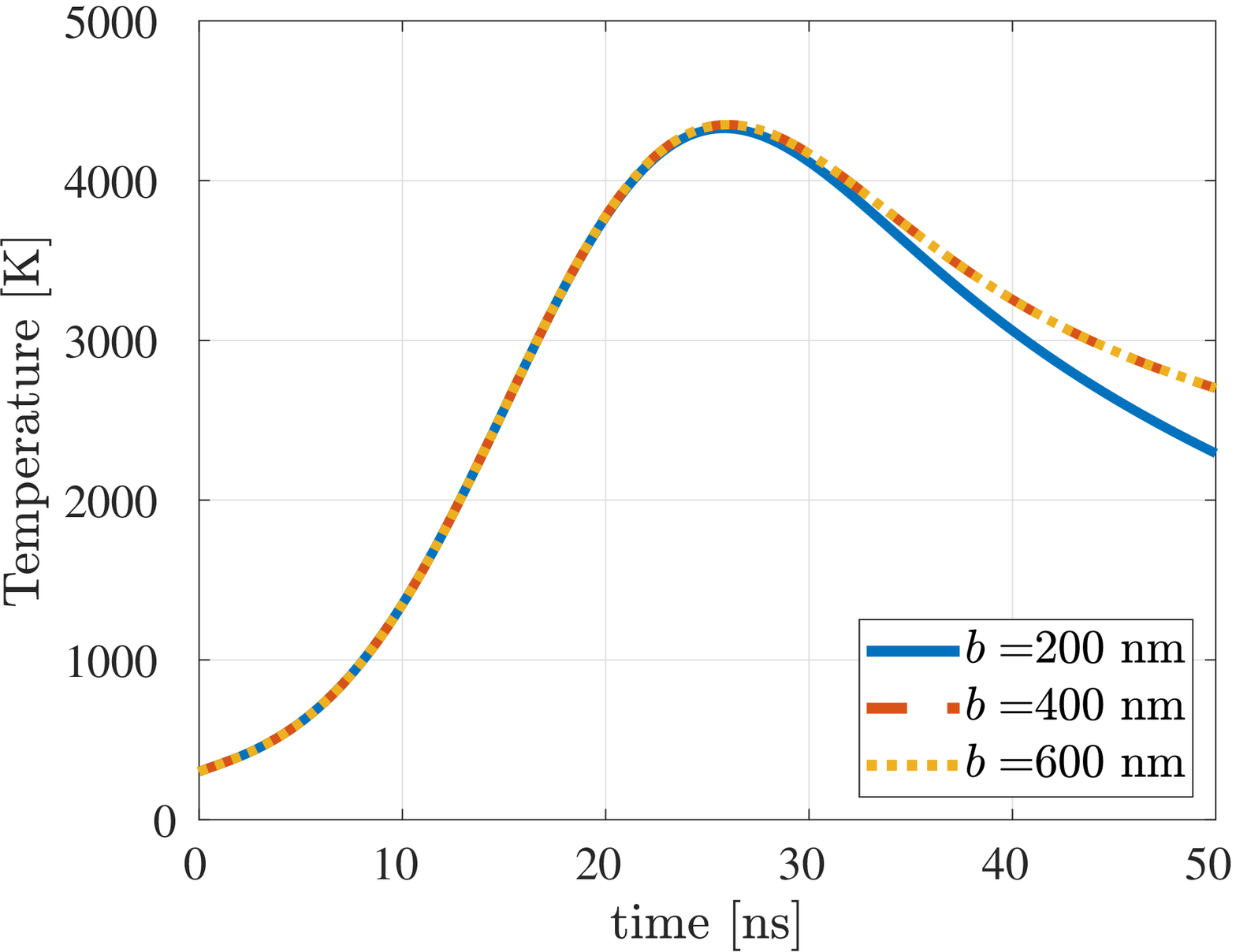}
  \caption{$h_0 = 20$ nm}
  \end{subfigure}
  \vspace{-0.1in}
  \caption{The convergence of the average temperature in the metal film using the 
  solution of the complete model with increasing substrate size, $b$.  }
  \label{fig:dns_substrate}
\end{figure*}
\begin{figure*}[thb]
\centering
  \begin{subfigure}{0.40\textwidth}
  \includegraphics[width = \textwidth]{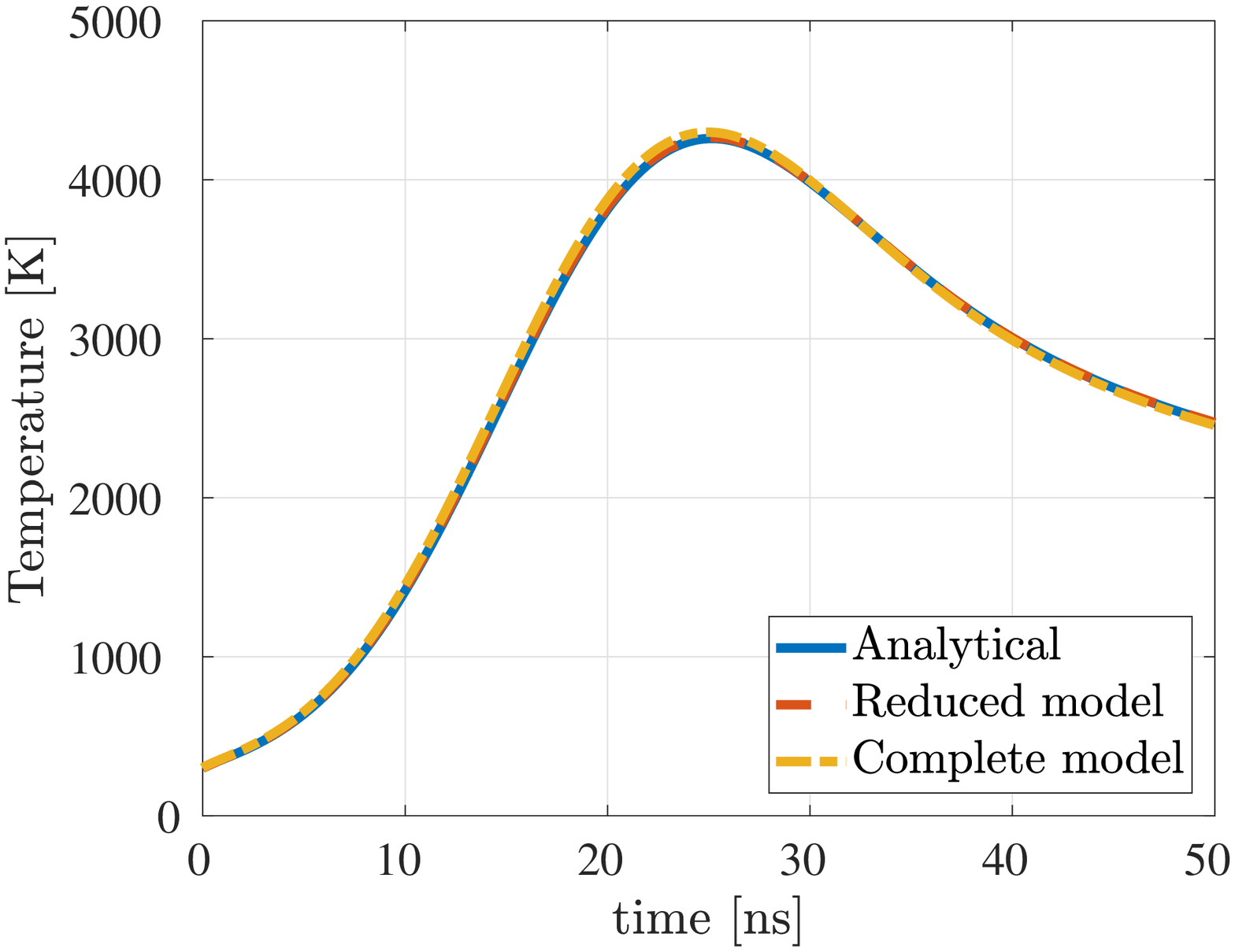}
  \caption{$h_0 = 10$ nm} 
\end{subfigure}\hspace{0.1in}
\begin{subfigure}{0.40\textwidth}
  \includegraphics[width = \textwidth]{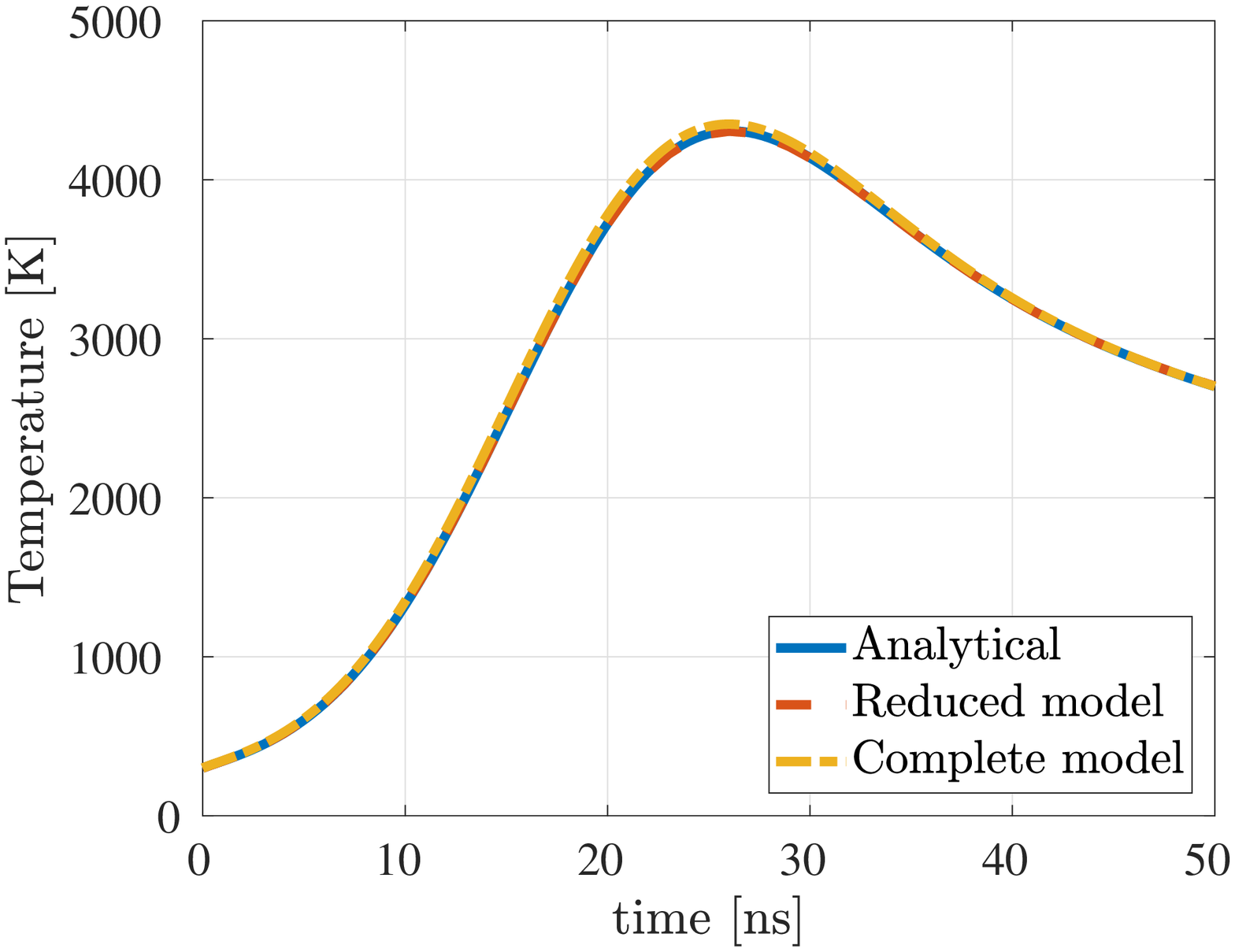}
   \caption{$h_0 = 20$ nm}
\end{subfigure}
    \caption{The comparison of the average temperatures of a flat film for the reduced, the analytical, and the complete model. 
    Here, the substrate thickness is $ b = 400\, $nm, and the heat transfer coefficient is $\alpha =  50\, \text{W~m}^{-2} \text{K}^{-1}$. }
    \label{fig:temp_methods}
\end{figure*}

The coupling of the temperature solution between the liquid and solid domains is
accomplished using Newton's law of cooling, which we impose on the top and bottom 
of the solid plate as follows
\begin{align}
\label{eq:bc_solid_top}
 k_m \frac{\partial T_f}{\partial y} &= \alpha \left( T_f - T_s \right) \,\,\,\,\,\,\, \text{at the top of the solid}, \\
\label{eq:bc_solid_bottom}
 k_s \frac{\partial T_s}{\partial y} &= \alpha \left( T_f - T_s \right) \,\,\,\,\,\,\, \text{at the bottom of the solid}, 
\end{align}
where $\alpha$ is the heat transfer coefficient. Note that the right hand sides
of Eqs.~\eqref{eq:bc_solid_top} and \eqref{eq:bc_solid_bottom} are equal,
implying the continuity of the flux between the liquid and substrate.
Furthermore, in the limit $\alpha \to \infty$, the boundary conditions
\eqref{eq:bc_solid_top} and \eqref{eq:bc_solid_bottom} both imply $T_f = T_s$.
Hence, the boundary conditions specified by Eqs.~\eqref{eq:bc_solid_top} and
\eqref{eq:bc_solid_bottom} effectively encapsulate both the continuity of flux,
Eq.~\eqref{eq:heatBC2}, and the continuity of  temperature, Eq.~\eqref{eq:heatBC3}.  Additionally, in Appendix \ref{app:exact}, see 
Fig.~\ref{fig:analytical_alpha_conv},  we confirm that the analytical solution, given 
in Section~\ref{sec:exact}, with the Newton's cooling law 
boundary condition, converges for large $\alpha$ to the solution specified by requiring continuity of temperature.
We note that the presented computational approach can be used for arbitrary metal-air interface 
shape.

In the remainder of this section, we verify that the numerical solutions to Eqs.~\eqref{eq:heat1_full} and \eqref{eq:heat2_full}, along with the boundary 
conditions, Eqs.~\eqref{eq:bc_solid_top} and \eqref{eq:bc_solid_bottom}, 
converge with the increasing heat transfer coefficient, $\alpha$, and the 
substrate thickness, $b$. To do this, we consider the following test problem. 
Assume the metal-air interface is 
flat and the solution is $x$-independent. Hence, 
the temperature for the 2D problem satisfies the 1D heat equation at any fixed
position $x$, and the 1D analytical solution, presented in Section~
\ref{sec:exact}, holds. We compare 
the temperature solution obtained from the complete model to the 1D analytical 
solution, where we average the temperature over the film thickness for the purpose
of this comparison.

Figure \ref{fig:dns_alpha} shows the temperature solution of the 
complete model for a flat film geometry and vanishing velocity in Eq.~(\ref{eq:heat1_full})
compared to the analytical solution for increasing values of $\alpha$. 
Clearly, as $\alpha$ increases, the results of the complete model approach the analytical solution.
The largest relative difference in the temperature between $\alpha = 50 \, \text{W~m}^{-2} \text{K}^{-1}$
and $\alpha = 70 \, \text{W~m}^{-2} \text{K}^{-1}$ is $0.7\%$ for a $h_0 = 10\,$nm, and
$0.6\%$ for a $h_0 = 20\,$nm. Since larger $\alpha$ decreases the required time-step,
we use $\alpha = 50\, \text{W~m}^{-2} \text{K}^{-1}$ in order to
reduce the computational time. 
Figure \ref{fig:dns_substrate} shows the convergence of the solution of the complete model
for the averaged temperature for increasing substrate size, $b$. The 
largest relative difference in the temperature between $ b = 400 \, $nm and 
$ b = 600 \, $nm is $0.12\%$ for a $h_0 = 10\,$nm, and $0.08\%$ for a 
$h_0 = 20\,$nm. We conclude that it is sufficient to use  $ b = 400\, $nm. 

So far, we have shown that the solution to the complete model described in this section converges 
with the increased heat transfer coefficient, $\alpha$, and the substrate thickness, $b$, to
the analytical solution with continuity of temperature boundary condition at the fluid-substrate interface.
Next, Fig.~\ref{fig:temp_methods} shows the comparison of the average
temperature in the metal film obtained using the reduced model given in Section
\ref{sec:Trice}, the analytical solution outlined in Section~\ref{sec:exact}, and the 
complete model described in this section.   
The average temperature of a flat metal film in both the reduced and the complete 
model agrees with the analytical solution. 
Hence, we are confident in the numerical implementation of the temperature models 
in our numerical solver. 

We point out this agreement 
between different models, in particular the reduced and complete model holds only for flat 
films for which the heat flow in the in-plane direction is not relevant.  For perturbed films, 
the two models produce different results, as we will discuss in the next section.

\section{Results and Discussion}
\label{sec:results}

We are now ready to discuss the influence of thermal effects on the film stability.   First in
Sections~\ref{sec:LSA}~-~\ref{sec:discussion} we focus on film geometry, and then consider 
filaments in Section~\ref{sec:filaments}.   
We start by discussing briefly in Section~\ref{sec:LSA}  the results of linear stability analysis carried out within
the long wave limit in a simplified setting (that assumes film 
temperature to be dependent only on the current value of the film thickness).  
Then, we  follow in Section~\ref{sec:film}   by presenting the results using the models for temperature computation 
outlined in the preceding section.   We will see that the outcomes of the models considered differ
significantly.  The main finding is that the reduced model overestimates  the 
Marangoni effect, and the particular reason for this is the omission of the in-plane heat 
conduction. We will also see that the temporal temperature variations lead to a change in the surface 
tension, which can in turn affect the stability of the perturbed interface 
during the evolution.  As we will see,  the interplay of the stabilizing/destabilizing Marangoni effect and temporal
variation of surface tension leads to a complex form of  film evolution.   In Section~\ref{sec:viscosity} we consider also
temperature variation of film viscosity, and its influence on the film stability.   Section~\ref{sec:filaments}
discusses the influence of thermal effects on stability of metal filaments.

For both films and for filaments, we focus on developing basic understanding of the influence of thermal 
effects on their stability.   Therefore, we focus on relatively simple computational domains 
and initial conditions - in particular, in simulations we will consider films and filaments perturbed by a single 
wavelength, and defer considering more complex domains and initial conditions to future work.  As we will
see, even for simple setups the influence of variation of material parameters (surface tension and viscosity)
is rather complex, and the simplicity of the considered computational domains and initial conditions helps in focusing
on the basic questions.  In addition, we focus on the question of stability versus instability of a given perturbation; 
for this reason in our simulations we choose initial conditions that are close to the critical ones where stability 
changes: such a choice helps to further simplify the problem considered and reach answers to the basic stability
questions.

\subsection{Linear Stability Analysis of a Thin Film in Two Dimensions}
\label{sec:LSA}

\begin{figure*}[thb]
 \centering
  \begin{subfigure}{0.40\textwidth}
  \includegraphics[width = \textwidth]{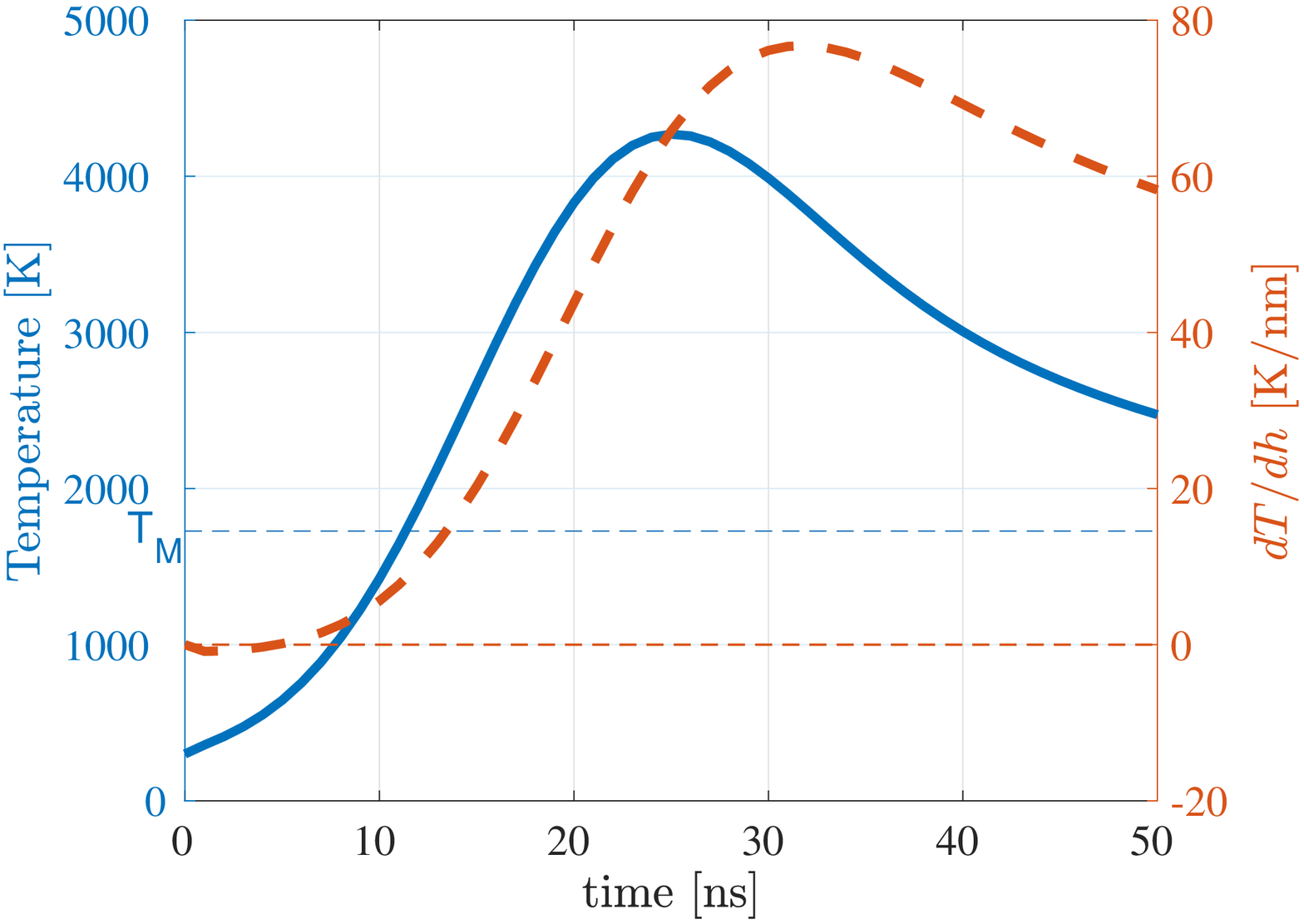} 
      \caption{$h_0 = 10$ nm} 
\end{subfigure}\hspace{0.1in}
\begin{subfigure}{0.40\textwidth}
  \includegraphics[width = \textwidth]{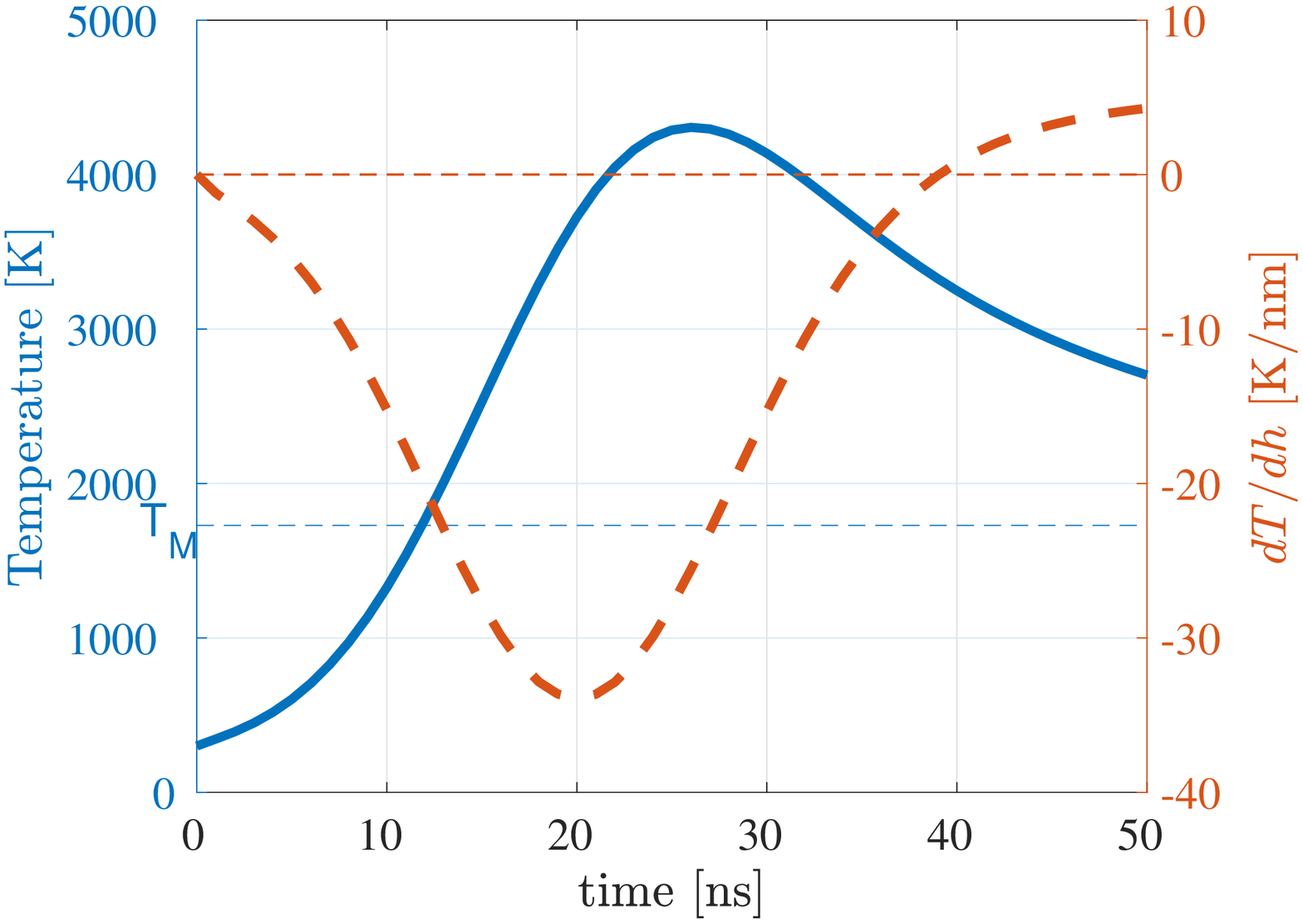}
    \caption{$h_0 = 20$ nm}
\end{subfigure}
\vspace{-0.1in}
    \caption{ The average temperature, $T_m^*$ (thick solid blue line) , and $\partial T_m^*/ \partial h$ (thick dashed
    orange line)  of a metal film  as a function 
    of time, using the reduced model. The thin horizontal  blue and orange dashed lines indicate the
    meting temperature, $T_M$, and the line $\partial T_m^*/ \partial h = 0$, respectively.
    }  
    \label{fig:Trice_Ni_10_20} 
\end{figure*}

First, to gain basic insight, we present the results of linear stability analysis (LSA) carried out 
within the long wave limit. While the LSA provides only approximate results since 
(i) it is valid only for early stages of instability, and (ii) corresponds to the long wave limit, we 
still expect it to provide a useful insight. 

The long wave limit \cite{oron_rmp97}, for a Newtonian film with Marangoni effect and the fluid-substrate
interaction in the form of the disjoining pressure \cite{Diez2007a} leads to 
the following 4th order nonlinear partial differential equation
\begin{eqnarray}
\label{eq:long_wave}
&&  3 \mu \frac{\partial h}{\partial t}  +  \\
&&  \nabla \cdot  \left[ \sigma_0 h^3 \nabla \nabla^2 h +
  \frac{3}{2} h^2 \nabla \sigma (T)  +    K_\pi h^2 \nabla \left( \frac{h_*^n}{h^n} - 
  \frac{h_*^m}{h^m} \right)  \right]   = 0, \nonumber
\end{eqnarray}
where $K_{\pi}$ is given by Eq.~\eqref{eq:K_pi}.
We assume the film thickness is perturbed around the equilibrium one, $h_0$, as
\begin{gather}
\label{eq:pert}
h(x,t) = h_0 \left(1  + \varepsilon  e^{\beta t + ikx}\right)\, ,
\end{gather}
where $\varepsilon$ is a small parameter, $\beta$ is the growth
rate of the perturbation and $k$ is the wavenumber. 
Note that temperature is not an independent variable here, instead it is a function of $h$. 
An alternative approach is to consider both $h$ and $T$ as independent variables, 
and perturb each of them separately, however, we are not doing this here for simplicity. 
Keeping only the leading order terms in $\varepsilon$ we obtain the dispersion relation:
\begin{gather}
\label{eq:dispersion_thermal}
    \beta = -\frac{h_0^2 \,k^2}{3\mu}\left[ \sigma_0 h_0 k^2 - 
    \frac{3}{2} \sigma_T \frac{\partial T}{\partial h}
    - K_\pi\left( n \frac{h_*^n}{h_0^n} - m\frac{h_*^m}{h_0^m} \right)  \right].
\end{gather}

To illustrate the expected influence of Marangoni effect on stability, in 
Fig.~\ref{fig:Trice_Ni_10_20} we plot the temperature gradient, 
$\partial T_m^*/ \partial h$, for a fixed film thickness computed using the reduced 
model (see Eq.~\eqref{eq:dT_dh}). The temperature of the film, $T_m^*$ (see Eq.~\eqref{eq:triceIntegral}), is 
plotted to show the film melting time.  We assume that before the temperature of the film 
rises above $T_m^*$, the film does not evolve.   
The value of the gradient changes as a function of time, and in order to obtain an estimate for 
the stability of a perturbed film using Eq.~\eqref{eq:dispersion_thermal}, 
we approximate ${\partial T}/{\partial h}$ by the largest absolute value of 
${\partial T^*_m}/{\partial h}$, during the time the film is melted. 
Hence,  the obtained dispersion curve provides the upper bound on the influence of the Marangoni effect.

\begin{figure*}[thb]
\centering
  \begin{subfigure}{0.40\textwidth}
    \includegraphics[width = \textwidth]{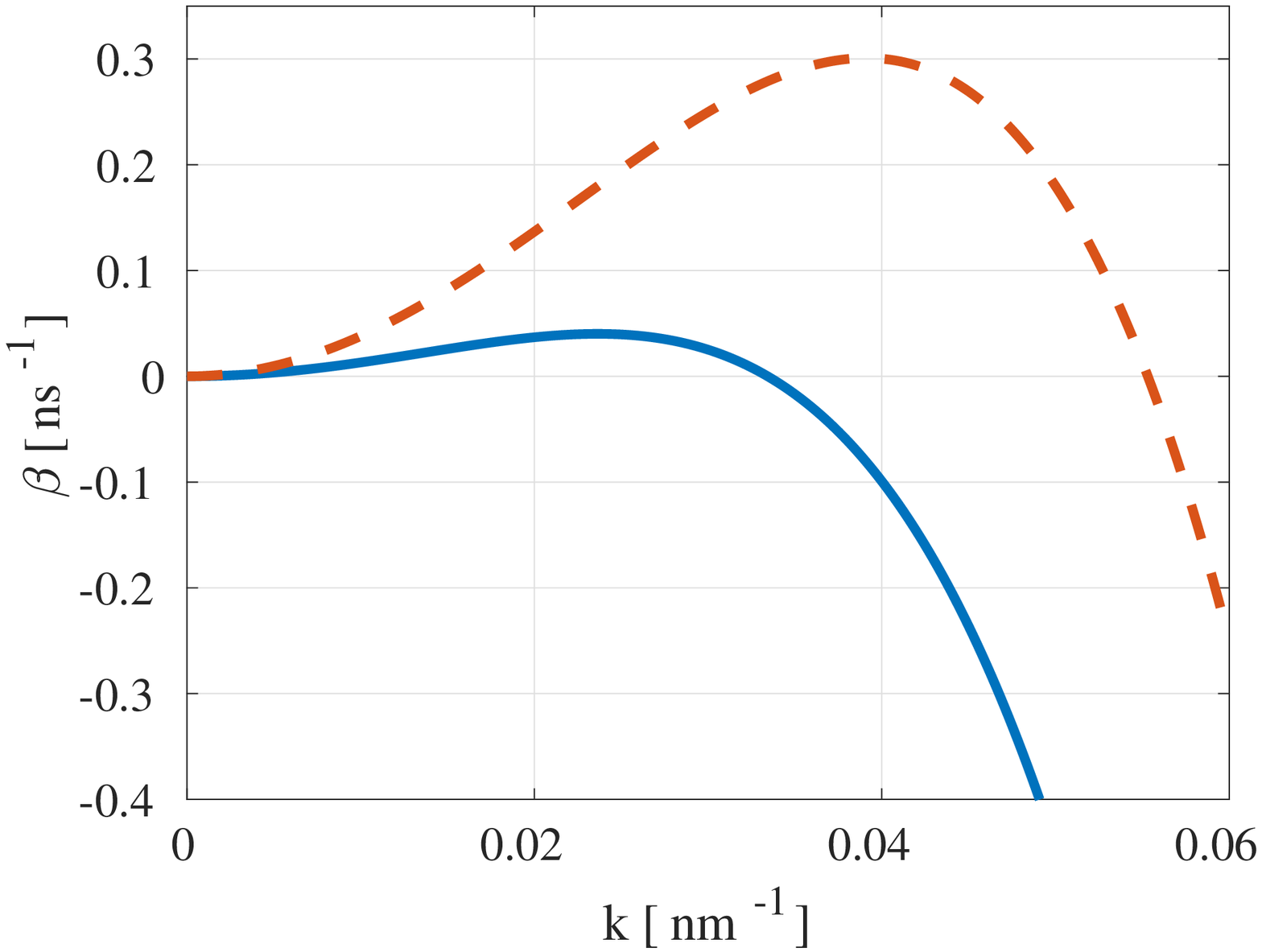}
    \caption{$h_0 = 10$ nm, ${\partial T}/{\partial h} = 76.7\, \text{K~nm}^{-1}$    
     } 
  \end{subfigure}
  \hspace{0.1in}
  \begin{subfigure}{0.40\textwidth}
    \includegraphics[width = \textwidth]{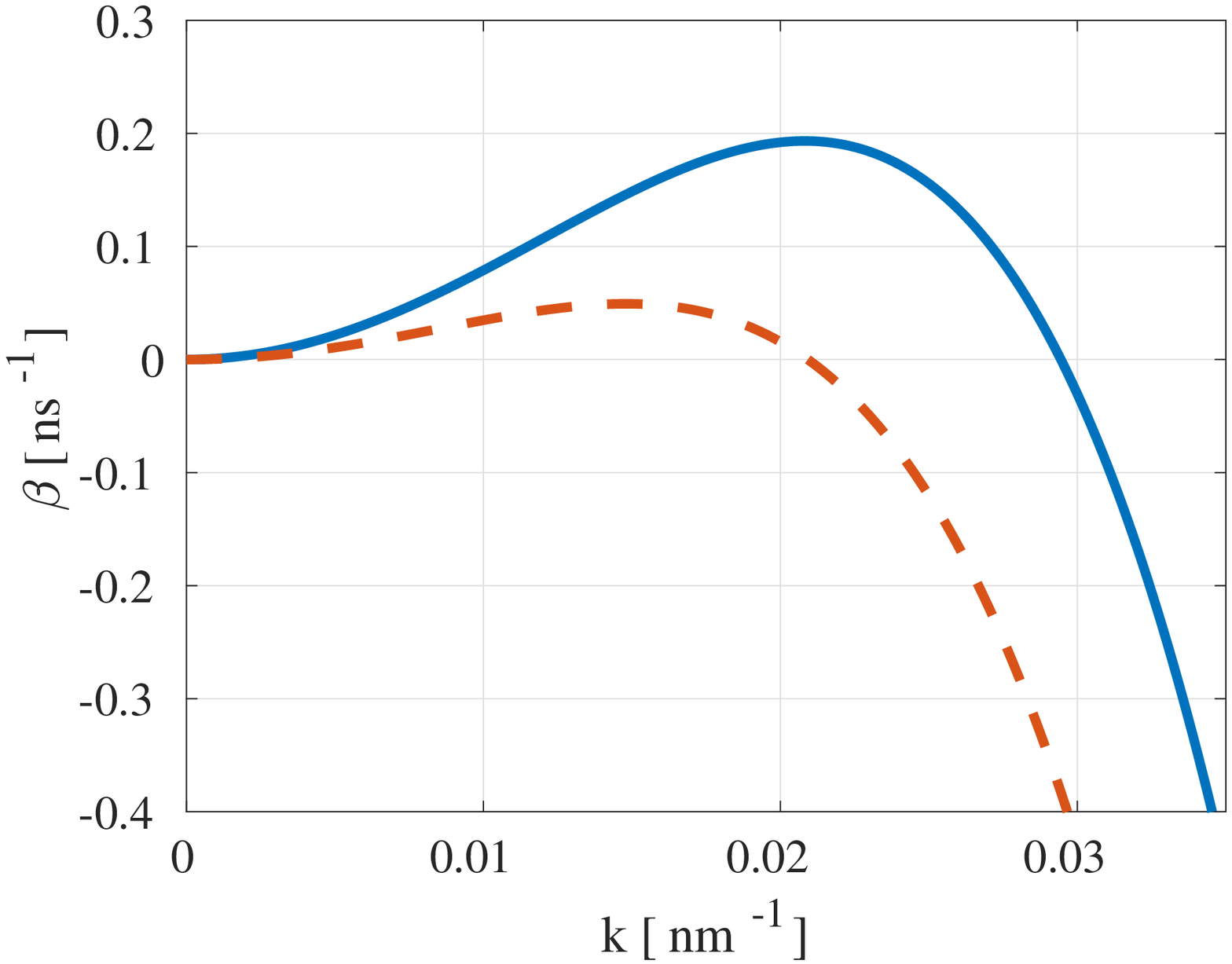}
  \caption{$h_0 = 20$ nm, ${\partial T}/{\partial h} = -33.9\, \text{K~nm}^{-1}$  
   }
  \end{subfigure}
  \caption{ The growth rate for a perturbed film resulting from the LSA, where $\partial T/\partial h$ 
  is approximated by the specified maximum values, as explained in the text,  with (blue solid) and without (orange dashed) Marangoni effect.}
  \label{fig:Growth_fixed_h}
\end{figure*}
Figure \ref{fig:Growth_fixed_h} shows the dispersion curve for $10$ and $20$ nm thick films.  
For the $h_0=10 \, $nm thick film, Fig.~\ref{fig:Growth_fixed_h}(a), the Marangoni effect
 is stabilizing, since ${\partial T}/{\partial h} > 0$, for all the times at which 
the film is melted.
Conversely, for $h_0 = 20\, $nm film, Fig.~\ref{fig:Growth_fixed_h}(b), the 
Marangoni effect is destabilizing for a significant period after melting, 
since ${\partial T}/{\partial h} < 0$. 

In the rest of this section, we will use the outlined LSA results to rationalize the results computed
using the reduced and complete models for the flow of thermal energy.

\subsection{Evolution of a Thin Film Interface in Two Dimensions}
\label{sec:film}

In this section, we examine the stability of the films by solving the Navier-Stokes 
equations including the thermal effects. We also include the fluid-substrate interaction
in the form of a disjoining pressure (see Eq.~\eqref{eq:Fma_vdw}).
We compare the influence of thermal effects on the film breakup using
the temperature solution from the reduced and from the complete model, described 
in Sections~\ref{sec:Trice} and \ref{sec:dns_model}, respectively.

The initial geometry of the film in the simulations is as described by Eq.~(\ref{eq:pert}) where $\varepsilon = 0.1 $.
At the time $t = 0$, the metal is in solid state (the pulse maximum occurs at $t_p = 28$ ns).   
For the simulations that use the reduced temperature model, we simulate only the times 
after the film is melted.  In the case of 
simulations such that the complete temperature model is implemented, we keep the film stationary until
melted (by putting the fluid velocity to zero).   

\subsubsection{Film of Thickness $10$ nm; $dT/dh > 0$}

\begin{figure*}[thb]
\centering
  \begin{subfigure}{0.40\textwidth}
    \includegraphics[width = \textwidth]{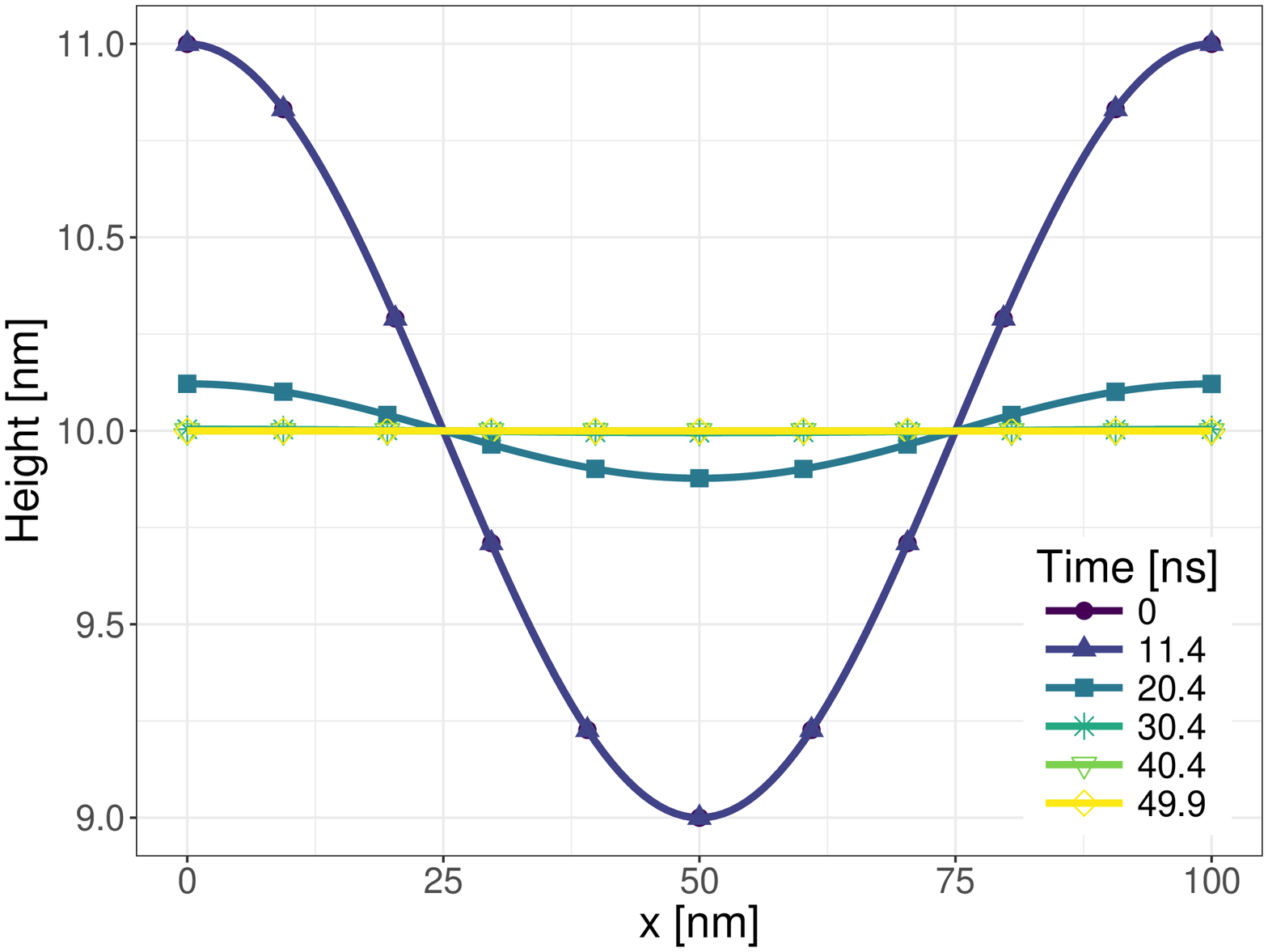}
    \caption{Reduced model} 
  \end{subfigure}\hspace{0.1in}
  \begin{subfigure}{0.40\textwidth}
    \includegraphics[width = \textwidth]{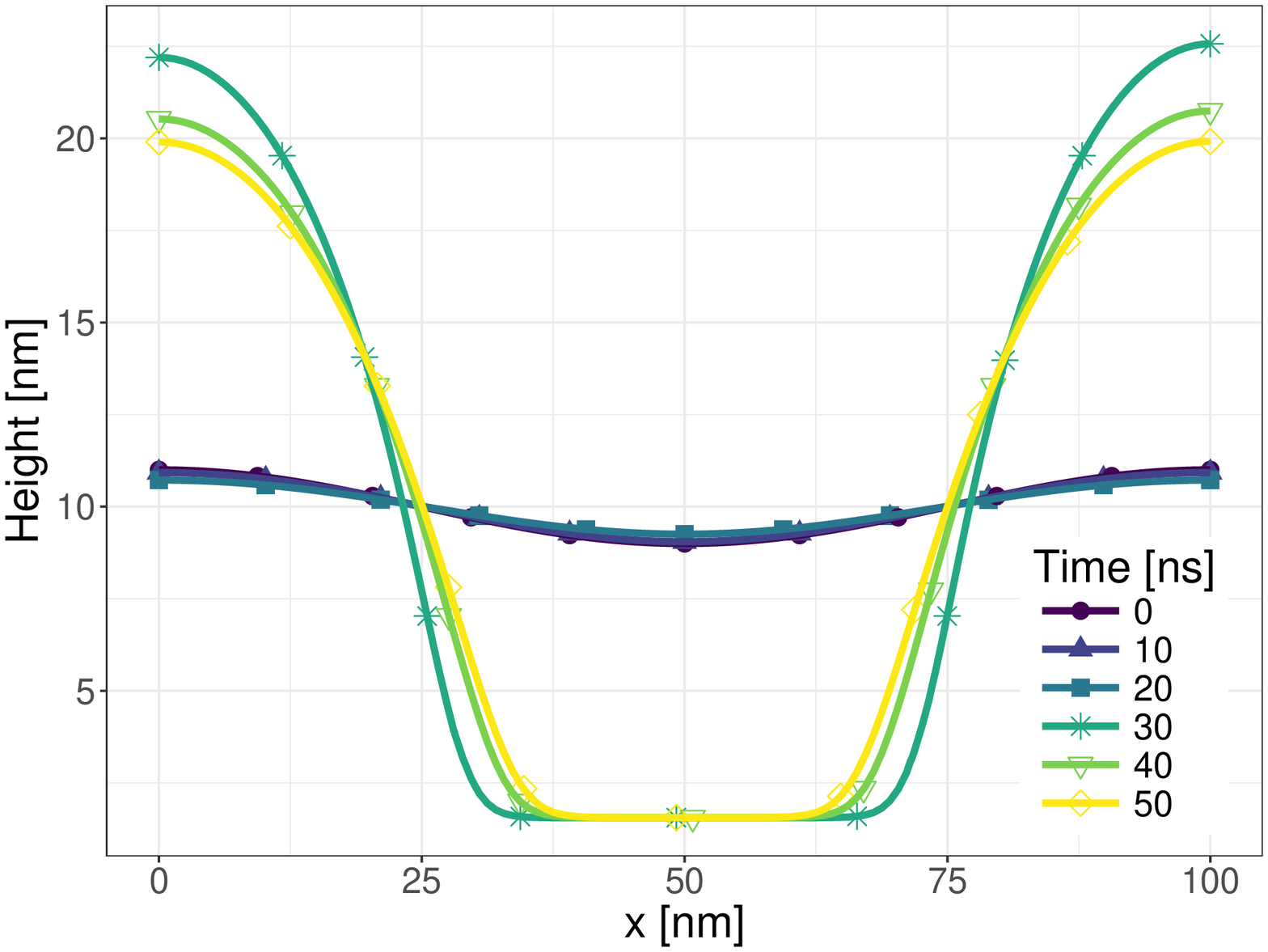}
  \caption{Complete model}
  \end{subfigure}
  \caption{The comparison of the evolution of the interface for the two models 
  considered,   for $h_0 = 10\, $nm and  $\lambda = 100\, 
  $nm.  The film is stable in (a) but goes through oscillatory instability in (b) 
  (for early times, until $t = 20$ ns, the imposed perturbation decays slightly in (b)).   
  Note also that for this and the following figures, the 
  evolution starts at $t \approx 11.4$ ns, the time  at which the film temperature rises above the melting temperature, $T_M$.
  Note different thickness scales in (a) and (b).  
   }
  \label{fig:024_compare}
\end{figure*}

Figure \ref{fig:024_compare} shows the evolution of the interface for a 
$10\, $nm film. We use stable perturbation wavelength, $\lambda = 100\, $nm, that is
slightly smaller than the critical one, 
$\lambda_c = 114\, $nm 
found from the LSA in Section~\ref{sec:LSA}. 
Hence, we expect the perturbation to be stable. 
Figure \ref{fig:024_compare}(a) shows the evolution of the interface with
temperature solution from the reduced model. The perturbation of the interface 
decays, as expected.
Figure \ref{fig:024_compare}(b) shows the evolution with the
temperature solution from the complete model, 
where initially (see $t = 20\, $ns) the perturbation decays, then grows for all following times, and the film eventually breaks into drops. 
Hence, the two temperature models, which agree for a flat
film, produce different evolution for a perturbed interface. 

Understanding the difference in stability resulting from the models requires considering 
both normal and tangential stress balances at the liquid-air interface, that is, both spatial 
gradients of the surface tension leading to Marangoni effect, and the temporal evolution of 
surface tension due to the evolution of temperature.    We discuss both of these effects 
next, for both reduced and complete models.

\begin{figure*}[thb]
\centering
  \begin{subfigure}{0.40\textwidth}
    \includegraphics[width = \textwidth]{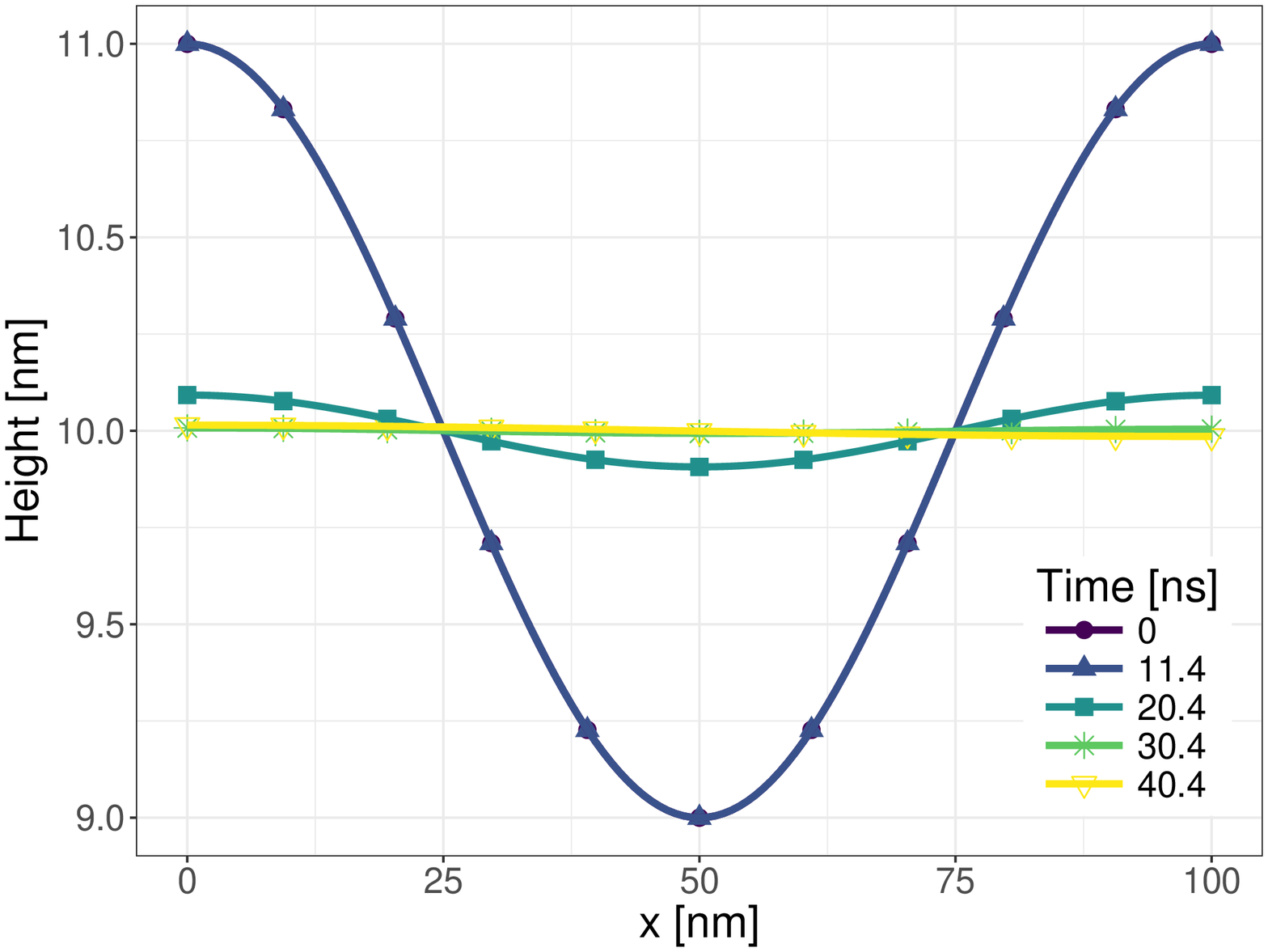}
    \caption{$ \sigma = \sigma(T_M)$} 
  \end{subfigure} \hspace{0.1in}
  \begin{subfigure}{0.40\textwidth}
    \includegraphics[width = \textwidth]{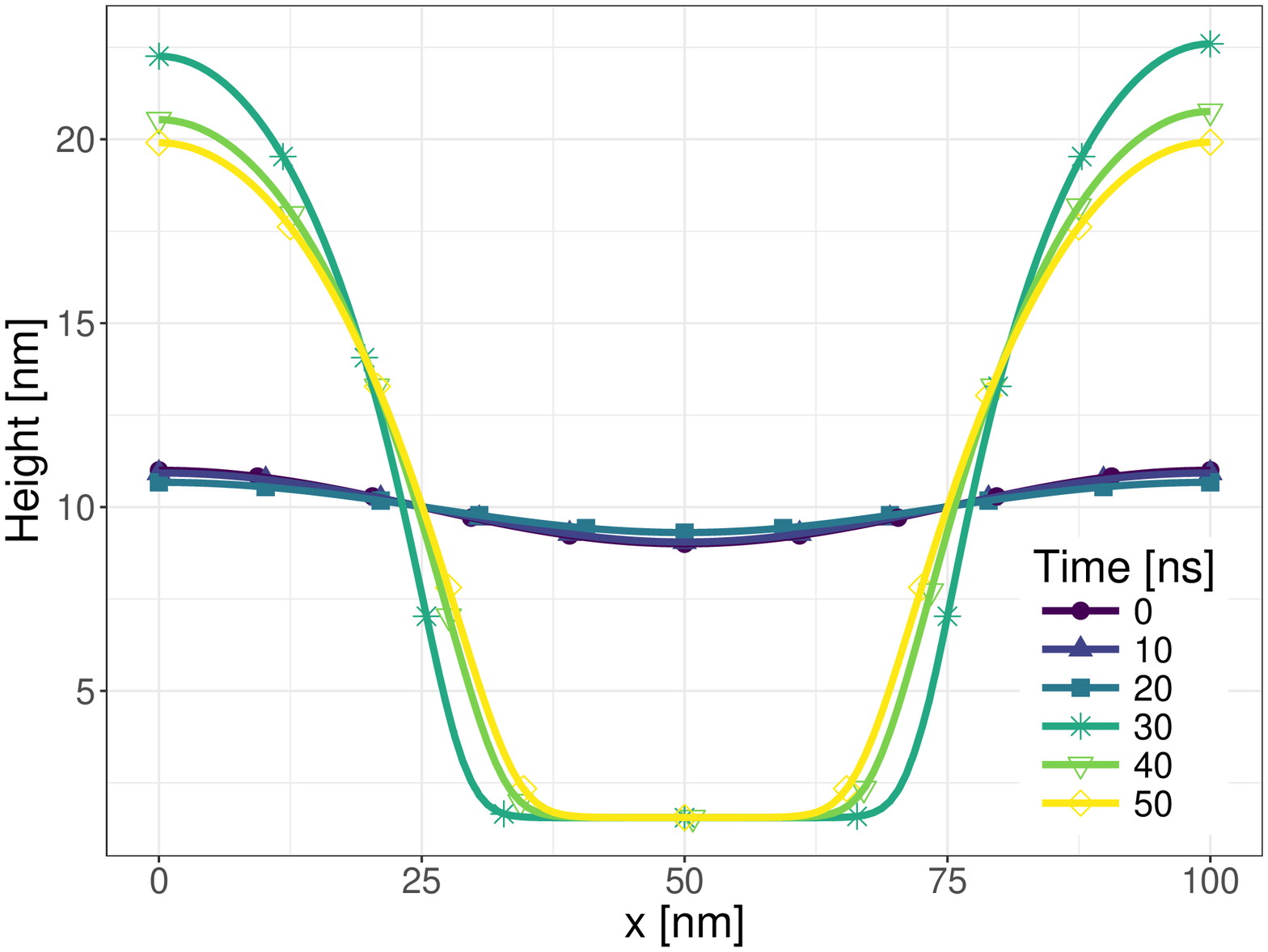}
  \caption{$\sigma = \sigma (T)$}
  \end{subfigure}
  \caption{The evolution of the interface with the temperature  solution from the complete model for
  $h_0 = 10\, $nm and $\lambda = 100\, $nm, and for the fixed (a) and temperature dependent (b) surface tension as noted.   
  Marangoni effect is not considered.   Note different thickness scales in (a) and (b).  
  }
  \label{fig:025_compare}
\end{figure*}

First, let us consider the influence of the temperature on the 
normal component of the surface force, and ignore the tangential component (therefore ignoring the Marangoni effect). 
Figure \ref{fig:025_compare} shows the evolution of the interface using the complete model, for  (a) constant 
surface tension, $\sigma = \sigma_0$, and (b) 
temperature dependent, $\sigma = \sigma (T)$, but the Marangoni effect is not 
included. 
In the simulation that use constant surface tension 
 the perturbation is stable, as expected from the dispersion relation, 
Eq.~(\ref{eq:dispersion_thermal}). 
However, when the surface tension dependence on the temperature is included, the 
perturbation initially decays (see $t = 20$~ns), but grows for all 
later times. 
Note in particular that the results shown in Figs.~\ref{fig:024_compare}(b) and  \ref{fig:025_compare}(b) are almost identical, showing 
that the Marangoni effect is essentially irrelevant in the present context.
Therefore, the stability change is not due to the 
spatial variations of $\sigma$, but due to the change 
of the normal component of the surface tension force due to temporal change of $\sigma$.

\begin{figure*}[tbh]
\centering
\begin{subfigure}{0.40\textwidth}
  {\includegraphics[width = \textwidth]{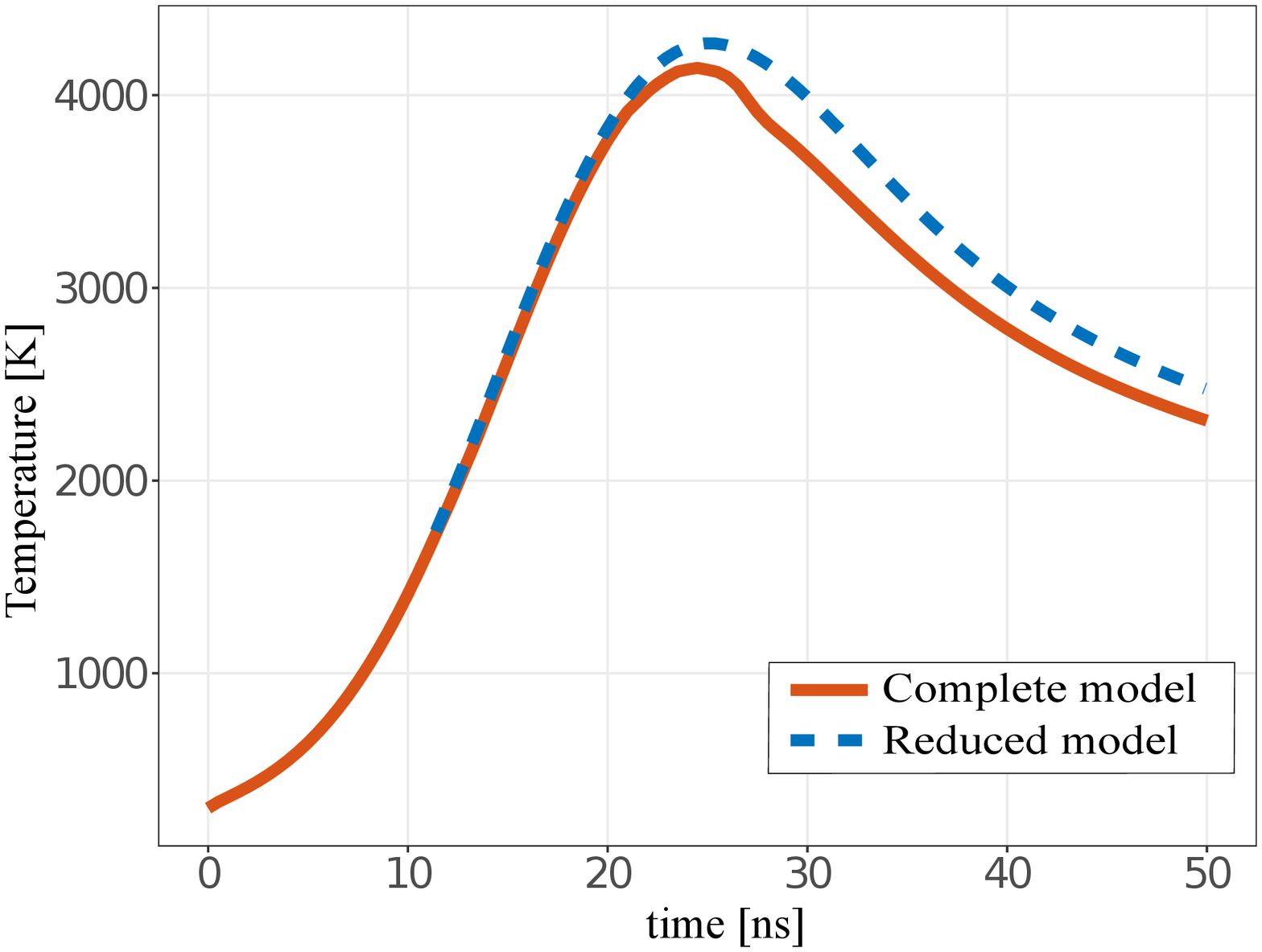}}
	\caption{}
\end{subfigure}\hspace{0.1in}
\begin{subfigure}{0.40\textwidth}
  {\includegraphics[width = \textwidth]{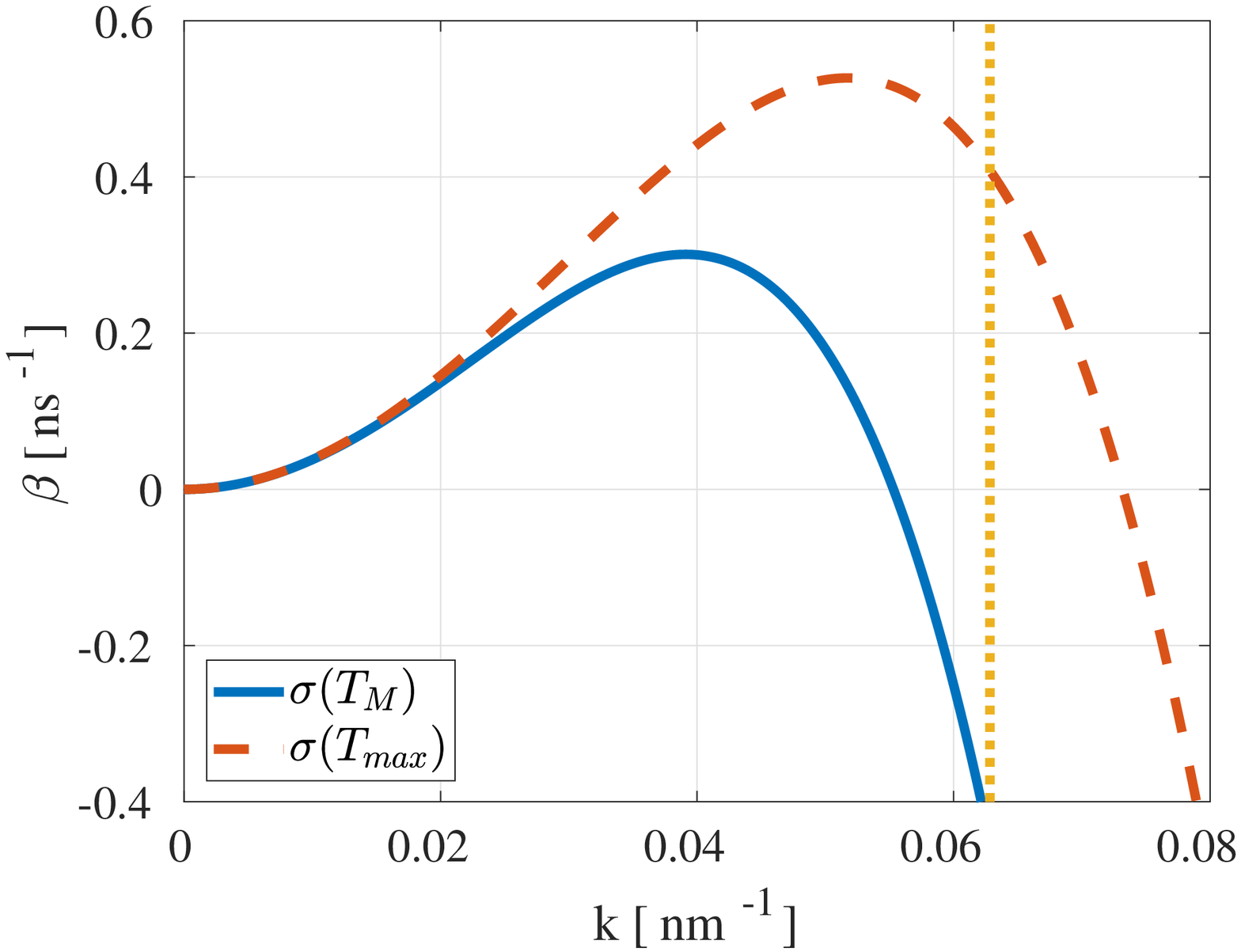}}
	\caption{}
  \end{subfigure}
    \caption{ (a) The average temperature of the metal film from Fig.~\ref{fig:024_compare}.   (b) Growth rate given by Eq.~\eqref{eq:dispersion_thermal}, for $h_0 = 10\, $nm  and for $\sigma$ at the melting temperature, $T_M$, and at the maximum temperature predicted by the reduced model, $T_{max}$.   The vertical dashed line in part (b) shows the value of $k$ used in the simulations.     Marangoni effect is not considered. }    \label{fig:growth_shift} 
\end{figure*}

We still need to explain why they the results shown in the parts (a) and (b) of Fig.~\ref{fig:024_compare} differ.
For this purpose, Fig.~\ref{fig:growth_shift}(a) shows the average temperature of the film in the 
reduced and complete models for the parameters as used in Fig.~\ref{fig:024_compare}. 
According to both models, from the melting time at 
$t = 11.4\,$ns, the temperature of the film increases to $T \gtrsim 4000\, $K, which 
corresponds to the decrease in the surface tension  from 
$\sigma (T_M) = 17.78\, \text{N~m}^{-1}$ to
$\sigma(T_{max}) = 10.15\, \text{N~m}^{-1}$. Figure \ref{fig:growth_shift}(b) shows the dispersion curve
computed using Eq.~\eqref{eq:dispersion_thermal}, with $\sigma_0 = \sigma (T_M)$ 
and $\sigma_0 = \sigma (T_{max})$, and $\partial T/ \partial h = 0$. The change in 
$\sigma_0$ shifts the critical wavenumber, $k_c$, and the stable perturbation ($\lambda = 100\, nm$ which corresponds to $k \approx 0.0628\, nm^{-1} $) becomes
unstable. This explains why the stable mode in simulations in Figs.~\ref{fig:024_compare}(b) 
and \ref{fig:025_compare}(b) becomes unstable as the film temperature increases, 
eventually leading to the film breakup.  Note that we have already shown that Marangoni effect is not
relevant for the simulations that use the complete model for temperature calculation.   

\begin{figure*}[tbh]
\centering
\begin{subfigure}{0.40\textwidth}
 \includegraphics[width = \textwidth]{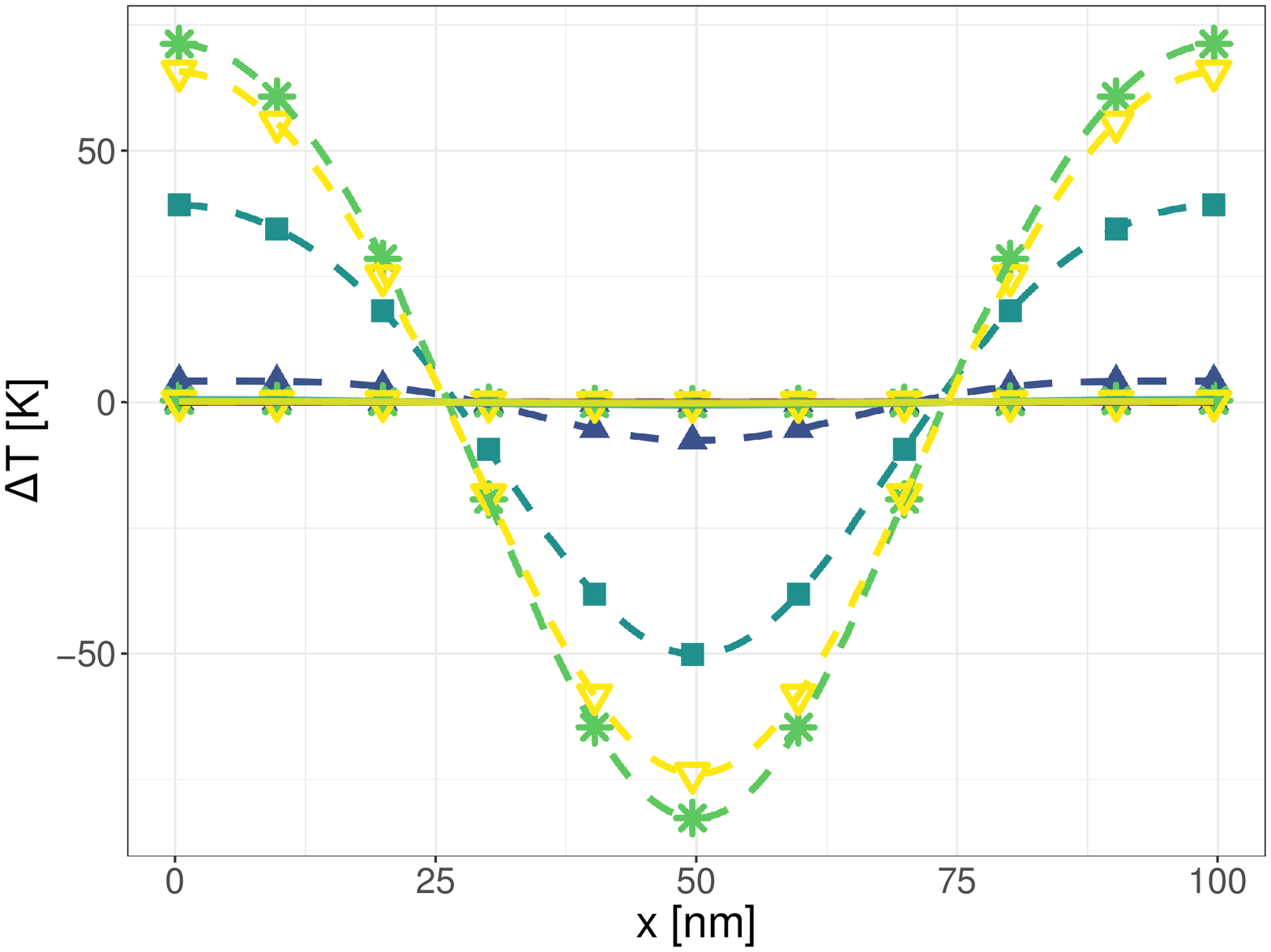}
\caption{$h_0 = 10\,$nm,  $\lambda = 100\, $nm}
\end{subfigure}\hspace{0.1in}
\begin{subfigure}{0.40\textwidth}
   \includegraphics[width = \textwidth]{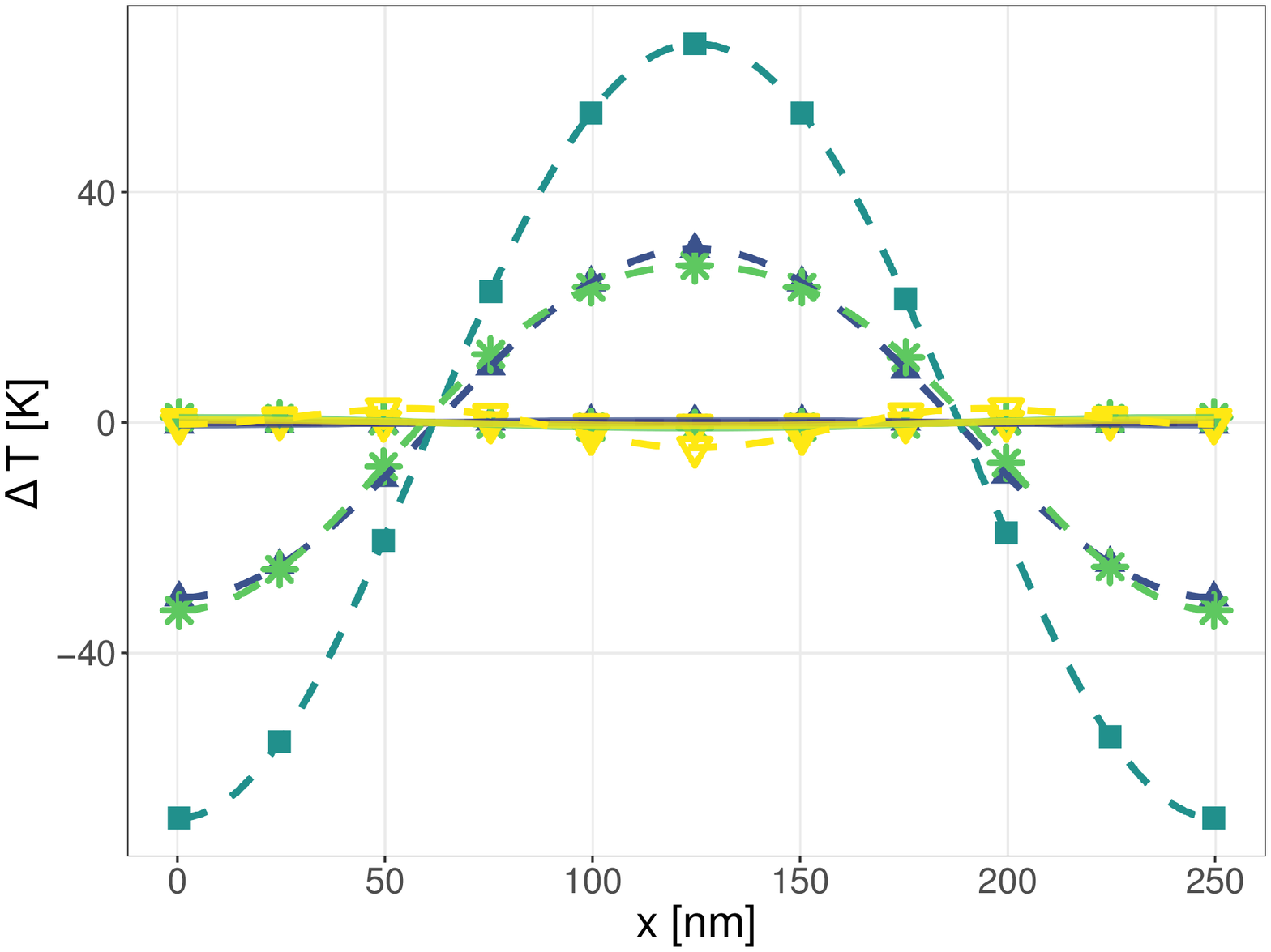}
  \caption{$h_0 = 20\,$nm, $\lambda = 250\, $nm}
\end{subfigure}
\begin{subfigure}{0.07\textwidth}
   \includegraphics[width = \textwidth]{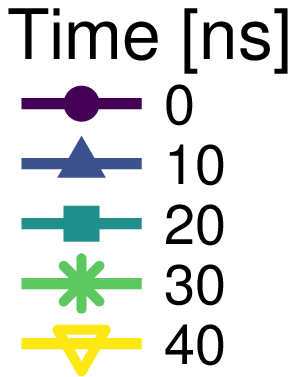}
\end{subfigure}
 \caption{The difference, $\Delta T$, between the computed and average temperatures at the liquid metal-air interface for a static perturbed film showing 
 the results for the reduced (dashed) and the complete (solid) models.  Note that for the temperature scale shown, the temperature gradients for the 
 complete model are almost invisible.
 } 
 \label{fig:temp_horiz_grad}
\end{figure*}
In contrast to the results obtained by implementing the complete temperature model shown in 
Figs.~\ref{fig:024_compare}(b) and \ref{fig:025_compare}(b), the  results obtained with the temperature 
from the reduced model, Fig.~\ref{fig:024_compare}(a), show stability, despite 
the fact that the average temperature of the film increases similarly as in the complete 
model, see Fig.~\ref{fig:growth_shift}(a). 
To gain better understanding of this finding, we examine the difference in the 
temperature solutions along the 
liquid-air interface for a perturbed {\it stationary} film, corresponding to the 
initial condition used in Figs.~\ref{fig:024_compare} and~\ref{fig:025_compare}.    (The motivation for considering a stationary 
film is the source term dependence on the film thickness - the film evolution 
would affect the source term, and we prefer to avoid this effect for simplicity of the argument.)
Figure \ref{fig:temp_horiz_grad}(a) shows the differences of the computed temperature along the 
interface from the average temperature. The temperature varies significantly more for the reduced model compared to the complete one.
Thus, the temperature gradients at the liquid-air interface are significantly larger for the reduced model,
and the stabilizing Marangoni effect prevents the interface in Fig.~\ref{fig:024_compare}(a) 
from becoming unstable.    This finding explains the different film evolution between the two models.  

To summarize the results for the film thickness of $h_0 = 10$ nm: If the temperature is computed using the complete
model, the Marangoni effect turns out to be irrelevant; however the temporal change of surface tension due to evolving 
laser pulse and film thickness may influence film stability, leading to instability in the case considered here.  If the temperature
is computed using the reduced model, then (stabilizing) Marangoni effect may compete with the destabilizing effect of the overall surface
tension decrease, leading to stability.   Therefore, computing temperature carefully is crucial for understanding the film stability.

\subsubsection{Film of Thickness $20$ nm; $dT/dh < 0$}

\begin{figure*}[thb]
\centering
  \begin{subfigure}{0.40\textwidth}
    \includegraphics[width = \textwidth]{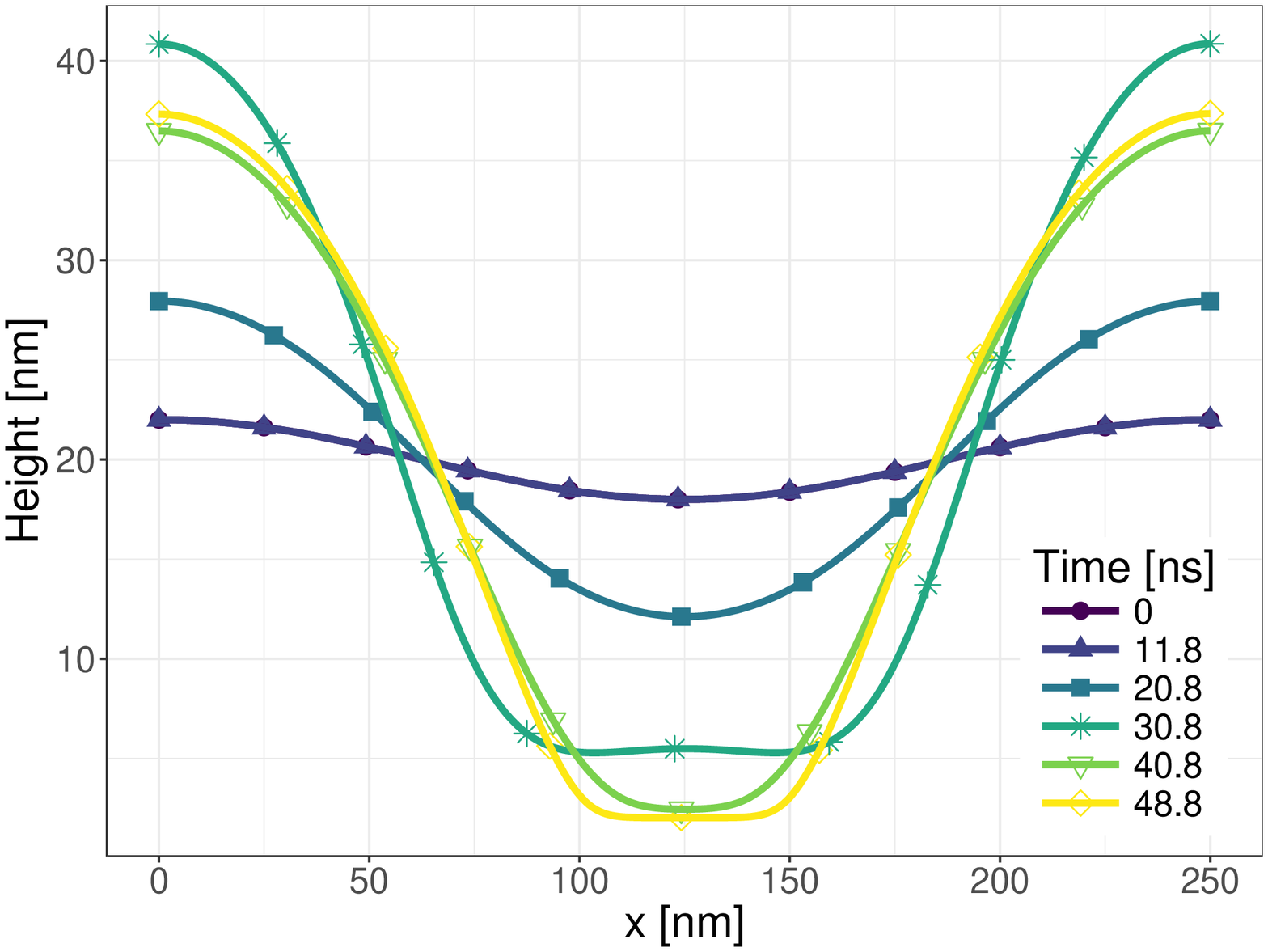}
    \caption{Reduced model} 
  \end{subfigure}\hspace{0.1in}
  \begin{subfigure}{0.40\textwidth}
    \includegraphics[width = \textwidth]{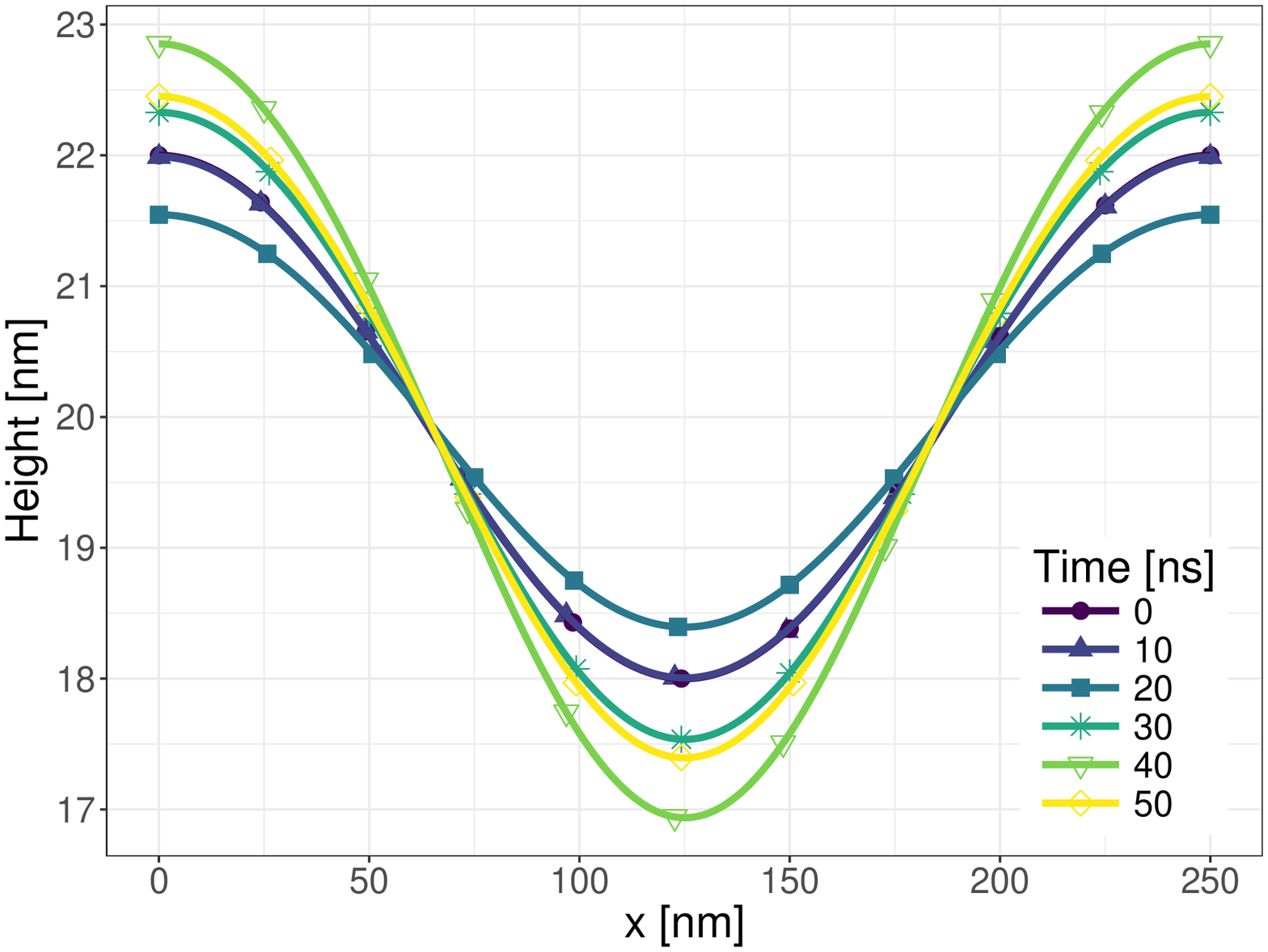}
  \caption{Complete model}
  \end{subfigure}
  \caption{The comparison of the evolution of the interface for the two models 
  considered,   for $h_0 = 20$ nm and  $\lambda = 250$  
  nm.  Here, the film is unstable in (a), but goes through 
  oscillatory instability in (b).     Note that the 
  evolution starts at $t \approx 11.8$ ns, the time at which the film temperature rises above the melting temperature, $T_M$.
  Note also different thickness scales in (a) and (b).    
  }
  \label{fig:042_compare}
\end{figure*}
Next, we consider thicker film, with the average thickness of $h_0 = 20\, $nm.   
Recall that  for $h_0  = 20$ nm, the reduced model predicts essentially the opposite direction of 
the Marangoni effect relative to $h_0 = 10$ nm, see Fig.~\ref{fig:Trice_Ni_10_20}. 
Here  we impose a perturbation of the wavelength, $\lambda = 250\, $nm, which is stable when Marangoni effect 
is ignored, see the dispersion relation, Eq.~\eqref{eq:dispersion_thermal}. 
Figure \ref{fig:042_compare} shows the comparison of the evolution of the interface using 
the reduced and complete temperature models. 
Using the reduced model,  Fig.~\ref{fig:042_compare}(a), we find instability,  
and the perturbation grows until the film breaks into drops. 
This is not surprising since the reduced temperature model predicts 
$\partial T_m^*/ \partial h < 0$ which 
destabilizes the film (see Fig.~\ref{fig:Growth_fixed_h}(b)). 
When  the temperature is computed using the complete model however, see 
Fig.~\ref{fig:042_compare}(b), the evolution is {\it oscillatory}:
at $t = 20\, $ns the perturbation decays; at 
$t = 30\, $ns and $t = 40\, $ns the perturbation grows; and at $t = 50\, $ns
the perturbation decays again. 
Similarly as for $h_0 = 10\, $nm, to  explain these dynamics, we examine 
the simulations without Marangoni effect,
and investigate the influence of the temperature on the normal component
of the surface force. 

\begin{figure*}[tbh]
\centering
  \begin{subfigure}{0.40\textwidth}
    \includegraphics[width = \textwidth]{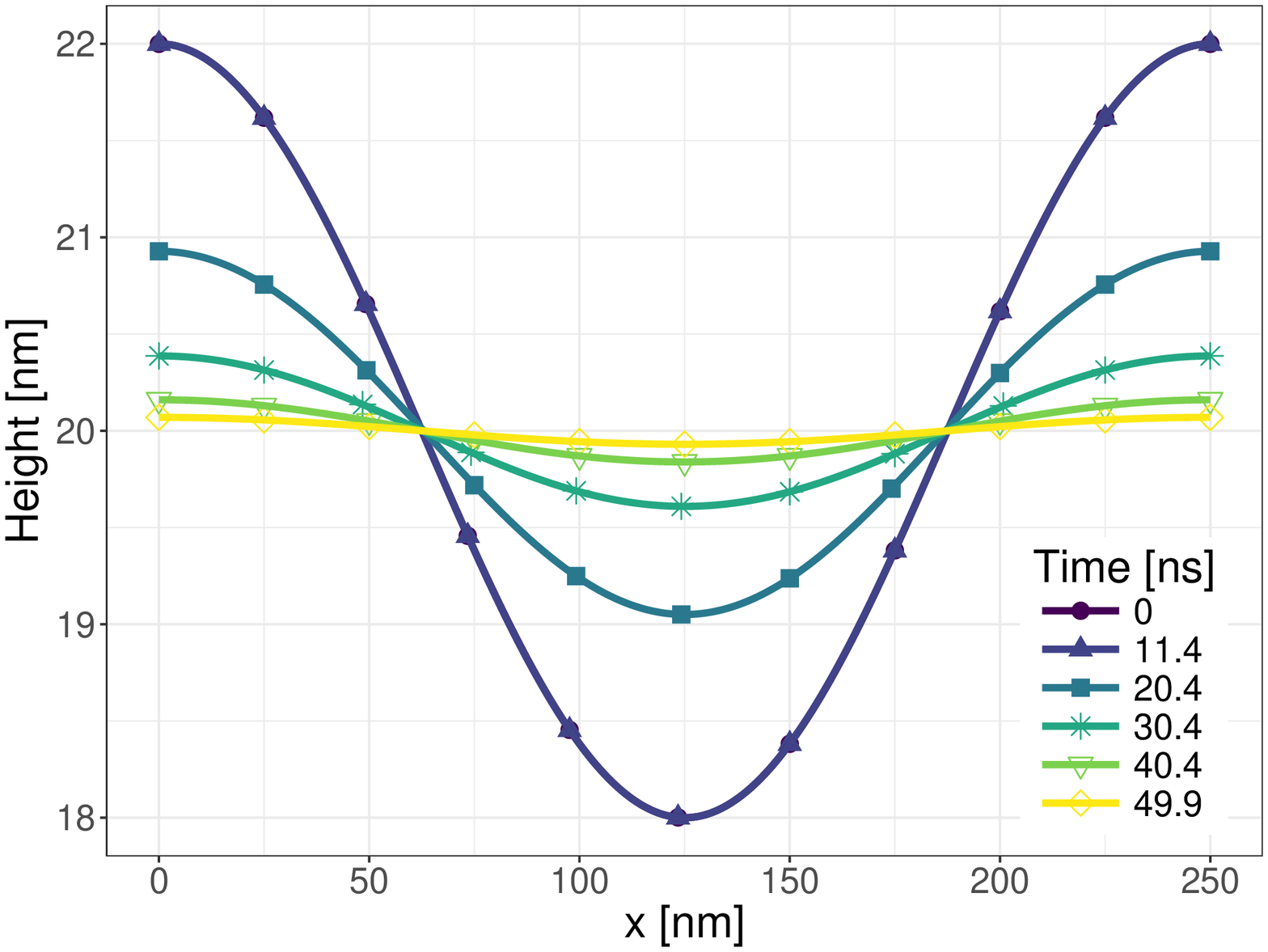}
    \caption{$\sigma = \sigma (T_M)$} 
  \end{subfigure}\hspace{0.1in}
  \begin{subfigure}{0.40\textwidth}
    \includegraphics[width = \textwidth]{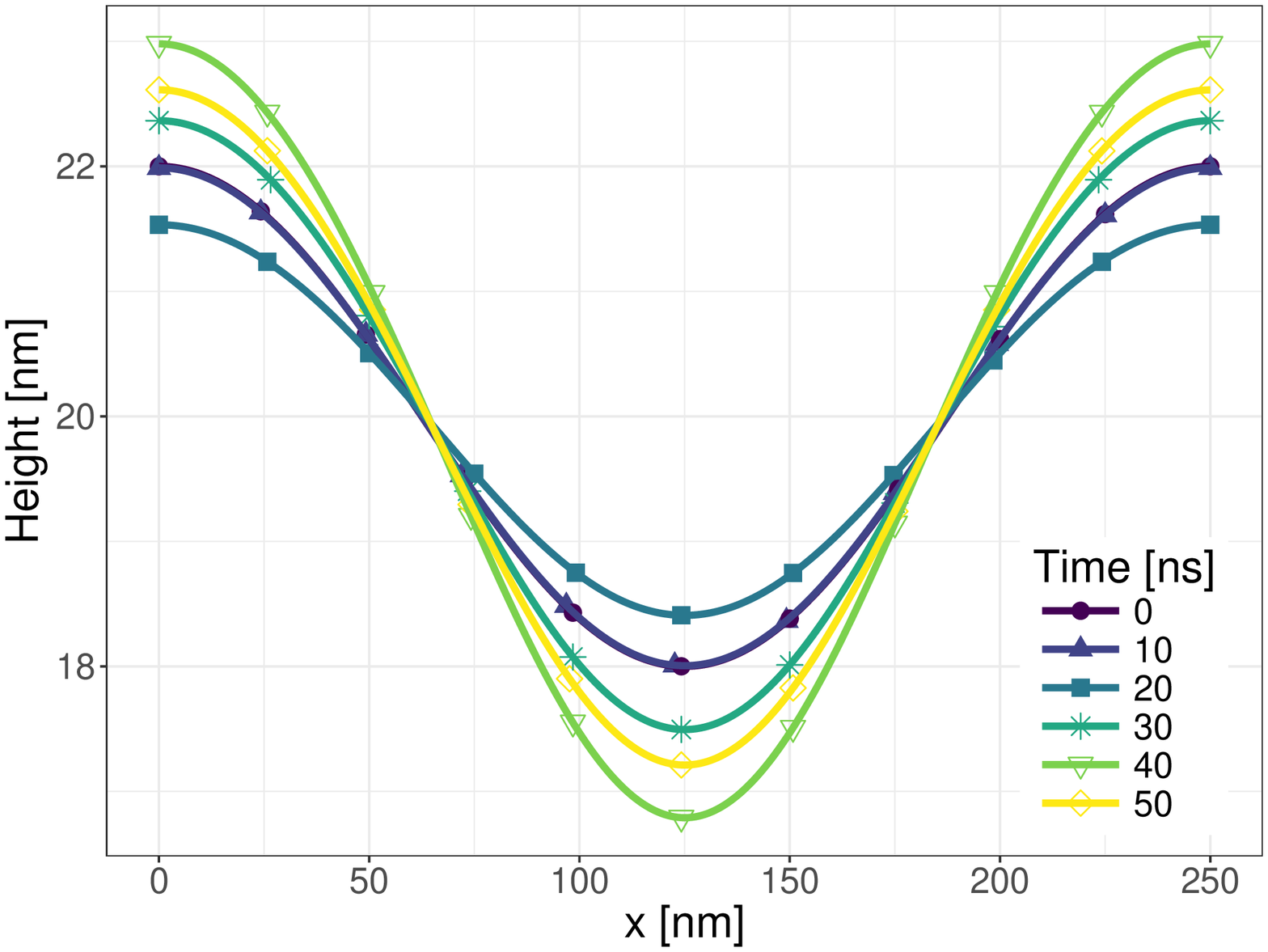}
  \caption{$\sigma = \sigma (T)$    }
  \end{subfigure}
    \caption{The evolution of the interface with the temperature  solution from the complete model for
  $h_0 = 20$ nm and $\lambda = 250$ nm, and for the two values of surface tension as noted.   
  Marangoni effect is not considered.   Note different thickness scales between (a) and (b).  
}  
  \label{fig:043_compare}
\end{figure*}

Figure \ref{fig:043_compare} shows the evolution of the interface when
the surface tension  is (a) constant,  $\sigma = \sigma_0$, and (b) 
temperature dependent, $\sigma = \sigma (T)$, but Marangoni effect is ignored. 
When constant  surface tension is considered, the perturbation is stable, as expected from the LSA. 
However, for temperature dependent surface tension, we uncover the 
same dynamics as  in Fig.~\ref{fig:042_compare}(b). Thus, we see again that the Marangoni force
is negligible.  We show next that the changes in the stability in the complete model are due to the 
temporal changes of the surface tension.

\begin{figure*}[tbh]
\centering
\begin{subfigure}{0.40\textwidth}
  {\includegraphics[width = \textwidth]{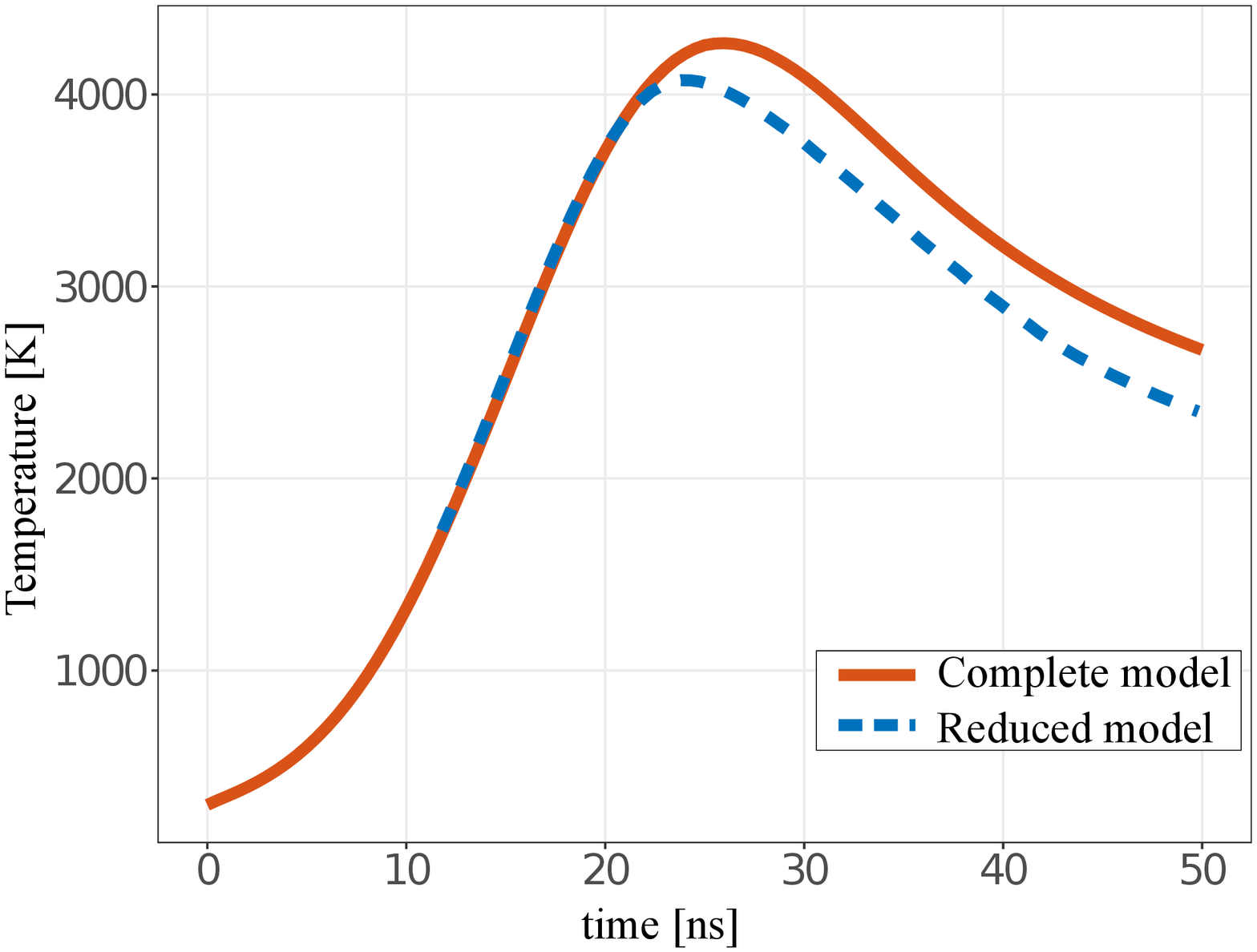}}
	\caption{}
\end{subfigure}\hspace{0.1in}
\begin{subfigure}{0.40\textwidth}
  {\includegraphics[width = \textwidth]{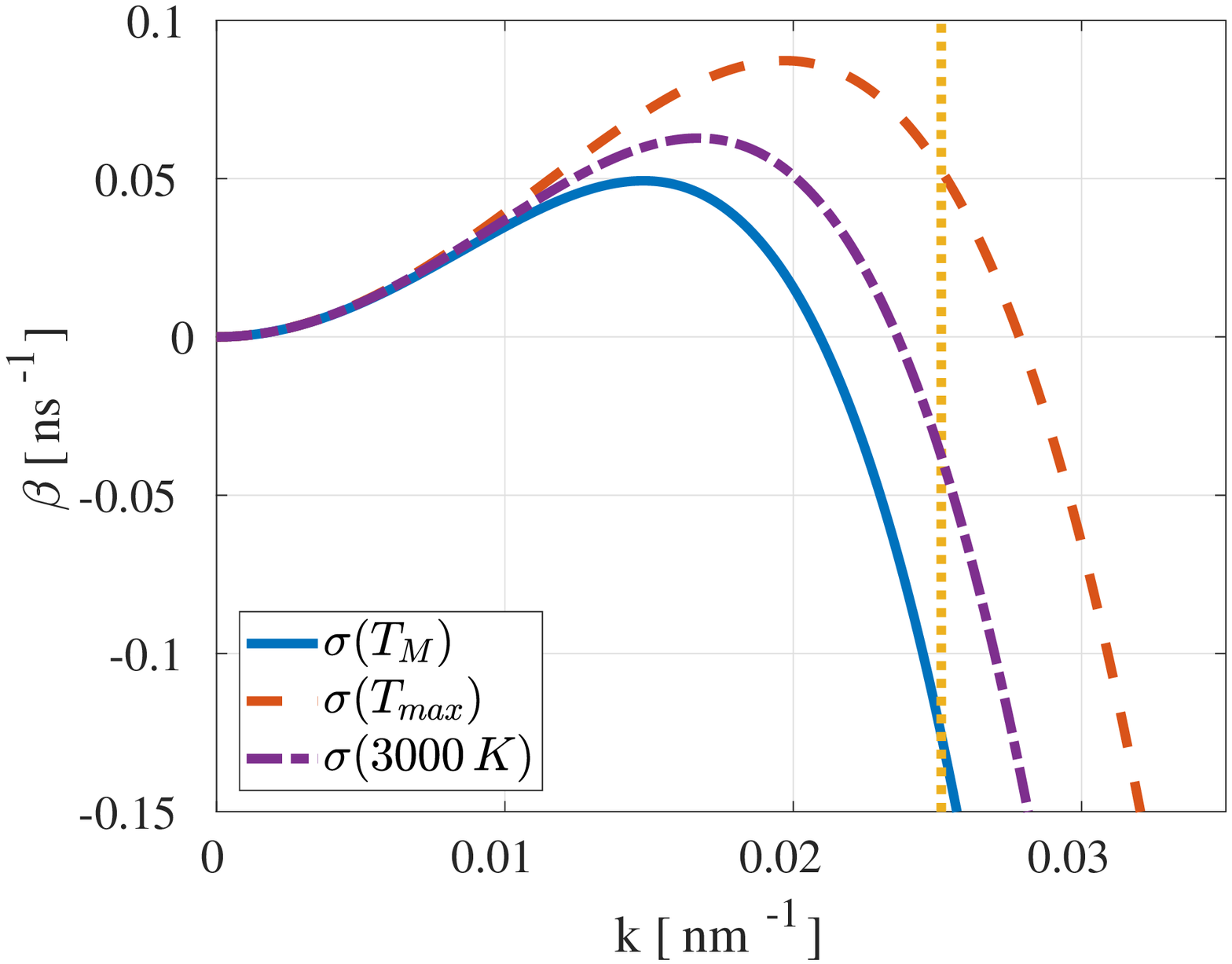}}
	\caption{}
  \end{subfigure}
    \caption{ (a) The average temperature of the metal film from Fig.~\ref{fig:042_compare}.   (b) Growth rate given by 
    Eq.~\eqref{eq:dispersion_thermal}, for $h_0 = 20\, nm$, $\partial T/ \partial h = 0$, and $\sigma$ at the melting temperature, $T_M$, and the maximum temperature 
    predicted by the reduced model, $T_{max}$.    The vertical dashed line in part (b) shows the wave number used in the simulations. 
        }  
    \label{fig:growth_shift_20} 
\end{figure*}

Figure \ref{fig:growth_shift_20}(a) shows the average temperature of the film using 
the reduced and complete models for the parameters as used in Fig.~\ref{fig:042_compare}.
From the melting time, at $t = 11.8 \, $ns, the temperature evolution (and the surface tension 
change) are similar as for the $h_0 = 10$ nm films.    
Figure \ref{fig:growth_shift_20}(b) shows the dispersion curve
computed using Eq.~\eqref{eq:dispersion_thermal}, with $\sigma_0 = \sigma (T_M)$ 
and $\sigma_0 = \sigma (T_{max})$, and $\partial T/ \partial h = 0$. The change in 
$\sigma_0$ shifts the critical wavenumber, $k_c$, and the (linearly) stable perturbation ($\lambda = 250\, $nm which corresponds to $k \approx 0.02512\, \text{nm}^{-1} $) becomes
unstable, as we see in Figures \ref{fig:042_compare}(b) and \ref{fig:043_compare}(b)
after $t = 20\, $ns. After $t = 40\, $ns, the temperature decreases again to 
$T \approx 3000$ K, hence (see Fig.~\ref{fig:growth_shift_20}(b)), the perturbation becomes (linarly) stable again.
In summary, similarly to the $h_0 = 10\, $nm film, the stability of the 
interface using the complete model is governed by the temporal variations of $\sigma$. 

Similarly as for the $h_0 = 10\, $nm film, the temporal changes 
of the surface tension do not explain the film instability 
for the temperature from the reduced model in Fig.~\ref{fig:042_compare}(a).  
Therefore, we compare again the temperature at the liquid-air 
interface of a {\it stationary} film using the reduced and complete models. 
Figure \ref{fig:temp_horiz_grad}(b) shows the deviation from 
the average temperature at the interface of a stationary film corresponding to the 
initial condition for the simulations shown in 
Fig.~\ref{fig:042_compare}. We see once again that the effect of the 
Marangoni effect is augmented significantly by  the reduced model.

To summarize, we find that for both thin and thick films (relative to the critical thickness at which $d T/dh$ changes
sign), the complete and reduced model produce different results, showing clearly that careful computation of 
heat flow is required to accurately describe the evolution.   Using the reduced model, or in other words ignoring
the heat conduction in the in-plane direction, leads to qualitatively different results compared to 
the ones obtained if this assumption is not made. 

\subsection{The Influence of the Temperature Dependent Viscosity}
\label{sec:viscosity}

Here we focus the influence of the viscosity variations with temperature 
on the stability and breakup dynamics.
During the metal heating and melting, the viscosity of the metal changes 
several orders of magnitude \cite{Smithells}. The viscosity of most 
metals can be modeled by an exponential as
\begin{equation}
\label{eq:visc_temp}
    \mu \left( T \right) = \mu_0 \exp\left(\frac{E}{RT} \right)
\end{equation}
where $\mu_0 = 0.1663 \, \text{mN~s~m}^{-2}$ and $E = 50.2\, \text{kJ~mol}^{-1}$ are constants dependent 
on the material, and $R = 8.3144\, \text{J~K}^{-1}\text{mol}^{-1}$ is the gas constant \cite{Smithells}. 
\begin{figure*}[thb]
\centering
\begin{subfigure}{0.40\textwidth}
 \includegraphics[width = \textwidth]{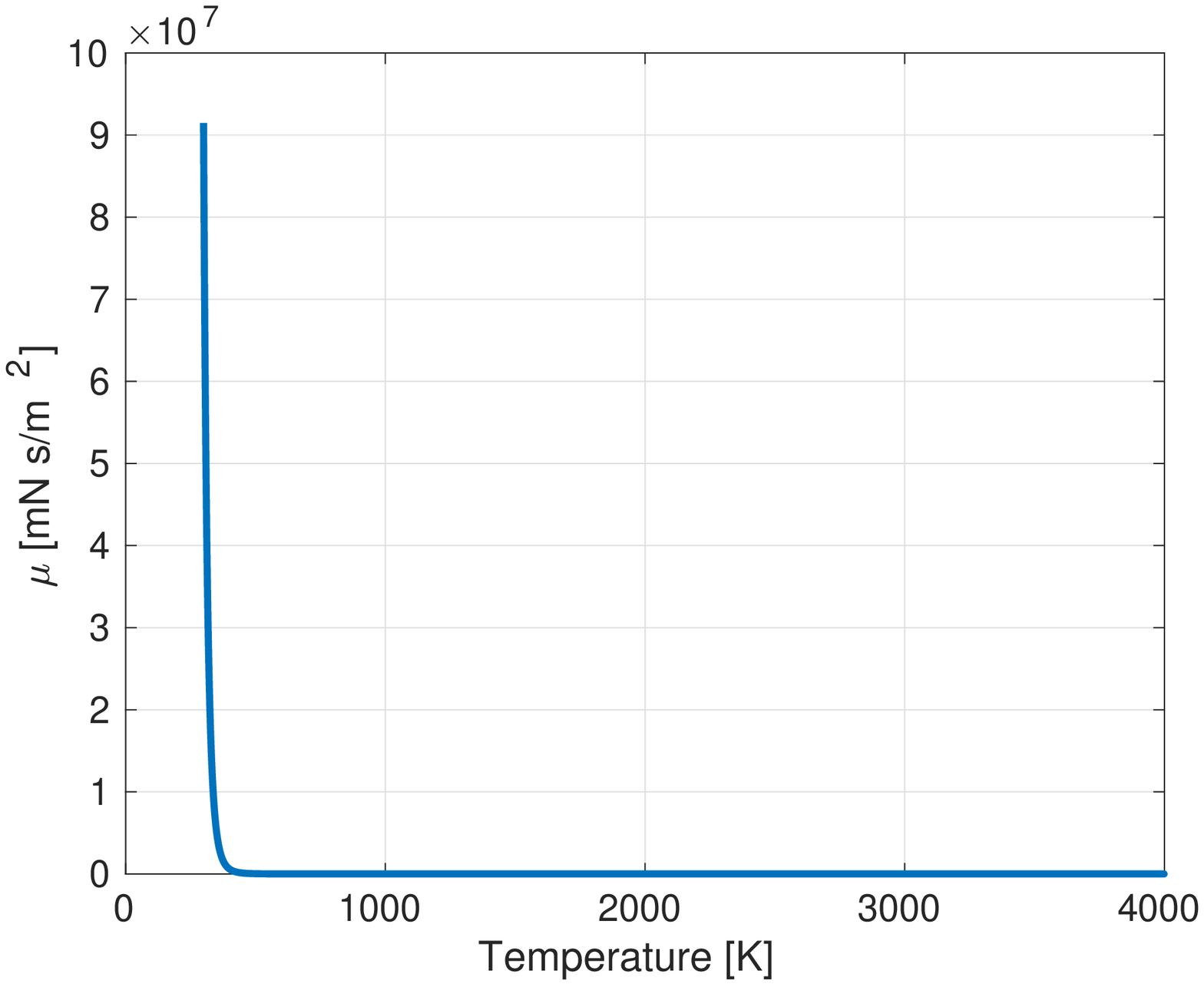}
\caption{}
\end{subfigure}\hspace{0.1in}
\begin{subfigure}{0.40\textwidth}
   \includegraphics[width = \textwidth]{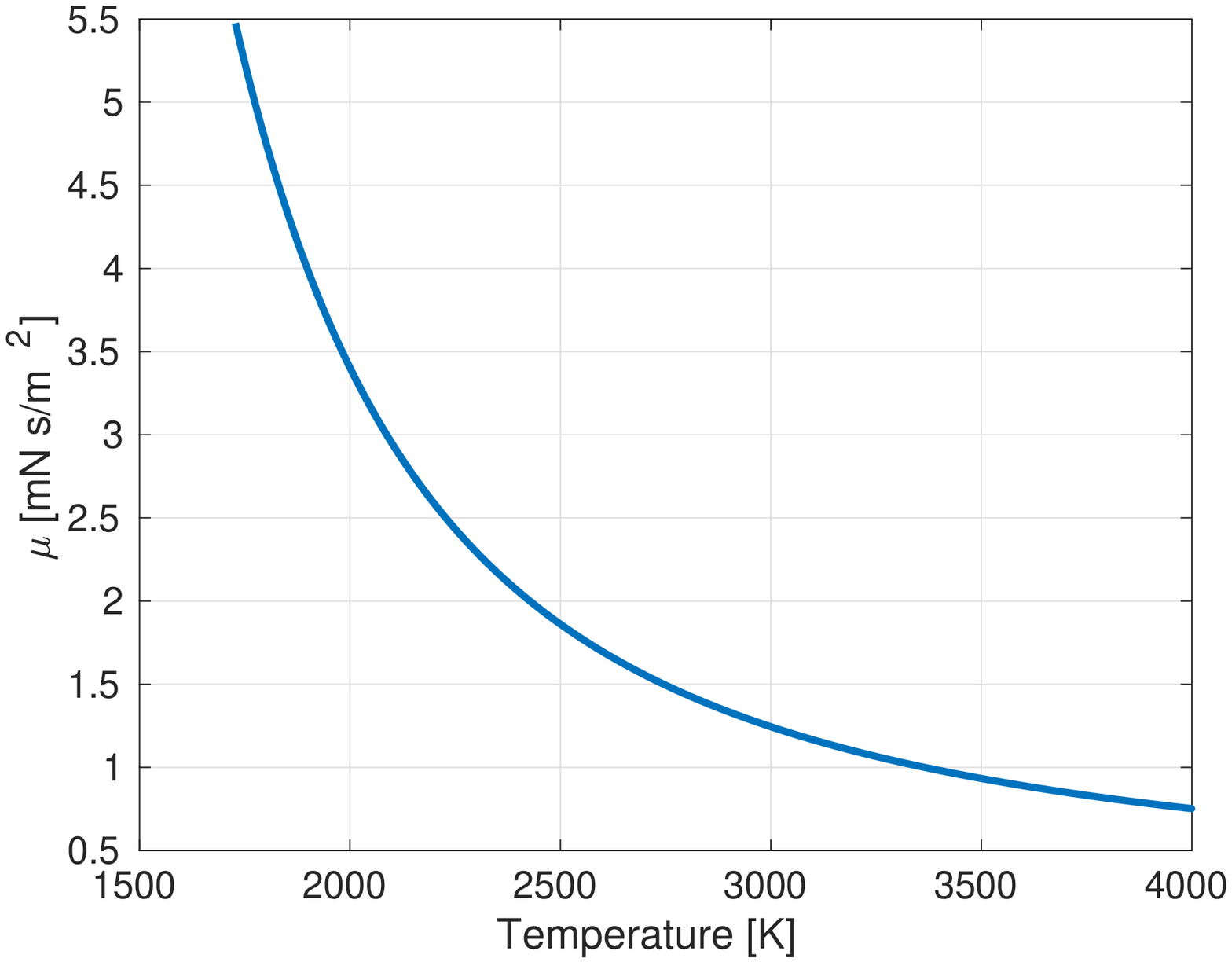}
  \caption{}
\end{subfigure}
 \caption{ Viscosity of nickel as a function of temperature given by Eq.~\eqref{eq:visc_temp}, for temperature range starting from (a) room temperature and (b) melting temperature of nickel.  } 
 \label{fig:viscosity} 
\end{figure*}
Figure  \ref{fig:viscosity} shows the viscosity of nickel as a function of
temperature.

The influence of the temperature dependent viscosity on the film breakup can be
estimated based on the dispersion relation specified by Eq.~\eqref{eq:dispersion_thermal}. 
This relations says that the stability of the film, and in particular the critical wave number, $k_c$, do not depend on 
the viscosity. 
However, the growth rate, $\beta$, is inversely proportional to 
$\mu$ and, as we will see, this may be sufficient to influence the stability. 

\begin{figure*}[thb]
\centering
\begin{subfigure}{0.40\textwidth}
 \includegraphics[width = \textwidth]{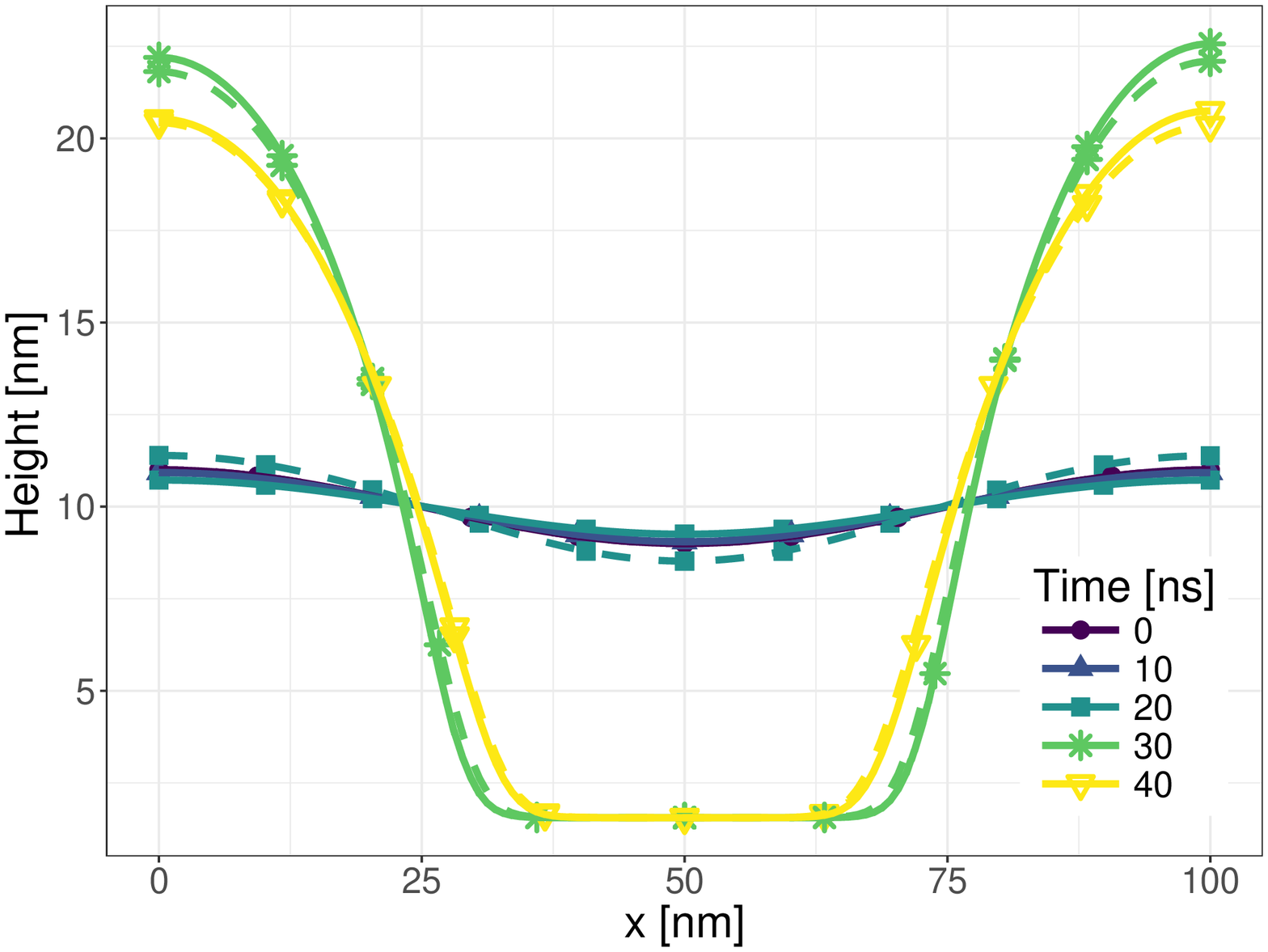}
\caption{$h_0 = 10$ nm   }
\end{subfigure}\hspace{0.1in}
\begin{subfigure}{0.40\textwidth}
   \includegraphics[width = \textwidth]{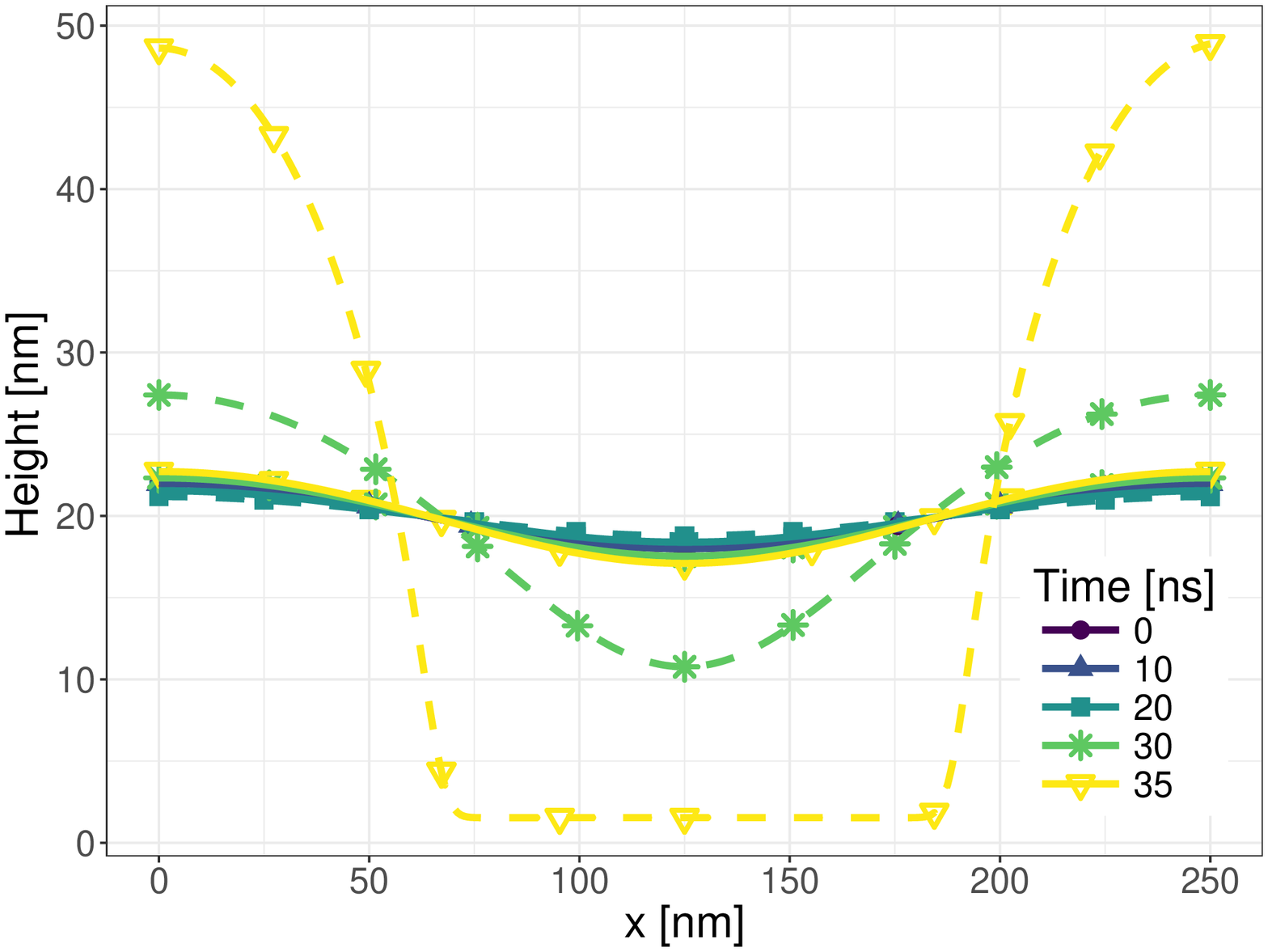}
  \caption{$h_0 = 20$ nm }
\end{subfigure}
 \caption{ The comparison of the evolution of the film interface with constant 
 (full line) and temperature dependent viscosity (dashed line) using the complete
 temperature model.  
  }  
 \label{fig:visc_film} 
\end{figure*}

To study the influence of  the variable viscosity on the breakup dynamics, 
we use the same initial geometry as in Section~\ref{sec:film} within the 
framework of the 
complete model described in Section~\ref{sec:dns_model}.  Here, surface tension is taken as temperature
dependent, but for simplicity we do not include Marangoni effect (which is essentially irrelevant for the complete
model).   Figure \ref{fig:visc_film} shows the evolution of 
the film interface with temperature dependent viscosity compared to the evolution 
for constant viscosity, $\mu = \mu (T_M)$.  For the $10 \, $nm film, the same
evolution dynamics are present in both cases: perturbations initially decay, but 
start growing as the film temperature rises (see the discussion
related to Figs.~\ref{fig:024_compare}(b) and~\ref{fig:025_compare}(b)).
As expected based on the LSA, the stability of the perturbations is not 
affected by the variable viscosity, but the growth rate is faster, and therefore 
the breakup time occurs $\approx 5 \, $ns faster with variable viscosity compared to the constant one (see Section~\ref{sec:film}). 
For the $20\,$nm film, again, the decay and the growth of the perturbations for 
the variable viscosity follows the same direction as the constant viscosity
in Figures \ref{fig:042_compare}(b) and \ref{fig:043_compare}(b). 
During the time of the perturbation growth, as in 
$t = 20\, $ns to $t = 40\, $ns for $\mu = \mu (T_M)$, 
the perturbation for $\mu = \mu (T)$ grows fast enough so that the film breaks. 
Recall that for $\mu = \mu (T_M)$, the stability changes after $t = 40\, $ns, 
and the film stabilizes due to the decrease of the film temperature. 
Therefore, inclusion of temperature dependent viscosity can strongly influence the film evolution.

\subsection{Discussion of the Influence of Thermal Effects on Film Stability}
\label{sec:discussion}

At the beginning of Section~\ref{sec:results}, we motivated the focus of this work on simple 
computational domains and initial conditions.   In principle, more complex setups 
would be needed to reach precise understanding regarding the influence of thermal 
effects in physical experiments where the relevant domains are large and the 
film and temperature perturbations are more complicated.    However, 
based on the existing results, we are already in the position to develop a basic insight. 

To be specific, let us ask for the following question: What is the influence of the
fact that the temperature of metal films rises significantly above melting temperature 
on the film stability?  Focusing first on the temperature dependence of surface 
tension, we start by noting that Marangoni effect is not relevant.  However, the overall
decrease of surface tension as the temperature of a film raises suggests that shorter
wavelengths are expected, see the dispersion curves in Figs.~\ref{fig:growth_shift}(b) and~\ref{fig:growth_shift_20}(b).  
Possibly, such an increase of surface tension may be responsible for observation of 
shorter wavelengths in experiments~\cite{trice_prl08}.   
Clearly, as the film temperature changes as a function of time due to time-dependent 
source term (laser pulse) and heat loss through the substrate, the effect of the surface
tension change will be time dependent as well.   Turning now to temperature dependence
of film viscosity, we note that viscosity influences the time scale of instability growth and 
could therefore influence the film stability strongly, as discussed in Section~\ref{sec:viscosity}. 

To conclude this brief discussion, a variation of material parameters with temperature clearly 
influences film stability in a manner which may be complex in particular due to the fact 
that the relevant time scales, related to the source term and to the temporal evolution 
of the film itself, are comparable.   In general, based on the results presented so far, one
expects that heating of the films above melting will result in a decrease of the emerging
wavelengths (such as the distance between drops that form eventually), compared to the ones expected if one assumes that the material parameters
(surface tension, viscosity) are given by their values at the melting temperature.  The details of the instability 
evolution however may depend on the particular choice of metals, substrates, and 
laser pulse energy and duration.

\subsection{Breakup of Liquid Metal Filaments}
\label{sec:filaments}

In a recent work \cite{hartnett2017}, that included both experimental and computational study,
we considered {\it concentration} Marangoni effect in a two-metal setup focusing on 
metal filament geometry.   In that study, 
it was found that {\it concentration} Marangoni effect played a significant role.  In particular, 
the Marangoni induced flow led to inversion of instability development, in the sense that 
initially thicker filament parts (nickel covered by a thin copper film) ended up thinning, while initially thinner filament parts 
increased in thickness due to {\it concentration} induced Marangoni flow. 
The question that we will consider in the present paper, is whether a similar effect could
be observed for {\it thermal} Marangoni effect. 

To answer this question, we consider the following setup: 
 the initial geometry is a flat filament with  superimposed rectangular perturbations (similarly as in~\cite{hartnett2017}
 just with a single metal (nickel).   
 Since we know from Section~\ref{sec:film} that the temperature gradients
are small in the metal film, we increase the thickness of the rectangular perturbations 
compared to~\cite{hartnett2017} in an attempt to increase the temperature
gradients.    Therefore we take the base 
filament thickness as $h_0 = 8\, $nm, and superimpose the perturbations of 
$\Delta h = 8 \, $nm (so that the film thickness vary between $8$ and $16$ nm), and the width of the filament is $w = 185\, $nm. 
The average filament thickness is kept at $12\,$nm as in~\cite{hartnett2017}.
We note that the filament simulations are carried out using the approach described
in~\cite{hartnett2017}; briefly, we do not consider disjoining pressure here but instead
specify the contact angle ($90^\circ$ for simplicity), and use Navier-slip boundary 
condition with the slip length of $20$ nm.   The simulations consider 
one half of the perturbation wavelength and half 
of the filament width, and impose symmetry boundary conditions at all in-plane 
directions.  The fluid is stationary until the melting time of the filament (see Section~
\ref{sec:film}).   

\begin{figure}[tbh]
  \centering
{\includegraphics[width = 0.45\textwidth]{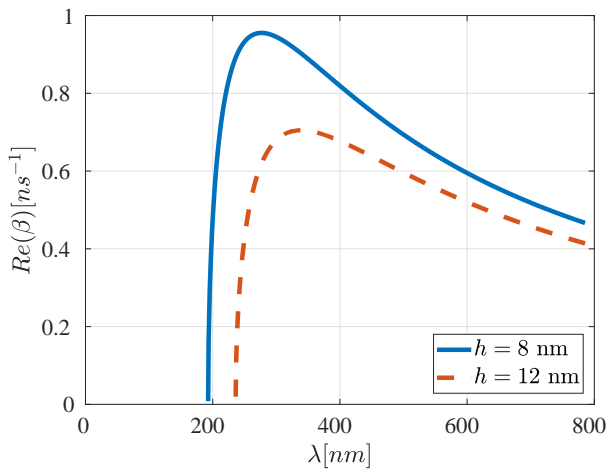}}
    \caption{Growth rate as a function of the wavelength for filament width 
    $w = 185\, $nm. Since the perturbation is large, the growth rate is computed both with $h_0 = 8\, $nm and $h_0 = 12\, $nm (average thickness).} \label{fig:RP_heights} 
\end{figure}

For early times, the initial geometry quickly evolves to a 
cylindrical filament on a substrate.  The basic idea regarding the 
stability of such a filament could be reached by considering the 
Rayleigh-Plateau type of instability of a free standing cylindrical jet.   
Within this model, the growth rate of the perturbations, $\beta$,
is given by~ \cite{rayleigh1878,Eggers1997}
\begin{equation}
\label{eq:RP}
 \beta^2 = \frac{\sigma}{\rho R^3}\left[ kR \left( 1 - k^2 R^2 \right) \frac{I_1 (kR)}{I_0 (kR)}  \right]
\end{equation}
where $R$ is the radius of the jet, and $I_0$ and $I_1$ are the modified Bessel functions. 
Hence the stability of a jet depends on its radius, $R$: the modes, $k$, for which $kR < 1$ 
are unstable and the modes for which $kR > 1$ are stable. Here, $k$ is the wavenumber related to the 
perturbation wavelength, $\lambda$, by $k = 2 \pi/\lambda$. The fastest growing mode corresponds to 
$k_m R \approx 0.7$.  In the present context, $R$ corresponds to the radius of a filament characterized
by the equilibrium contact angle $\theta$, and of the same cross-sectional area as the initial rectangular geometry
of the thickness $h$ and width $w$, i.e.,
\begin{equation}
 \label{eq:RP_radius}
 R = \sqrt{ \frac{h w}{\theta - \cos \theta \sin \theta}}.
\end{equation}
Figure \ref{fig:RP_heights} shows the growth rate for 
a filament.   We consider both the filament thickness without the perturbation, 
$h_0 = 8\, $nm and the average filament thickness including the perturbations, 
$h_0 = 12\, $nm. 
The Rayliegh-Plateau stability curve gives us an approximation for the critical 
wavelength, $\lambda_c \sim 236\, $nm using $h_0 = 12\, $nm; including the presence
of substrate is known to make $\lambda_c$ slightly larger~\cite{Diez2009}.    Our numerical
results show consistently that $\lambda_c$ value is in the range $[240,250]$ nm. 
In what follows, we  show the results 
for two filaments: one with a stable and one with an unstable perturbation. 
We compare the results for simulations with and without the thermal effects. The
temperature is governed by the complete model described in Section~\ref{sec:dns_model}.

\begin{figure*}[ptbh]
    \centering
\begin{subfigure}{0.9\textwidth}
 {\includegraphics[width = \textwidth]{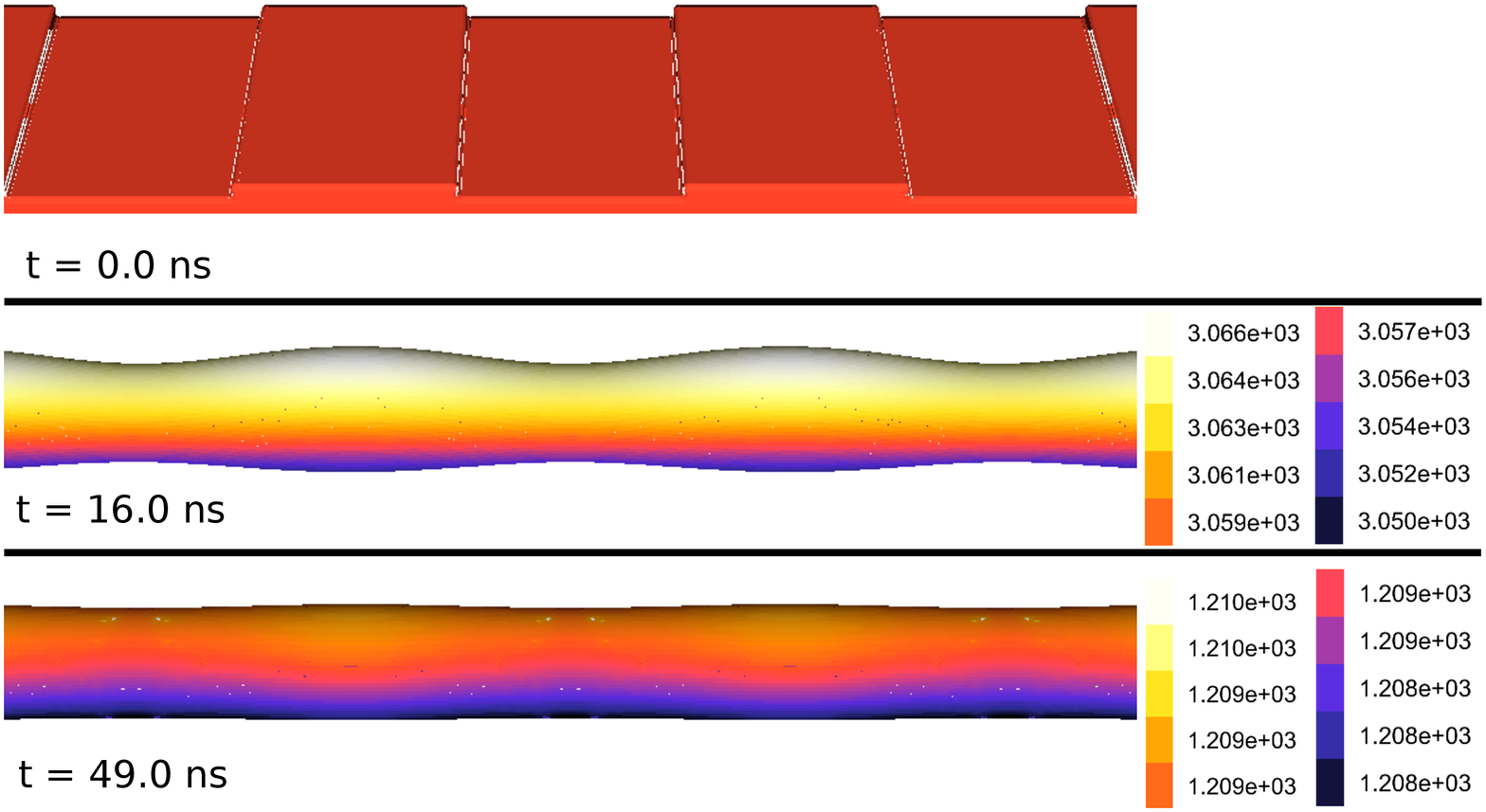}}
   \caption{}
\end{subfigure}\\
\begin{subfigure}{0.7\textwidth}
 {\includegraphics[width = \textwidth]{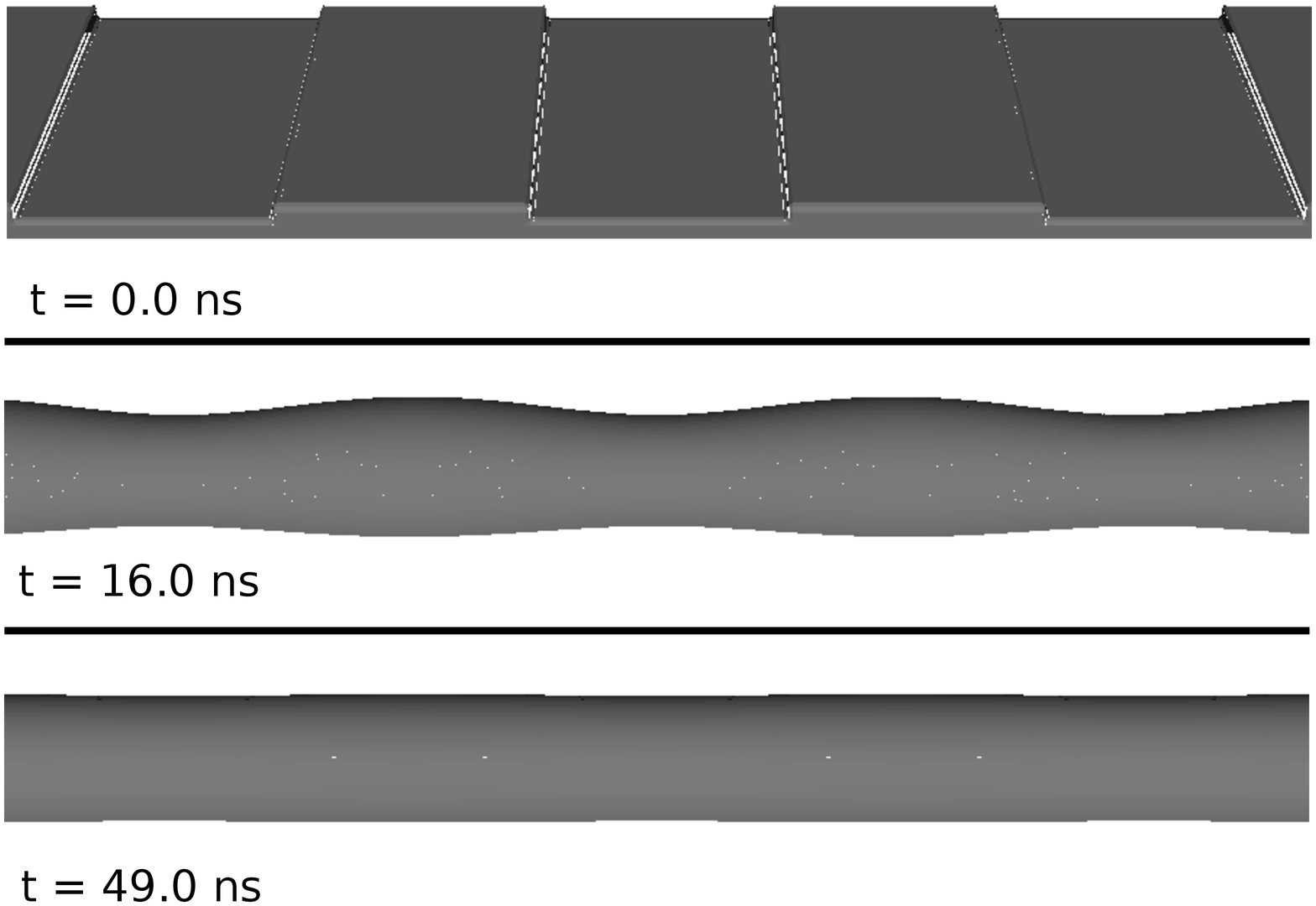}}
  \caption{}
\end{subfigure}\hspace{0.2\textwidth}
    \caption{ Evolution of a stable filament with wavelength $\lambda = 240\, nm$, 
    (a) surface tension dependent on the temperature and
    (b) surface tension fixed at $\sigma_0$. The color in part (a) represents the 
    temperature at the interface in degrees Kelvin.}  
    \label{fig:thermal_filam_027_040}
\end{figure*}

Figure \ref{fig:thermal_filam_027_040} shows the evolution of a linearly stable filament,
with the wavelength of the perturbations close to the critical one. 
The results are similar, independently of whether the surface tension is treated as a constant or temperature dependent.
To understand this result, we note that the filament setup differs in a significant manner from the
the film one: here, a change of the surface tension 
does not change the stability of the filament; it only modifies the growth rate. 
Hence, the thermal variation of the surface tension does not change the qualitative 
behavior.   

\begin{figure*}[ptbh]
    \centering
\begin{subfigure}{0.9\textwidth}
 {\includegraphics[width = \textwidth]{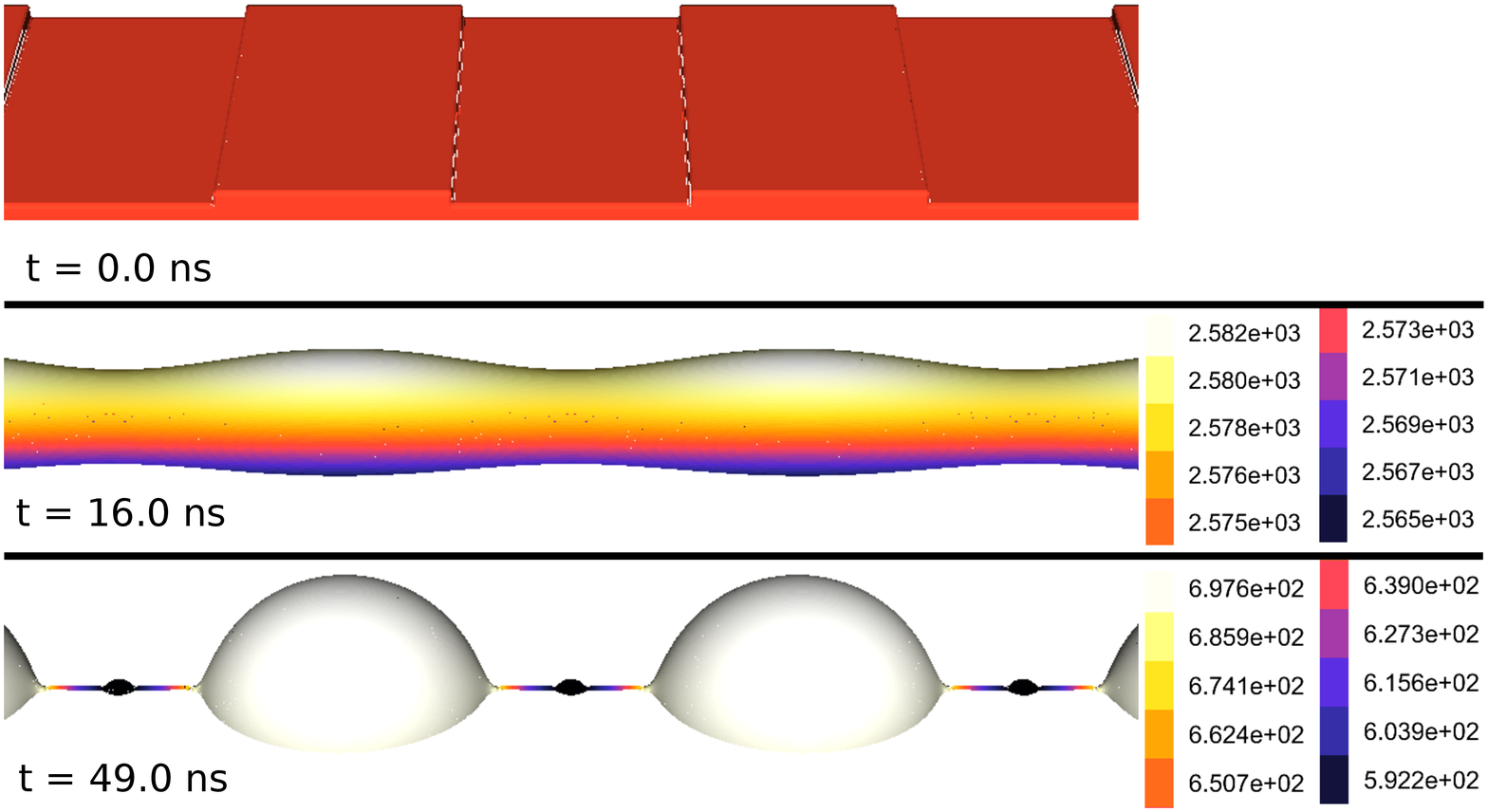}}
   \caption{}
\end{subfigure}\\
\begin{subfigure}{0.7\textwidth}
 {\includegraphics[width = \textwidth]{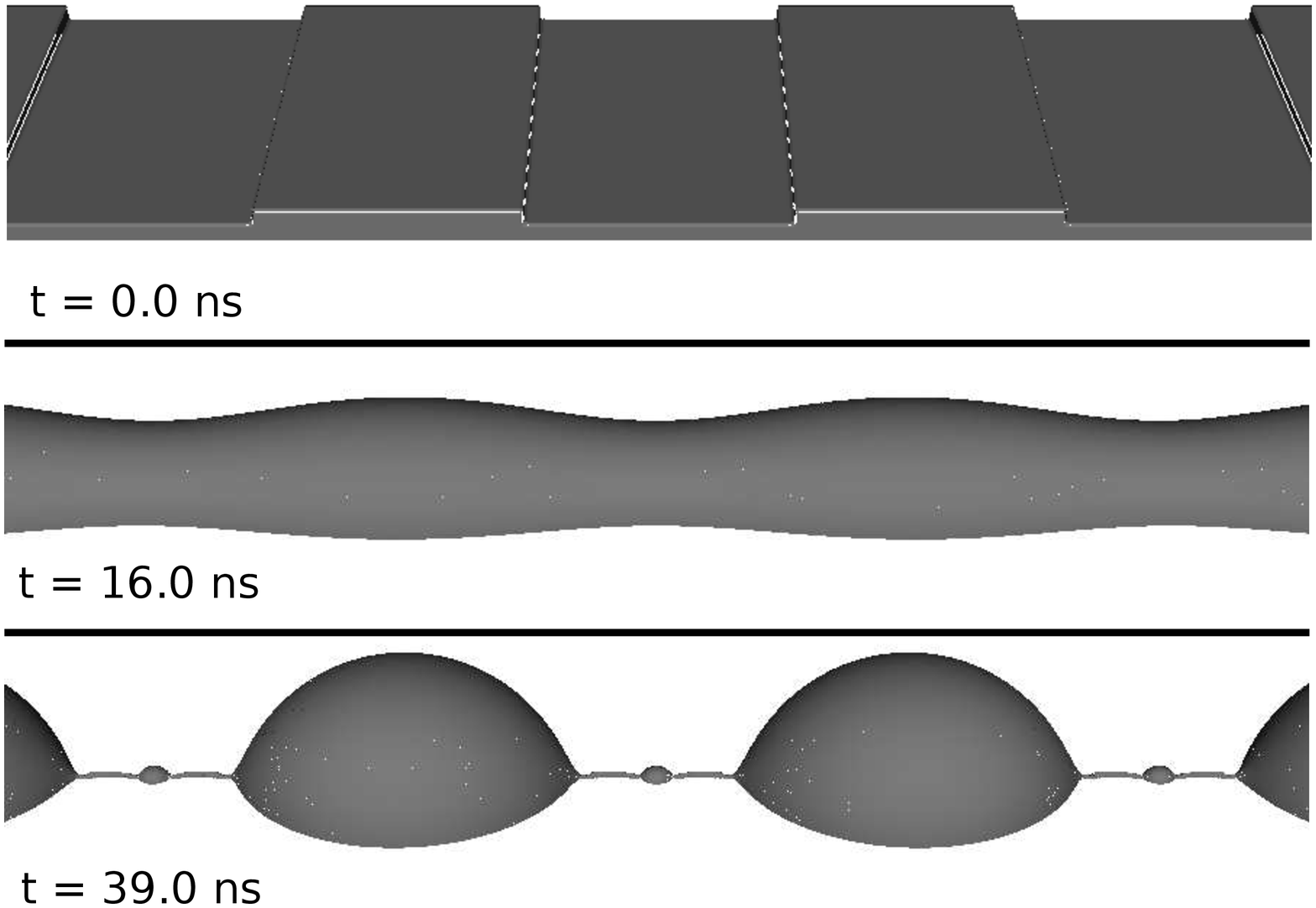}}
  \caption{}
\end{subfigure}\hspace{0.2\textwidth}
    \caption{ Evolution of an unstable filament with wavelength $\lambda = 250\, nm$, 
    (a) surface tension dependent on the temperature and
    (b) surface tension fixed at $\sigma_0$. The color in part (a) represents the 
    temperature at the interface in degrees Kelvin.  Note slower breakup in (a) due to a decrease
    of surface tension with temperature.
        }
    \label{fig:thermal_filam_028_041}
\end{figure*}
Figure \ref{fig:thermal_filam_028_041} shows the evolution of an unstable filament. 
Again, the thermocapillary force does not change the qualitative breakup dynamics. 
However, the breakup with temperature dependent surface tension happens 
about $10\, $ns slower compared to the constant surface tension. 
This is expected, since the increase in the temperature leads to a decrease in the surface tension, which in turn leads to a decrease of the growth rate. 

An obvious question to ask is why {\it concentration} Marangoni effect, discussed in~\cite{hartnett2017} is so much more
prominent compared to the {\it thermal } Maranogni effect discussed here.   There are at least two sources of the 
difference: first, the thermal Marangoni effect is much weaker compared to the concentration one  (one needs the temperature
difference of $\approx 1,600$ degrees  to produce the change of surface tension of nickel corresponding to the difference of 
the surface tensions of nickel and copper considered in~\cite{hartnett2017}, see Tab.~\ref{tab:thermal_params}):
and  second, for the setup considered in the present work, the thermal Marangoni effect is {\it induced}: the filament needs to heat up 
for the  thermal Marangoni effect to be established; in the setup considered in~\cite{hartnett2017}, the concentration Marangoni 
effect is present from the very beginning of the evolution.   We note in passing that the present results also suggest that thermal
Marangoni effect can be safely ignored in two (or multi) metal setting: concentration Marangoni effect is expected to play a 
dominant role.

\subsection{Conclusions}
\label{sec:conclusions}

In this paper we have studied the influence of the thermal effects on the evolution
of thin metal films and filaments. For the films, we have shown that the dynamics of the evolution can change 
due to the surface tension dependence on temperature. Perhaps surprisingly, the influence of
the temperature is not manifested through the Marangoni effect, but through the capillary force (balance of normal stresses). 
In other words, the thermal effects influence the interface evolution 
due to the time-dependent changes of the surface tension during a laser pulse. 

We have the reached the main conclusion outlined in the preceding paragraph by considering 
two models for computing the film temperature. The reduced 1D temperature model (Section~\ref{sec:Trice}) 
is found to overestimate the temperature gradients along the free interface, since the in-plane heat conduction is not considered.
The complete model, based on the numerical computation of the temperature 
(Section~\ref{sec:dns_model}) shows that the temperature gradients along the interface are in fact not strong enough to
influence the breakup of the films. 
The changes in the viscosity during the metal film heating can accelerate the growth 
of the perturbations, leading to a breakup of films that would not break if a constant 
value of viscosity at the melting temperature were used.   This finding is also to a certain degree unexpected 
since it is known (at least within the long wave limit) that viscosity influences only the growth rate, and does not
modify the range of unstable wave numbers.  However, an interplay of the time scales responsible for heating and
for instability development modifies the evolution in a nontrivial manner.   Our results suggest that the fact that 
temperature rises significantly over melting will result in the emergence of shorter length-scales compared to 
the ones expected if the material parameters considered were assumed to be fixed at their values corresponding to the melting 
temperature.    

In the case of filaments, the temperature dependence of surface tension has
only rather minor influence on the qualitative behavior of the breakup of the liquid metal 
filaments. At least for the parameters considered here, the stability of a (single metal) 
filament is not influenced by surface tension variation.  
This is in contrast to the two-metal filaments, considered in~\cite{hartnett2017},  
where concentration dependence of surface tension can qualitatively change the dynamics.

The presented results open new avenues of research.  Some of the questions that one may ask are as follows:
\begin{itemize}
\item
What is the influence of the substrate thickness and thermal properties on the findings reported here?   Presumably
for sufficiently thin substrate, heat diffusion in the in-plane direction may be less important, modifying the results, 
and perhaps bringing the findings of the complete model (that includes the heat diffusion in the in-plane direction)
and the reduced model (that does not include the in-plane heat diffusion) closer together.   
\item 
The result presented here show in some cases oscillatory instability, with the film evolution evolving in a 
non-monotonous manner.  Can one understand the conditions required for such oscillatory evolution more
precisely?
\item
The results of the present paper, combined with the analysis of the propagation of the melting front considered
in a stationary setup~\cite{font2017}, should allow to analyze evolution of metal films on the substrates that 
may go through the phase transitions themselves.  What is the influence of substrate melting on the film 
stability?
\item 
How significant are thermal effects for the evolution and stability of multimetal films and alloys?
\end{itemize}
We expect that the results presented here will serve as a basis for answering some of the outlined questions.

\begin{acknowledgments}
We acknowledge many useful discussions regarding metal films with Ryan Allaire, William Batson III, Linda Cummings, Javier Diez, Francesc Font, 
Jason Fowlkes, Alejandro Gonz\`alez, Kyle Mahady and Philip Rack.   This work was supported by the NSF grant  
No. CBET- 1604351.
\end{acknowledgments}

\appendix

\section{Laser Source Term}
\label{app:laser}

The absorption, reflectance and transmittance of a thin metal film can be computed from Maxwell's equations 
with appropriate boundary conditions. 
The equations are greatly simplified when considering a single film layer on a 
transparent (non-absorbing) substrate. The simplified expressions
for computing reflectance and transmittance given in \citet{Heavens} are
\begin{widetext}
\begin{align}
\label{eq:BookDef1}
    &{R}_1 = \frac{t_{12}^2 + u_{12}^2}{p_{12}^2 + q_{12}^2}, \\
\label{eq:BookDef2}
    &{T}_1 = \frac{n_2}{n_0}\, \frac{((1+g_1)^2 + h_1^2) ((1+g_2)^2 + h_2^2) }{ e^{2 \gamma_1}+ (g_1^2 + h_1^2) (g_2^2 + h_2^2) e^{-2 \gamma_1} + C \cos(2 \gamma_2) + D \sin(2 \gamma_2)} 
\end{align}
\end{widetext}
where the terms in Eqs.~\eqref{eq:BookDef1} and \eqref{eq:BookDef2} are defined as
\begin{align*}
    &\gamma_1 = \frac{ 2 \pi k_1 h}{\lambda_l},  \,\,\,\,\,\, \gamma_2 = \frac{2 \pi n_1 h}{\lambda_l} \\
    &g_1 = \frac{n_0^2 - n_1^2 - k_1^2}{(n_0 + n_1)^2 + k_1^2},    &g_2 = \frac{n_1^2 - n_2^2 + k_1^2}{(n_1 + n_2)^2 + k_1^2} \\
    &h_1 = \frac{2 n_0 k_1}{(n_0 + n_1)^2 + k_1^2},   &h_2 = \frac{-2 n_2 k_1}{(n_1 + n_2)^2 + k_1^2}\\ 
    &C = 2 (g_1 g_2 - h_1 h_2),     &D = 2 (g_1 h_2 + g_2 h_1) \\
    &p_2 = e^{\gamma_1} \cos(\gamma_2),  &p_{12} = p_2 + g_1 t_2 - h_1 u_2\\
    &q_2 = e^{\gamma_1} \sin(\gamma_2),  &q_{12} = q_2 + h_1 t_2 + g_1 u_2\\
\end{align*}
\begin{align*}
    &t_2 = e^{-\gamma_1} (g_2 \cos(\gamma_2) + h_2 \sin(\gamma_2)), \\  &t_{12} = t_2 + g_1 p_2 - h_1 q_2\\
    &u_2 = e^{-\gamma_1} (h_2 \cos(\gamma_2) - g_2 \sin(\gamma_2)),  \\  &u_{12} = u_2 + h_1 p_2 + g_1 q_2
\end{align*}
and $h$ is the metal film thickness, $\lambda_l$ is the wavelength of the incident radiation, $n_0$ is the refractive index of air, $n_1$ and $k_1$ are the metal refractive index and the extinction coefficient respectively, and $n_2$ is the refractive index of the substrate.
\begin{figure}[tbh]
  \begin{center}
    \includegraphics[width = 0.4\textwidth]{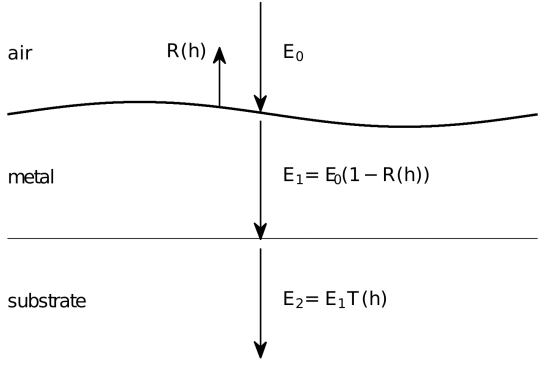}
  \end{center}
  \caption{$E_0$ is the intensity of the incident radiation. $R(h)$ and $T(h)$ 
are the thickness dependent reflectance and transmittance of the metal. }
\label{fig:RTAscheme}
\end{figure}
Figure \ref{fig:RTAscheme} shows a schematic of the laser energy absorption. The incident energy 
$E_0$ is perpendicular to the film surface. One part of the energy, $R(h)$, is reflected at the film 
surface, and the rest of the energy, denoted by $E_1$, penetrates the surface. 
Then, a part of the laser energy, denoted $E_2$ is transmitted through the metal. 
Hence the energy absorbed by the metal film is 
\begin{gather}
 \label{eq:A_trice}
    {A} = E_0 \left[ 1- {T}(h) \right] \left[ 1- {R}(h) \right].
\end{gather}
The expressions for ${R}$ and ${T}$ given in Eqs.~\eqref{eq:BookDef1} and 
\eqref{eq:BookDef2}, can be approximated by simpler functions, as it is done by 
Trice et al.~\cite{trice_prb07}
\begin{gather}
 \label{eq:RT_Trice}
    T_2(h) = e^{-\alpha_a h}, \,\,\,\,\,\,\,\,\,\, R_2(h) = r_0 \left( 1 - e^{-a_r h}  \right), 
\end{gather}
where $ \alpha_a = 4 \pi k_1/\lambda_l$, and $r_0$ and $a_r$ can be found by fitting $T_2$ and $R_2$ to  Eqs.~\eqref{eq:BookDef1} and \eqref{eq:BookDef2}.
\begin{figure*}[tbh]
\centering
    \includegraphics[width = 0.8\textwidth]{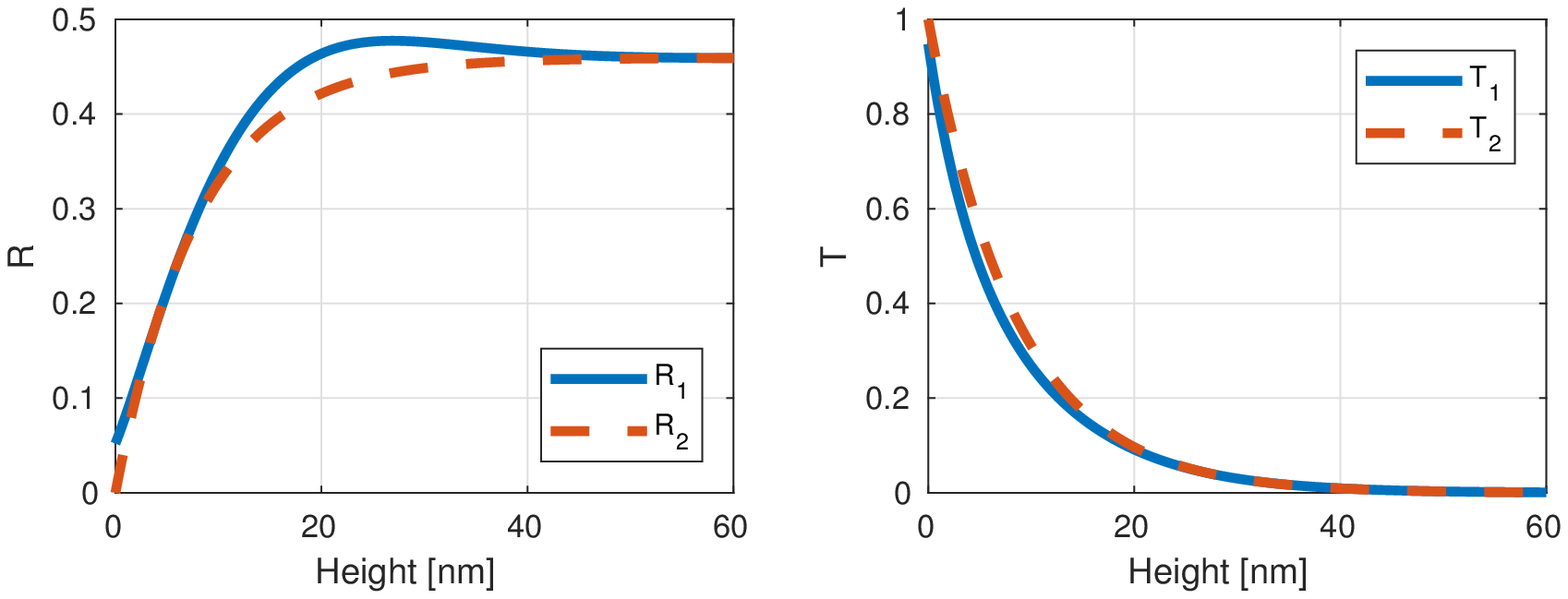}
  \caption{The comparison of the reflectance and transmittance given
by the Eqs.~\eqref{eq:BookDef1} and \eqref{eq:BookDef2} and Eq.~\eqref{eq:RT_Trice}.}
\label{fig:RTA_Nickel}
\end{figure*}
Figure \ref{fig:RTA_Nickel} shows the comparison of the reflectance and transmittance given
by Eqs.~\eqref{eq:BookDef1} and \eqref{eq:BookDef2} and Eq.~\eqref{eq:RT_Trice}. The parameters used 
here are $\lambda_l = 248\, $nm~\cite{hartnett2017}, $ n_0 = 1$, $ n_1 = 1.7167 $, $ k_1 = 2.3067$, $ n_2 = 1.59157$~\cite{johnson74}, $R_0 = 0.4594$ and
$ a_r^{-1} = 8\, $nm.

In Section~\ref{sec:temp_models} for simplicity we use Eqs.~(\ref{eq:RT_Trice}) for computing the absorption of 
the laser energy by a metal film.
\vspace{0.2in}

\section{The Temperature Solution of the Reduced Model in the Limit of Small Film Thickness}
\label{app:trice_small_h}
The solution to the reduced temperature model given in Section~\ref{sec:Trice} contains
integrals that pose numerical difficulties for small film thicknesses. 
Here we give the expressions that can be used for computing the 
temperature of the metal film, $T^*_m$, and the gradient of the temperature with 
respect to the film thickness, $\partial T_m^*/\partial h$, to alleviate those 
difficulties. 
In the limit of small film thickness, $T^*_m$ and $\partial T_m^*/\partial h$, can be expanded using asymptotic series as
\begin{widetext}
\begin{equation}
    T^*_m \left( h \to 0, t \right) = T_0 +
     S^*\, e^{-\frac{t_p^2 }{2 \sigma_{tp}^2}} \int_0^t \operatorname{exp}\left( - \frac{\left( t-u \right)^2}{2 \sigma_{tp}^2} +  \right. \\ \left.
     \frac{t_p}{\sigma_{tp}^2}\left( t- u \right)\right) \left[ \frac{h}{C_K \sqrt{\pi\, u}} - \frac{ h^3}{2 \sqrt{\pi}\left( C_K \sqrt{u}\right)^3} + ... \right] \, du,
\end{equation}
\begin{multline}
    \frac{\partial T}{\partial h} \left( h \to 0, t \right) = 
    \frac{\partial S^*}{\partial h} \, e^{-\frac{t_p^2 }{2 \sigma_{tp}^2}} \int_0^t \operatorname{exp}\left( - \frac{\left( t-u \right)^2}{2 \sigma_{tp}^2} + \frac{t_p}{\sigma_{tp}^2}\left( t- u \right)\right) \left[ \frac{h}{C_K \sqrt{\pi\, u}} - \frac{ h^3}{2 \sqrt{\pi}\left( C_K \sqrt{u}\right)^3} + ... \right] \, du + \\ 
     S^*\, e^{-\frac{t_p^2 }{2 \sigma_{tp}^2}} \int_0^t \operatorname{exp}\left( - \frac{\left( t-u \right)^2}{2 \sigma_{tp}^2} + \frac{t_p}{\sigma_{tp}^2}\left( t- u \right)\right) \left[ \frac{1}{C_K \sqrt{\pi\, u}} - \frac{ 3 \,h^2}{2 \sqrt{\pi}\left( C_K \sqrt{u}\right)^3} + ... \right] \, du.
\end{multline}
where
\begin{equation}
    S^*\left( h \to 0, t \right) =  \alpha_a \left[1   -  \left(\frac{\alpha_a}{2} + a_r r_0\right) h + 
         \frac{1}{6} \left(\alpha_a^2 + 3 \alpha_a a_r r_0 + 3  a_r^2 r_0 \right) h^2  + ...     \right], \,\,\,\,\, \text{ as } h \to 0.
\end{equation}
\end{widetext}
Hence, the integrals in Eqs.~\eqref{eq:triceIntegral} and \eqref{eq:dT_dh} 
are convergent as $h \to 0$. In our simulations, we use the expression given here for 
small film thicknesses, since the direct evaluation of the integrals in Eqs.~\eqref{eq:triceIntegral} and \eqref{eq:dT_dh} is numerically difficult.

\section{Analytical Temperature Solution}
\label{app:exact}

Here we provide the details of the analytical temperature solution in the 
fluid-substrate domain specified in Section~\ref{sec:exact}. 
Note that in order to simplify the presentation, we change the notation for the domain boundaries compared to Section~\ref{sec:temp_models}, 
and denote the bottom of the 
substrate as $y = 0$, the fluid-substrate interface as $y = a$, and the fluid-air interface as 
$y = b$.
The temperature in the fluid, $T_m$, and the temperature in the substrate, $T_s$, satisfy the diffusion equation
\begin{gather}
\label{eq:full_heat}
    \frac{\partial T_s}{\partial t} = \mathcal{D}_s \frac{\partial^2 T_s}{\partial y^2} 
    \,\,\,\,\,\,\,\,\,\,\, \text{in}\,\,\,\,\,\,  0 < y < a \\
    \frac{\partial T_m}{\partial t} = \mathcal{D}_m \frac{\partial^2 T_m}{\partial y^2}  + S \left( y, t \right) \,\,\,\,\,\,\,\,\,\,\, \text{in}\,\,\,\,\,\,  a < y < b 
\end{gather}
where
\begin{gather}
    \mathcal{D}_s = \frac{k_s}{\rho_s C_{eff_s}}, \,\,\,\,\,\,\,\,     \mathcal{D}_m = \frac{k_m}{\rho_m C_{eff_m}}, \nonumber
\end{gather}
along with the boundary conditions
\begin{align}
\label{eq:boundary_a_0}
    T_s  &= T_0    \,\,\,\,\,\,\,\, &
            &\text{at %the bottom of the substrate 
            }y = 0, \\
\label{eq:boundary_a_1}
    T_s &= T_m \,\,\,\,\,\,\,\,  & 
            &\text{at %the fluid-substrate interface 
            } y = a, \\
\label{eq:boundary_a_3}           
    k_s \frac{\partial T_s}{\partial y} &= k_m \frac{\partial T_m}{\partial y} \,\,\,\,\,\,\,\,  & 
            &\text{at %the fluid-substrate interface 
            }  y = a,\\ 
\label{eq:boundary_a_4}
    k_m \frac{\partial T_m}{\partial y} &= 0 \,\,\,\,\,\,\,\,  &
            &\text{at %the fluid-air interface 
            } y = b.
\end{align}
The source term, $S(y,t)$ is given by Eq.~\eqref{eq:source}. The solution to the above equations
can be written compactly in terms of Green's functions as given in Eqs.~\eqref{eq:exact_m} and \eqref{eq:exact_s}, 
where
\begin{align}
    G_{i,j} \left( y, t; \xi, \tau \right) &= \sum_{n = 1}^{\infty} e^{-\beta^2_n \left(t - \tau \right)} \frac{1}{N_n} \frac{k_j}{\alpha_j} \psi_{i,n} \left(y \right) \psi_{j,n} \left(\xi \right) \\
    N_n &= \frac{k_s}{\mathcal{D}_s} \int_0^a \psi^2_{1,n} \text{d} \xi + \frac{k_m}{\mathcal{D}_m} \int_a^b \psi^2_{2,n} \text{d} \xi\ .
\end{align}
Here $\psi_{i,n}$ and $\beta_n$ are eigenfunctions and eigenvalues computed using separation of variables, and 
\begin{align}
\label{eq:e_funcs}
    &\psi_{i,n} = A_{i,n} \Phi_{i,n} (y) + B_{i,n} \Theta_{i,n} (y) \,\,\,\, \text{ in }\,\,\,\, y_i < y < y_{i+1}, \\
    &\Phi_{i,n}(y) = \sin{\left( \frac{\beta_n}{\sqrt{\alpha_i}} y \right)}, \\
    &\Theta_{i,n} = \cos{\left(\frac{\beta_n}{\sqrt{\alpha_i}} y \right)},
\end{align}
where $y_0 = 0$, $y_1 = a$ and $y_2 = b$. In order to simplify the notation, let 
\begin{equation}
    \gamma = \frac{a \beta_n}{\sqrt{\mathcal{D}_s}}, \,\,\,\,
    \eta = \frac{b \beta_n}{\sqrt{\mathcal{D}_m}},\,\,\,\,
    \mathcal{K} = \frac{k_s}{k_m} \sqrt{ \frac{\mathcal{D}_m}{\mathcal{D}_s}}.
\end{equation}
The eigenfunctions $\psi_{i,n}$ satisfy the boundary conditions in Eqs.~\eqref{eq:boundary_a_0}~-~\eqref{eq:boundary_a_4}. Hence, it follows
\begin{align}
    &\psi_{1,n} = 0 \text{ at } y = 0  \nonumber \\
      &\rightarrow B_{1,n} = 0, \, A_{1,n} = 1 \text{ without loss of generality,}\\
\label{eq:continuity_bc_heat}
    &\psi_{1,n} = \psi_{2,n} \text{ at } y = a \nonumber \\
      &\rightarrow \sin{\gamma} = 
      A_{2,n} \sin{\left( \frac{a}{b} \eta \right)} + B_{2,n} \cos{ \left( \frac{a}{b} \eta \right) },  \\
\label{eq:flux_bc_heat}
    & \frac{k_s}{k_m} \frac{ \partial \psi_{1,n}}{\partial y} =  \frac{ \partial \psi_{2,n}}{\partial y} \text{ at } y = a \nonumber \\
      &\rightarrow \mathcal{K} \cos{ \gamma } = 
       A_{2,n} \cos{ \left( \frac{a}{b} \eta \right) } - B_{2,n} \sin{ \left( \frac{a}{b} \eta \right)},   \\  
\label{eq:zero_flux_bc}
    &\frac{\partial \psi_2}{\partial y} = 0 \text{ at } y = b \nonumber \\
      &\rightarrow  A_{2,n} \cos{ \eta } - B_{2,n} \sin{ \eta} = 0.
\end{align}
We can solve for the coefficients $A_{2,n}$ and $B_{2,n}$ using Eqs.~\eqref{eq:continuity_bc_heat} and \eqref{eq:flux_bc_heat}
\begin{align}
  A_{2,n} &= \frac{1}{\Delta} \left[ - \sin{\gamma} \sin {\left( \frac{a}{b}\eta \right)}  - 
      \mathcal{K} \cos{\gamma} \cos {\left( \frac{a}{b}\eta \right)} \right], \\
  B_{2,n} &= \frac{1}{\Delta} \left[ \mathcal{K} 
      \cos{\gamma} \sin {\left( \frac{a}{b}\eta \right)}  - 
      \sin{\gamma} \cos {\left( \frac{a}{b}\eta \right)} \right],
\end{align}
where 
\begin{equation}
    \Delta = - \sin^2{\left( \frac{a}{b} \eta \right)} - \cos^2{\left( \frac{a}{b} \eta \right)} = -1.
\end{equation}
In order to have a solution, we require vanishing determinant of the system of Eqs.~\eqref{eq:continuity_bc_heat}~-~\eqref{eq:zero_flux_bc}
% , which allows solving for the eigenvalues, $\beta_n$
\[ 
\begin{vmatrix}
\sin{\gamma} & -\sin{ \left( \frac{a}{b}\eta \right) } & -\cos{ \left( \frac{a}{b}\eta \right) } \\ 
\mathcal{K}\cos{\gamma} & -\cos{ \left( \frac{a}{b}\eta \right) } & -\sin{ \left( \frac{a}{b}\eta \right) } \\ 
0            & -\cos{\eta}                           & -\sin{\eta} \\ 
\end{vmatrix} = 0 . \]
The equation above leads to the following equation for the eigenvalues, $\beta_n$
\begin{equation}
\tan{\frac{a \beta_n}{\sqrt{\mathcal{D}_s}} } \tan{ \left(\frac{\beta_n}{\sqrt{\mathcal{D}_m}} \left( b - a \right)  \right)} = \mathcal{K},
\end{equation}
which can be solved numerically. 

\begin{figure*}[htb]
\centering
  \begin{subfigure}{0.40\textwidth}
    \includegraphics[width = \textwidth]{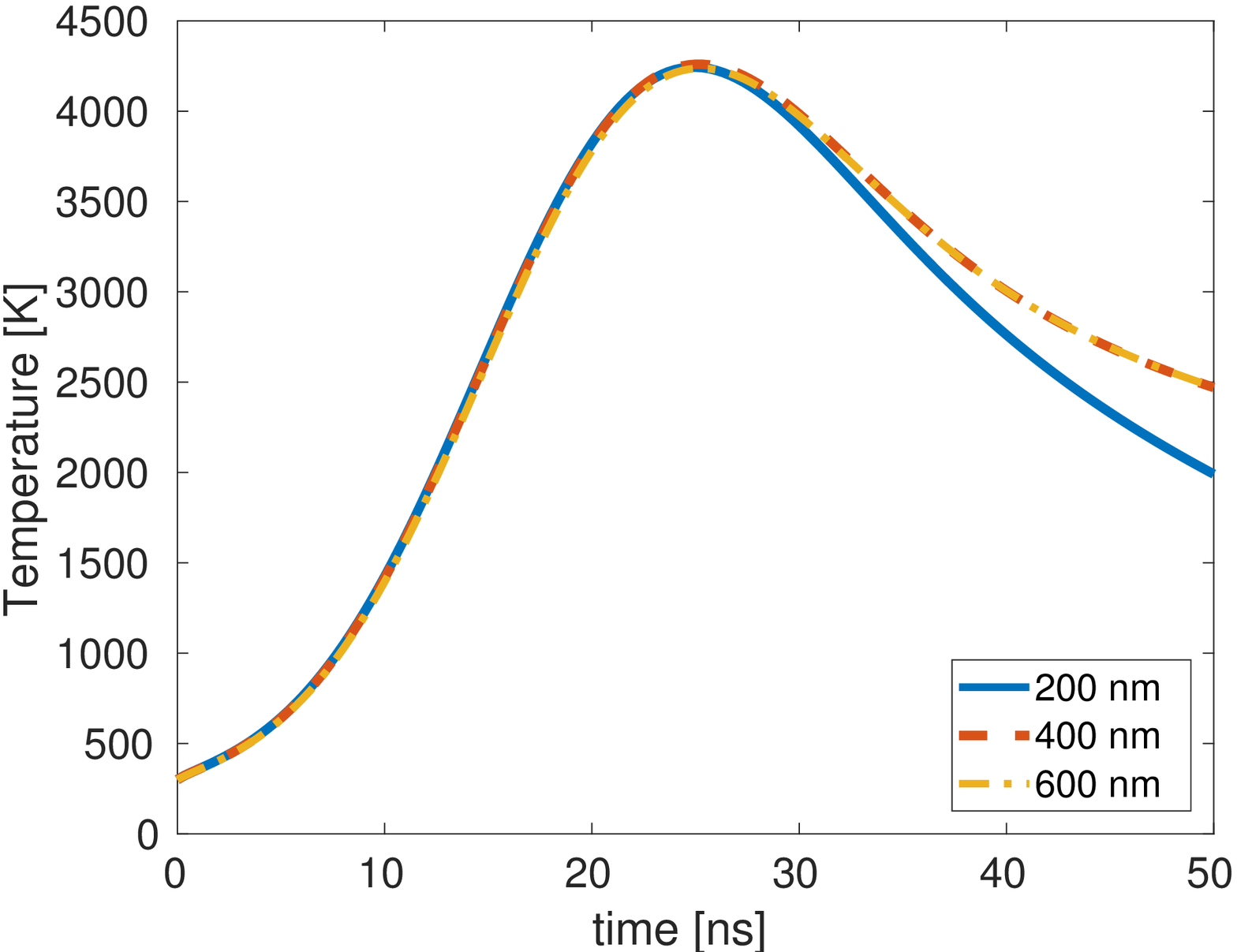}
    \caption{} 
  \end{subfigure} \hspace{0.1in}
  \begin{subfigure}{0.40\textwidth}
    \includegraphics[width = \textwidth]{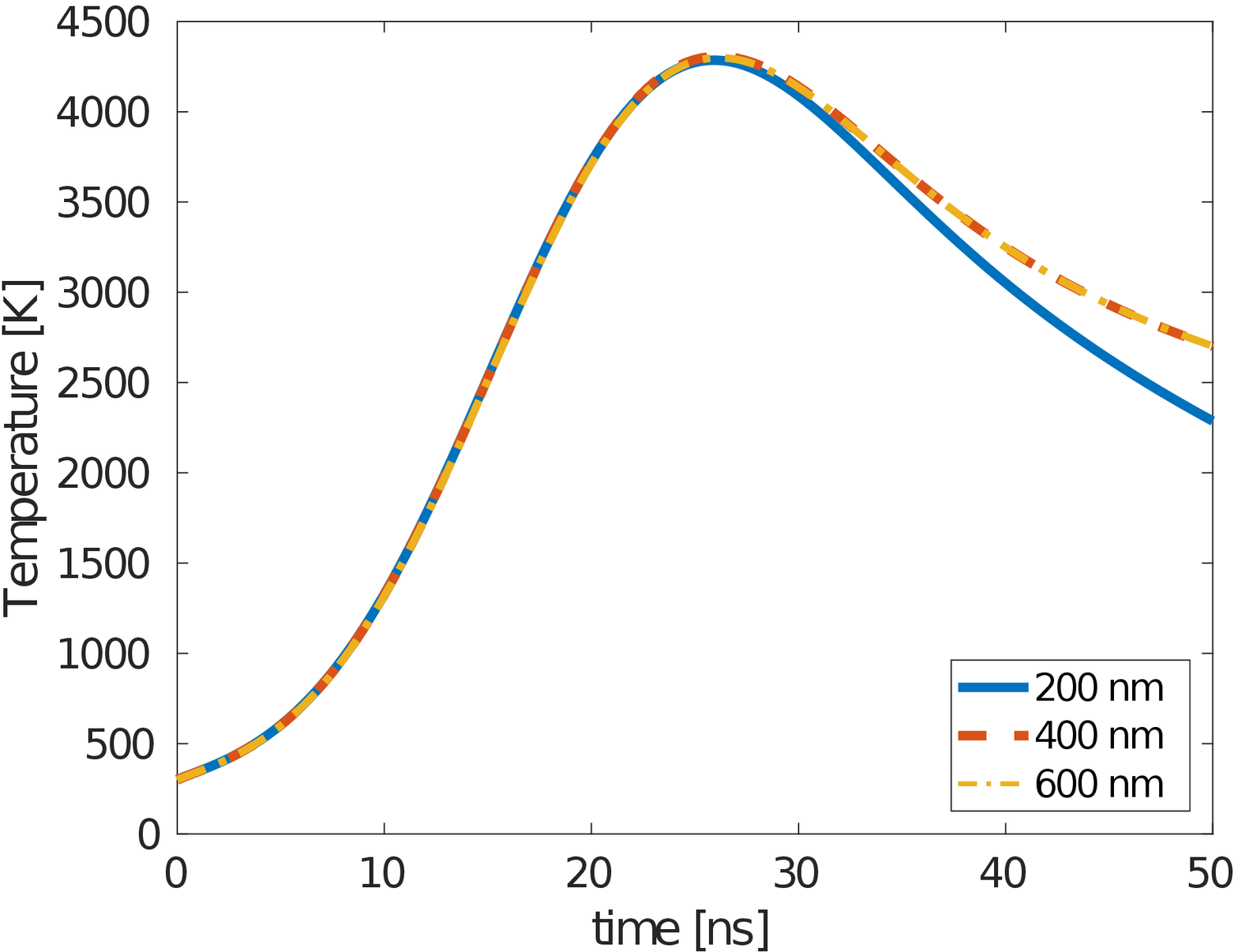}
  \caption{}
  \end{subfigure}
  \vspace{-0.1in}
  \caption{Convergence of the analytical solution with increased substrate depth for
  film thickness of (a) $h_0 = 10\, nm$ and (b) $h_0 = 20\, nm$. }
  \label{fig:analytical_subs_conv}
\end{figure*}
Figure \ref{fig:analytical_subs_conv} shows the average temperature in the metal 
as a function of time for different substrate thicknesses. We see that the solution converges 
as the substrate thickness increases. Hence, for a substrate thick enough, the solution is 
equivalent to the one obtained in the setup involving semi-infinite substrate. This result is used in 
Section~\ref{sec:temp_models} to justify comparing the temperature obtained using the reduced model 
and semi-infinite substrate, with the one obtained by using the complete model and finite substrate thickness.

\section{Newton's Law of Cooling}

We show that replacing the continuity of temperature boundary condition at the 
fluid-substrate interface with Newton's law of cooling yields an equivalent solution as long 
as the heat transfer coefficient is large enough. 

We replace the boundary condition \eqref{eq:boundary_a_1} by
\begin{equation}
 \label{eq:boundary_a_2_newt}
    -k_s \frac{\partial T_s}{\partial y} = \alpha \left( T_s - T_m\right) \,\,\,\,\,\,\,\,
            \text{at~}  y = a. 
\end{equation}
(the notation is the same as in the preceding appendix).  
Then, the eigenfunctions of the same form as given in Eq.~\eqref{eq:e_funcs} satisfy the 
boundary conditions \eqref{eq:boundary_a_0}, \eqref{eq:boundary_a_3}, \eqref{eq:boundary_a_4}
and \eqref{eq:boundary_a_2_newt}.  Hence, it follows
\begin{align}
    & \psi_{1,n} = 0 \text{ at } y = 0  \nonumber \\
      &\rightarrow B_{1,n} = 1, \, A_{1,n} = 1 \text{ without loss of generality,} \\
    & -k_s \frac{\partial \psi_{1,n}}{\partial y} = \alpha \left( \psi_{i,n} - 
        \psi_{2,n} \right) \text{ at } y = a  & \,\, & \nonumber \\
\label{eq:cooling_bc}
    & \rightarrow -k_s\cos{\gamma} = 
       \alpha \sin{\gamma}  -\alpha A_{2,n} \sin{\left( \frac{a}{b} \eta \right)} - 
       \alpha B_{2,n} \cos{ \left( \frac{a}{b} \eta \right) }, \\
\label{eq:newton_flux_bc}
    & k_s \frac{ \partial \psi_{1,n}}{\partial y} = 
      k_m \frac{ \partial \psi_{2,n}}{\partial y} \text{ at } y = a   \nonumber \\
      &\rightarrow \mathcal{K} \cos{ \gamma } = A_{2,n} \cos{ \left( \frac{a}{b} \eta \right) } - 
                                     B_{2,n} \sin{ \left( \frac{a}{b} \eta \right)}, \\
\label{eq:cooling_bc_2}
    & \frac{\partial \psi_2}{\partial y} = 0 \text{ at } y = b   \nonumber \\
     &\rightarrow  A_{2,n} \cos{ \eta } - B_{2,n} \sin{ \eta} = 0.
\end{align}
From the Eqs.~\eqref{eq:cooling_bc} and \eqref{eq:newton_flux_bc}, we can solve for the 
coefficients $A_{2,n}$ and $B_{2,n}$
\begin{align}
  A_{2,n} &= \frac{1}{\Delta} \left[ 
    \left( - H \cos{\gamma} -\sin{\gamma} \right) \sin {\left( \frac{a}{b}\eta \right)} -
    \mathcal{K} \cos{\gamma} \cos {\left( \frac{a}{b}\eta \right)} \right], \\
  B_{2,n} &= \frac{1}{\Delta} \left[ \mathcal{K} 
      \cos{\gamma} \sin {\left( \frac{a}{b}\eta \right)}  + 
      \left( -H \cos{\gamma} - \sin{\gamma} \right) \cos {\left( \frac{a}{b}\eta \right)} \right],
\end{align}
where
\begin{equation}
H = \frac{k_s \beta_n}{\alpha \sqrt{\mathcal{D}_s}}.
\end{equation}
The condition for existence of a solution is vanishing determinant as follows
\[ 
\begin{vmatrix}
-H \cos{\gamma} - \sin{\gamma} & -\sin{ \left( \frac{a}{b}\eta \right) } & -\cos{ \left( \frac{a}{b}\eta \right) } \\ 
\mathcal{K}\cos{\gamma} & -\cos{ \left( \frac{a}{b}\eta \right) } & -\sin{ \left( \frac{a}{b}\eta \right) } \\ 
0            & -\cos{\eta}                           & -\sin{\eta} \\ 
\end{vmatrix} = 0.  \]
The equation satisfied by the eigenvalues $\beta_n$ is now 

\begin{equation}
\left( H + \tan{\frac{a \beta_n}{\sqrt{\mathcal{D}_s}}} \right) \tan{ \left( \frac{ \beta_n}{\sqrt{\mathcal{D}_m}}\left( b - a \right)  \right) }  = \mathcal{K} 
\end{equation}
which can be solved numerically. 

\begin{figure*}[htb]
\centering
  \begin{subfigure}{0.40\textwidth}
    \includegraphics[width = \textwidth]{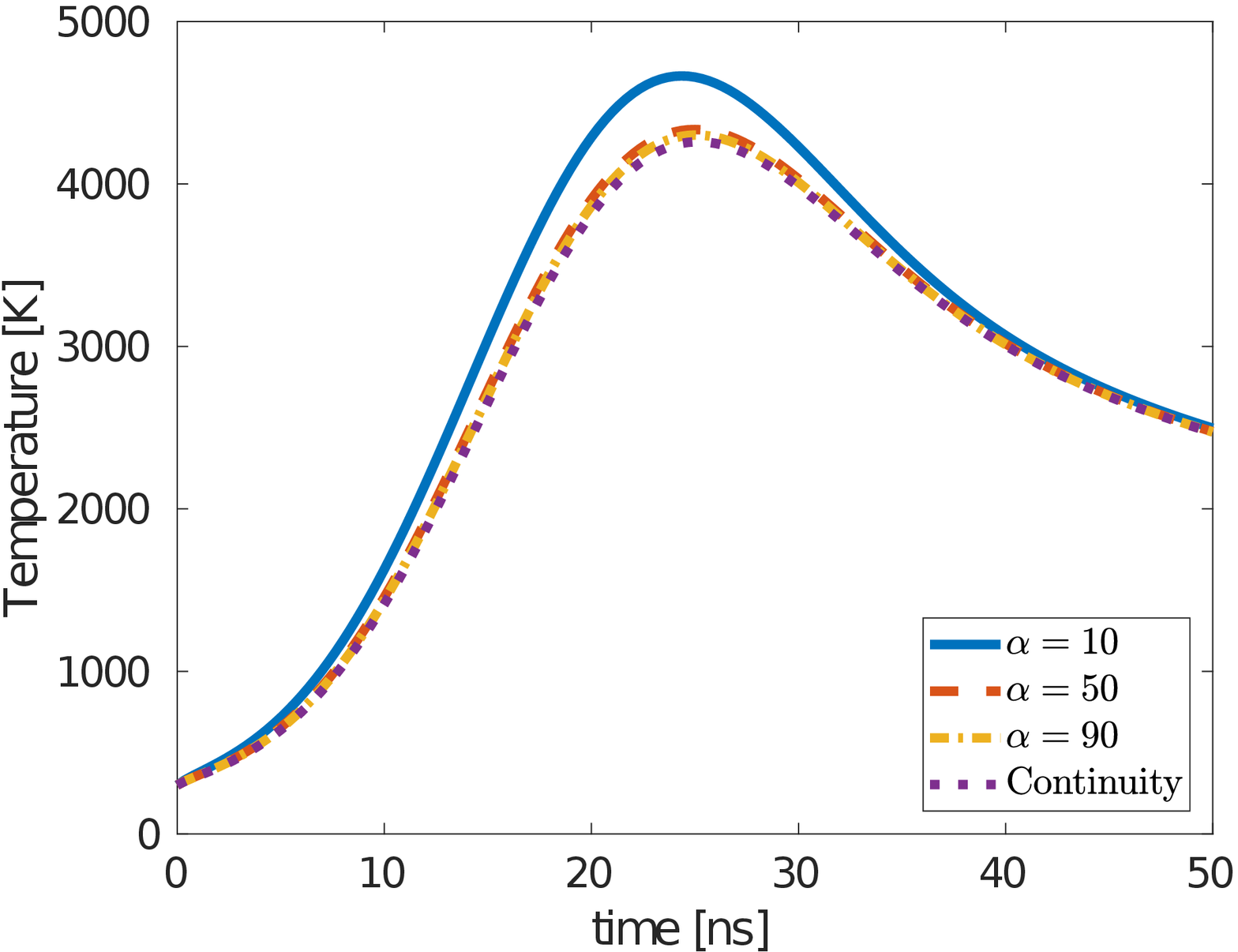}
    \caption{} 
  \end{subfigure} \hspace{0.1in}
  \begin{subfigure}{0.40\textwidth}
    \includegraphics[width = \textwidth]{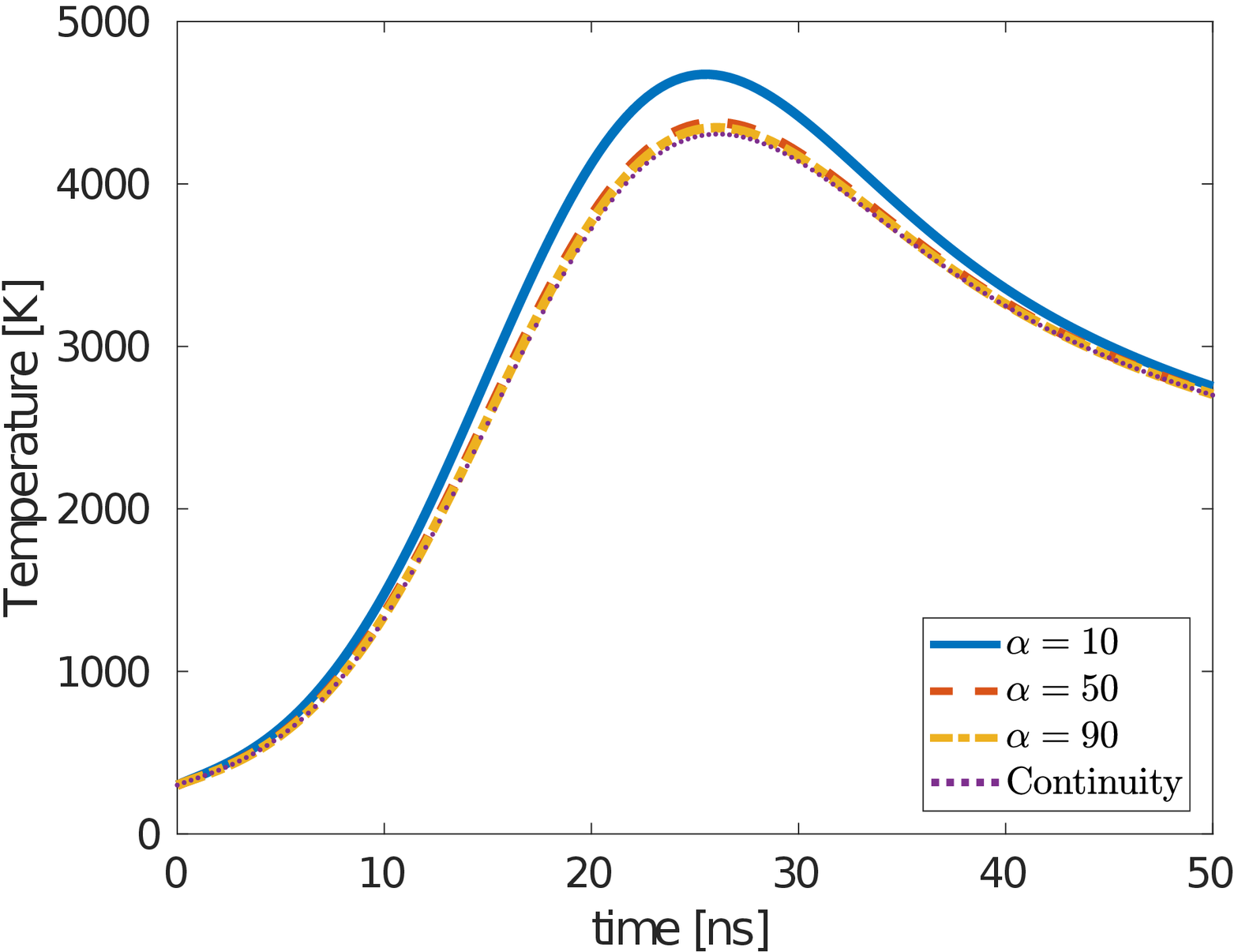}
  \caption{}
  \end{subfigure}
  \vspace{-0.1in}
  \caption{Convergence of the analytical solution with increased $\alpha$ for
  film thickness of (a) $h_0 = 10\, nm$ and (b) $h_0 = 20\, nm$. }
  \label{fig:analytical_alpha_conv}
\end{figure*}
Figure \ref{fig:analytical_alpha_conv} shows the average temperature in the metal 
as a function of time for different $\alpha$. We see that the solution with 
Newton's cooling law converges to the solution with continuity of temperature
for large $\alpha$. This result is used in 
Section~\ref{sec:temp_models} to justify comparing the reduced model that implements continuity of temperature with the complete model that uses Newton's law of cooling.

\end{document}